\documentclass[11pt]{article}
\usepackage[dvips]{graphicx}
\usepackage{graphics}
\usepackage{graphicx}
\usepackage{amssymb}
\usepackage{amsmath}
\usepackage{amssymb}
\usepackage{overcite}
\usepackage{epsf}
\input{epsf}
\numberwithin{equation}{section}
\begin{document}
\begin{titlepage}
\title{\bf Electronic Transport in Metallic Systems and Generalized Kinetic Equations\thanks{International Journal of Modern Physics B (IJMPB), Volume: 25, Issue: 23-24
(2011)  p.3071-3183 }}  
\author{A. L. Kuzemsky 
\\
{\it Bogoliubov Laboratory of Theoretical Physics,} \\
{\it  Joint Institute for Nuclear Research,}\\
{\it 141980 Dubna, Moscow Region, Russia.}\\
{\it E-mail:kuzemsky@theor.jinr.ru} \\
{\it http://theor.jinr.ru/\symbol{126}kuzemsky}}
\date{}
\maketitle
\begin{abstract}
This paper reviews some selected approaches to the description of transport properties,
mainly electroconductivity, in crystalline and disordered metallic systems. A detailed
qualitative theoretical formulation of the electron transport processes in metallic systems
within a model approach is given. Generalized kinetic equations which were derived
by the method of the nonequilibrium statistical operator are used. Tight-binding picture
and modified tight-binding approximation (MTBA) were used for describing the
electron subsystem and the electron-lattice interaction correspondingly. The low- and
high-temperature behavior of the resistivity was discussed in detail. The main objects
of discussion are nonmagnetic (or paramagnetic) transition metals and their disordered
alloys. The choice of topics and the emphasis on concepts and model approach makes it
a good method for a better understanding of the electrical conductivity of the transition
metals and their disordered binary substitutional alloys, but the formalism developed
can be applied (with suitable modification), in principle, to other systems. The approach
we used and the results obtained complements the existent theories of the electrical conductivity
in metallic systems. The present study extends the standard theoretical format
and calculation procedures in the theories of electron transport in solids.
\vspace{1cm}

%

%
%
\end{abstract}
\end{titlepage}
\newpage
\tableofcontents
\newpage
%

%
\section{Introduction}
%
Transport properties of matter constitute the transport of charge, mass, spin, energy and 
momentum~\cite{mott,zi,sol04,tri04,duan05,gold10,maek,dietl}.
It has not been our aim to discuss all the aspects of the charge and thermal transport in metals.
We are concerned in the present work mainly with some selected approaches to the problem  of electric charge 
transport (mainly electroconductivity) in crystalline and disordered metallic systems. Only the fundamentals of the subject are treated.
In the present work we aim to obtain a better understanding of  the electrical conductivity
of the transition metals and their disordered binary substitutional alloys both by themselves 
and in relationship to each other within the statistical mechanical approach. Thus our consideration will concentrate
on the derivation of generalized kinetic equations suited for the relevant models of metallic systems.\\  
The problem of the electronic transport in solids is an interesting and actual part of the physics of 
condensed matter~\cite{ol,pi,mea,sm,slat,bla,zim69,pear,ber,ros,smi,enz,hum,dug,lund,ger,mi,dive}. It includes the transport of 
charge and heat in crystalline and disordered metallic conductors of various nature. Transport of charge is connected with  an electric current. Transport of heat has many aspects,
main of which is the heat conduction. Other important aspects are the thermoelectric effects. The effect,
termed Seebeck effect, consists of the occurrence of a potential difference in a circuit composed of two distinct
metals at different temperatures. Since the earlier seminal attempts to construct  the  quantum theory of the
electrical, thermal~\cite{som,mres,bar,pei29} and thermoelectric and thermomagnetic transport phenomena~\cite{bloh}, there is a great 
interest in the calculation of transport coefficients in solids in order to explain the experimental results as 
well as to get information on the microscopic structure of  materials~\cite{sut,ha,sing,mar04}. \\
A number of physical effects enter the theory 
of quantum transport processes in solids at various density of carriers and temperature regions. A variety of theoretical
models has been proposed to describe these 
effects~\cite{mott,zi,sol04,tri04,duan05,gold10,ol,pi,mea,sm,slat,bla,pear,ber,ros,smi,enz,hum,dug,ger,mi,fu07,lei08,cjac10,pras80,matt94,kun09}. 
Theory of the electrical and heat conductivities of crystalline and disordered
metals  and semiconductors have been developed  by many authors during last 
decades~\cite{mott,zi,sol04,tri04,duan05,gold10,enz,fu07,lei08,cjac10,pras80,matt94,kun09}.  There 
exist a lot of 
theoretical methods for the calculation of transport coefficients~\cite{ros,enz,fu07,lei08,cjac10,zub,zub81,lib,kuz07,vliet08}, as a rule having a fairly restricted
range of validity and applicability. In the present work the description of the electronic and some aspects of 
heat transport  in metallic systems are briefly reviewed, and the theoretical approaches to the calculation of the resistance 
at low and high temperature are surveyed.  As a basic tool we use the method of the nonequilibrium statistical operator~\cite{zub,zub81}(NSO).  
It   provides a useful and compact  description of the transport processes. 
Calculation of  transport coefficients  within NSO approach~\cite{zub} was presented and discussed in the 
author's work~\cite{kuz07}. The present paper can be considered as the second part of the review article~\cite{kuz07}.
The close
related works on the study of electronic transport in metals are briefly summarized in the present work. 
It should be emphasized that the choice of generalized kinetic equations among all other methods of the theory of transport in
metals is related with its efficiency and compact form. They are an alternative (or complementary) tool for studying of transport processes,
which complement other existing methods.\\
Due to the lack of space  many interesting and actual topics must be omitted. An  important and extensive problem
of thermoelectricity was mentioned very briefly; thus it has not been possible to do justice to all the available
theoretical and experimental results of great interest. The thermoelectric and transport properties of the layered high-$T_{c}$ cuprates were
reviewed by us already in the extended review article~\cite{kuz03}.\\
Another interesting aspect of transport in solids which we did not touched is the spin transport~\cite{maek,dietl}.
 The spin degree of freedom of charged carriers in metals and semiconductors has attracted in last decades
big attention and continues to play a key role in the development of many applications,
establishing a field that is now known as spintronics. Spin transport and manipulation in not only
ferromagnets but also nonmagnetic materials are currently being studied actively in a variety of
artificial structures and designed new materials. This enables the fabrication of spintronic properties
on intention. A study on spintronic device structures   was reported as early as in late sixties. Studies of spin-polarized
internal field emission using the magnetic semiconductor $EuS$ sandwiched between two metal
electrodes opened a new epoch in electronics. Since then, many discoveries have been made using 
spintronic structures~\cite{maek,dietl}. Among them is giant magnetoresistance   in magnetic multilayers.   
Giant magnetoresistance has enabled the realization of   sensitive sensors for
hard-disk drives, which has facilitated  successful use of spintronic devices in everyday life.
There is big literature on this subject and any reasonable discussion of the  spin transport deserves a separate extended
review. We should mention here that some aspects of the spin transport in solids were discussed by us 
in Refs.~\cite{kuz07,kuz06}.\\
In the present study a   qualitative theory for 
conductivity in metallic systems is developed and applied to systems like transition metals and their disordered alloys.  
The nature of transition metals is discussed in details and the tight-binding approximation  and method of model Hamiltonians
are described. For the interaction of the electron 
with the lattice vibrations we use the modified tight-binding approximation (MTBA). Thus this approach can not be considered as
the first-principle method and has the same shortcomings and limitations as describing a transition metal within the Hubbard model.
In the following pages, we shall present a formulation of the theory of the electrical transport in the approach of
the nonequilibrium statistical operator. Because several other sections in this review require a certain background in
the use of statistical-mechanical methods, physics of metals, \emph{etc.}, it was felt that some space should be devoted to 
this background.
Sections  2 to 8 serves as an extended introduction to the core sections 9-12  of the present paper.
Thus those sections are intended as a brief summary and short survey of the most important notions and concepts of 
charge transport (mainly electroconductivity)
for the sake of a self-contained formulation. We wish to describe those concepts which have proven to be of value, 
and those notions which will be of use in clarifying subtle points.\\
First, in order to fix the domain of study, we must briefly consider
the various formulation of the subject and introduce the basic notions of the physics of metals and alloys. 
%
%
%
\section{ Metals and Nonmetals. Band Structure}
%
The problem of the fundamental nature of the metallic state is of
long standing~\cite{mott,zi,sol04,slat}. 
It is well known that 
materials are conveniently divided into two broad classes: insulators (nonconducting) and 
metals (conducting)~\cite{slat,sid,and,kuz08}. More specific classification divided materials into
three classes: metals, insulators, and semiconductors.
The most characteristic property of a metal is its ability to conduct electricity.
If we classify crystals in terms of the type of bonding between atoms,
they may be divided into the following five categories (see Table 1).\\
%
%
%
\begin{table}
\label{tb1}
\begin{center}
Table1. Five Categories of Crystals
\caption{ { \bf Five Categories of Crystals }}
\end{center}
\begin{center}
\begin{tabular}{|l|l|} \hline
Type of Crystal  &   substances\\
 \hline
 ionic & alkali halides, alkaline oxides, \emph{etc.}\\
 \hline
homopolar bounded (covalent)& diamond, silicon, \emph{etc.} \\
 \hline
metallic& various metals and alloys\\ 
 \hline
 molecular & $Ar$, $He$, $O_{2}$, $H_{2}$, $CH_{4}$, \emph{etc.}\\
 \hline
 hydrogen bonded & ice, $KH_{2}PO_{4}$, fluorides, \emph{etc.} \\ 
 \hline
\end{tabular}
\end{center}
\end{table}
%
%
%
%
Ultimately we are interested in studying all of the properties
of metals~\cite{mott}. At the outset it is natural to approach this problem through studies of the electrical conductivity
and closely related problem of the energy band structure~\cite{sut,ha,sing,mar04}.\\
The energy bands in solids~\cite{slat,ha,mar04} represent the fundamental electronic structure of a crystal just as the atomic term
values represent the fundamental electronic structure of the free atom. The behavior of an electron in one-dimensional
periodic lattice is described by  Schr\"{o}dinger equation
\begin{equation}\label{2.1}
  \frac{d^{2}\psi}{dx^{2}} + \frac{2m}{\hbar^{2}} (E - V)\psi = 0,
\end{equation}
where $V$ is periodic with the period of the lattice $a$. The variation of energy $E(k)$ as a function of quasi-momentum
within the Brillouin zones, and the variation of the density of states $D(E)dE$ with energy, are of considerable
importance for the understanding of real metals. The assumption that the potential $V$ is small compared with the total 
kinetic energy of the electrons (approximation of nearly free electrons) is not necessarily true for all metals. The theory
may also be applied to cases where the atoms are well separated, so that the interaction between them is small.
This treatment is usually known as the approximation of "tight binding"~\cite{slat}. In this approximation the behavior of an electron in the region
of any one atom being only slightly influenced by the field of the other atoms~\cite{ha,call}. Considering a simple 
cubic structure, it is found that the energy of an electron may be written as
\begin{equation}\label{2.2}
  E(k) = E_{a} - t_{\alpha} - 2t_{\beta} ( \cos (k_{x}a) + \cos (k_{y}a) + \cos (k_{z}a)),
\end{equation}
where $t_{\alpha}$ is an integral depending on the difference between the potentials in which the electron moves
in the lattice and in the free atom, and $t_{\beta}$ has a similar significance~\cite{ha,call} 
(details will be given below). Thus
in the tight-binding limit, when electrons remain to be tightly bound to their original atoms, the valence 
electron moves mainly about individual ion core, with rare hopping from ion to ion. This is the case for  the
$d$-electrons of transition metals. In the typical transition metal the radius of the outermost $d$-shell is less
than half the separation between the atoms. As a result, in the transition metals the $d$-bands are relatively narrow.
In the nearly free-electron limit the bands are derived from the $s$-  and $p$-shells which radii are significantly
larger than half the separation between the atoms. Thus, according this simplified picture simple metals have 
nearly-free-electron energy bands  (see Fig.1). 
\begin{figure}
\centerline{ \includegraphics[width=3.65in]{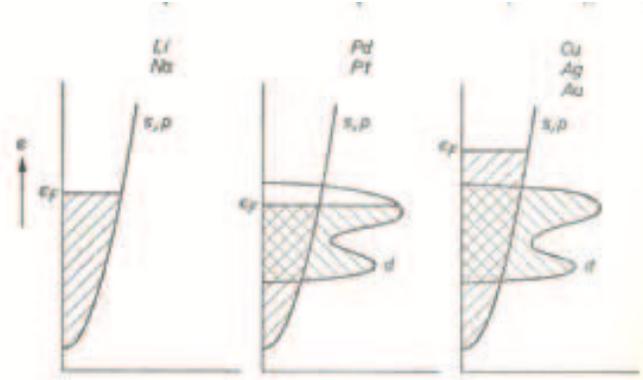} }
\vspace*{8pt}
\caption{Schematic form of the band structure of various metals} 
\label{f.1}
\end{figure}
Fortunately in the case of simple metals the combined results of the energy
band calculation and experiment have indicated that the effects of the interaction between the electrons and ions
which make up the metallic lattice is extremely weak. It is not the case for transition metals and their disordered
alloys~\cite{mottm,jfr}.\\
An obvious characterization of a metal is that it is a good electrical 
and thermal conductor~\cite{mott,zi,slat,zei,ki}. Without considering details it is possible to see how the simple Bloch
picture outlined above accounts for the existence of metallic properties, insulators, and semiconductors. When an electric
current is carried, electrons are accelerated, that is promoted to higher energy levels. In order that this may occur, there 
must be vacant energy levels, above that occupied by the most energetic electron in the absence of an electric field,
into which the electron may be excited. At some conditions there exist many vacant levels within the first zone into which
electrons may be excited. Conduction is therefore possible. This case corresponds with the noble metals. It may happen that the
lowest energy in the second zone is lower than the highest energy in the first zone. It is then possible for electrons
to begin to occupy energy contained within the second zone, as well as to continue to fill up the vacant levels in
the first zone and a certain number of levels in the second zone will be occupied. In this case the metallic
conduction is possible as well. The polyvalent metals are materials of this class.\\
If, however, all  the available energy levels within the first Brillouin zone are full and the lowest possible
electronic energy at the bottom of the second zone is higher than the highest energy in the first zone 
by an amount $\Delta E$, there exist no vacant levels into which electrons may be excited. Under these conditions
no current can be carried by the material and an insulating crystal results.\\
For another class of crystals, the zone structure is analogous to that of insulators but with a very small value
of $\Delta E$. In such cases, at low temperatures the material behaves as an insulator with a higher specific resistance.
When the temperature increases a small number of electrons will be thermally excited across the small gap and enter
the second zone, where they may produce metallic conduction. These substances are termed semiconductors~\cite{slat,zei,ki}, and their
resistance decreases with rise in temperature in marked contrast to the behavior of real metals (for a detailed review of semiconductors
see Refs.~\cite{ldsemi,yu10}).\\
The differentiation between metal and insulator can be
made by measurement of the low frequency electrical conductivity near $T = 0$ K. For the substance which we can refer as an
ideal insulator the electrical conductivity should be zero, and for metal it remains finite or even becomes infinite.
Typical values for the conductivity of metals and insulators differ by a factor of the order $10^{10} - 10^{15}$. So big
difference in the electrical conductivity is related directly to a basic difference in the structural and quantum chemical
organization of the electron and ion subsystems of solids. In an insulator the position of all the electrons are highly 
connected with each other and with the crystal lattice and a weak direct current field cannot move them. In a metal
this connection is not so effective and the electrons can be easily displaced by the applied electric field. Semiconductors
occupy an intermediate position due to the presence of the gap in the electronic spectra.\\
An attempt to give a comprehensive empirical classification of solids types was carried out 
by Zeitz~\cite{zei} and Kittel~\cite{ki}.
Zeitz reanalyzed  the generally accepted classification of materials into three broad classes: insulators, metals and
semiconductors and divided materials into five categories: metals, ionic crystals, valence 
or covalent crystals, molecular crystals, and semiconductors. Kittel added one more category: hydrogen-bonded crystals.
Zeitz also divided metals further  into two major classes, namely, monoatomic metals and alloys.\\
Alloys constitute an important class of the metallic systems~\cite{mi,zei,ki,sid,wat,ldtm,ldtm2}. This class of substances 
is very numerous~\cite{sid,wat,ldtm,ldtm2}. 
A metal alloy is a mixed material that has metal properties and is made by melting at least one pure metal along with 
another pure chemical or metal. 
Examples of metal alloys $Cu-Zn$, $Au-Cu$ and an alloy of carbon and iron, 
or copper, antimony and lead.  Brass is an alloy of copper and zinc, and bronze is an alloy of copper and tin.
Alloys of titanium, vanadium, chromium and other metals are used in many applications.
The titanium alloys (interstitial solid solutions) form a big variety of equilibrium phases.
Alloy metals are usually formed to combine properties of metals and the exact proportion of 
metals in an alloy will change the characteristic properties of the alloy.
We confine ourselves to those alloys which may be regarded essentially as very close to pure metal 
with  the properties intermediate to those of the constituents.\\  
There are different types of monoatomic metals within the Bloch model for the electronic structure of a crystal:
simple metals, alkali metals, noble metals,
transition metals, rare-earth metals, divalent metals, trivalent metals, tetravalent metals, pentavalent semimetals,
lantanides, actinides and their alloys. The classes of metals according to crude Bloch
model provide us with a simple qualitative picture of variety of metals. This simplified classification takes into
account the state of valence atomic electrons when we decrease the interatomic  separation towards its bulk metallic
value. 
Transition metals have narrow $d$-bands in addition to the 
nearly-free-electron energy bands of the simple metals~\cite{mottm,jfr}.  In addition, the correlation of electrons
plays an essential role~\cite{mottm,jfr,kuz09}.
The Fermi energy lies within the $d$-band so that the $d$-band
is only partially occupied.  Moreover the Fermi surface have much more complicated form and topology. The  concrete calculations of the band 
structure of many transition metals ($Nb$, $V$, $W$, $Ta$, $Mo$, \emph{etc.}) can be found in 
Refs.~\cite{slat,sut,ha,sing,mar04,mottm,oand,nbta,hod,laur} and in Landolt-Bornstein reference books~\cite{ldtm,ldtm3}.\\
The noble metal atoms have one $s$-electron outside of a just completed $d$-shell. The
$d$-bands of the noble metals lie below the Fermi energy but not too deeply. Thus they influence many of the physical
properties of these metals. It is, in principle, possible to test the predictions of the single-electron band
structure  picture by comparison with experiment. In semiconductors it has been performed with the measurements of the
optical absorption, which gives the values of various energy differences within the semiconductor bands. In metals
the most direct approach is related to the experiments which studied the shape and size of the Fermi surfaces. In spite
of their value, these data represent only a rather limited scope in comparison to the many properties of metals 
which are not so directly related to the energy band structure. 
Moreover, in such a picture there are many weak points: there is no sharp boundary between insulator and
semiconductor, the theoretical values of $\Delta E$ have discrepancies with experiment,  the metal-insulator 
transition~\cite{mot} cannot be described correctly, and the notion "\emph{simple}" metal  have no single meaning~\cite{nwiz}.
The crude Bloch model even met more serious difficulties when it was applied to insulators. The improved theory of
insulating state was developed by Kohn~\cite{ins} within a many-body approach. He proposed a new and more comprehensive
characterization of the insulating state of matter. This line of reasoning was continued 
further in Refs.~\cite{mot,ed,ins1}  on a more precise and firm theoretical and experimental basis.\\
Anderson~\cite{and} gave a critical analysis of the Zeitz and Kittel classification schemes. He concluded that "in every real
sense the distinction between semiconductors and metals or valence crystals as to type of binding, and 
between semiconductor and
any other type of insulator as to conductivity, is entirely artificial; semiconductors do not represent in any real sense
a distinct class of crystal"~\cite{and} (see, however Refs.~\cite{slat,zei,ki,lund,cjac10}). Anderson has pointed also the  extent to which the standard classification
falls. His conclusions were confirmed by further development of solid state physics. During the last decades a lot of new substances 
and materials were synthesized and tested. Their conduction properties and temperature behavior of the resistivity 
are differed  substantially and constitute  a difficult task for consistent classification~\cite{ka} (see Fig.1). 
\begin{figure}[bt]
\centerline{\includegraphics[width=3.65in]{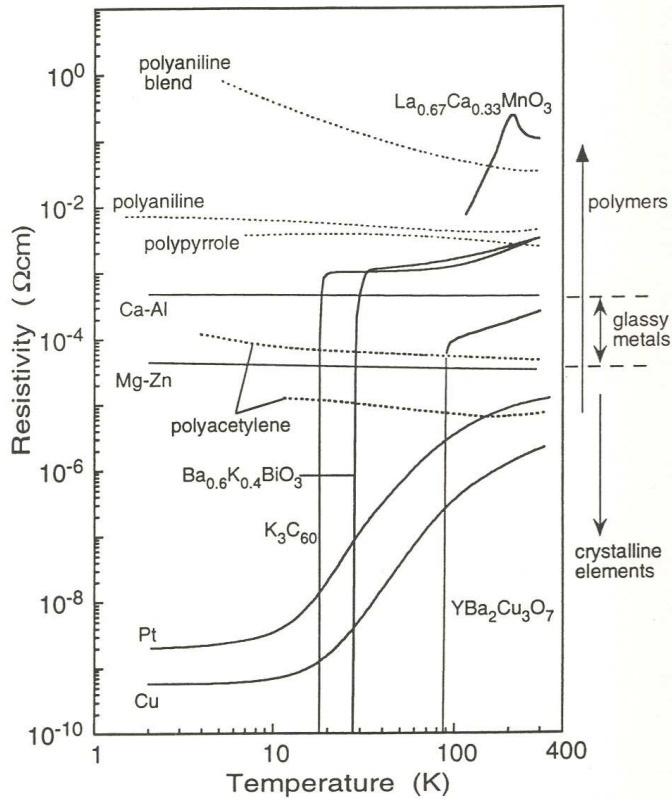}}
\vspace*{8pt}
\caption{Resistivity of various conducting materials (from Ref.~\cite{ka})} 
\label{f.2}
\end{figure}
Bokij~\cite{bok} carried out an interesting analysis of notions 
"metals" and "nonmetals" for chemical elements. According to him, there are typical metals ($Cu$, $Au$, $Fe$) and typical
nonmetals ($O$, $S$, halogens), but the boundary between them and properties determined by them are still an open question. The
notion "metal" is defined by a number of specific properties of the corresponding elemental substances, e.g. by 
high electrical conductivity and thermal capacity, the ability to reflect light waves(luster), plasticity, and ductility.
Bokij emphasizes~\cite{bok}, that when defining the notion of a metal, one has also to take into account the crystal structure. 
As a rule, the structure of metals under normal conditions are characterized by rather high symmetries and high 
coordination numbers (c.n.) of atoms equal to or higher than eight, whereas the structures of crystalline
nonmetals under  normal conditions are characterized by lower symmetries and coordination numbers of 
atoms (2-4). \\
It is worth noting that
such topics like studies of the strongly correlated electronic systems~\cite{kuz09}, high-$T_c$ superconductivity~\cite{proz07}, 
colossal magnetoresistance~\cite{duan05} and multiferroicity~\cite{duan05} have
led to a new development of solid state physics during the last decades. 
Many transition-metal oxides show very large ("colossal") magnitudes of
the dielectric constant and thus have immense potential for applications in modern
microelectronics and for the development of new capacitance-based energy-storage
devices. These and other
interesting phenomena to a large extend first have been revealed and intensely investigated
in transition-metal oxides. The complexity of the ground states of these materials arises from
strong electronic correlations, enhanced by the interplay of spin, orbital, charge and lattice
degrees of freedom~\cite{kuz09}. These phenomena are a challenge for basic research and also bear big
potentials for future applications as the related ground states are often accompanied by so-called
"colossal" effects, which are possible building blocks for tomorrow's correlated electronics.
The measurement of the response of transition-metal oxides to $ac$ electric fields is one of the
most powerful techniques to provide detailed insight into the underlying physics that may comprise
very different phenomena, e.g., charge order, molecular or polaronic relaxations, magnetocapacitance,
hopping charge transport, ferroelectricity or density-wave formation. 
In the recent work~\cite{lunk}, authors thoroughly discussed the mechanisms that can lead
to colossal values of the dielectric constant, especially emphasizing effects generated
by external and internal interfaces, including electronic phase separation. 
The authors of the 
work~\cite{lunk} studied the materials showing so-called colossal dielectric constants (CDC), i.e. values
of the real part of the permittivity $\varepsilon'$ exceeding 1000. Since long, materials with high dielectric
constants are in the focus of interest, not only for purely academic reasons but also because new
high-$\varepsilon'$ materials are urgently sought after for the further development of modern electronics.
In addition, authors of the work~\cite{lunk} provided a detailed overview and discussion of the dielectric properties
of $CaCu_{3}Ti_{4}O_{12}$ and related systems, which is today's most investigated material
with colossal dielectric constant. Also a variety of further transition-metal oxides
with large dielectric constants were treated in detail, among them the system
$La_{2-x}Sr_{x}NiO_{4}$ where electronic phase separation may play a role in the generation
of a colossal dielectric constant.
In general, for the miniaturization of capacitive electronic elements materials with high-$\varepsilon'$ are
prerequisite. This is true not only for the common silicon-based integrated-circuit technique
but also for stand-alone capacitors.\\
Nevertheless, as regards to metals, the workable practical definition of Kittel can be adopted: metals are characterized 
by high electrical conductivity, so that a portion of electrons in metal must be free to move about. The electrons available
to participate in the conductivity are called conduction electrons. Our picture of a metal, therefore, must be that it
contains electrons which are free to move, and which may, when under the influence of an electric field, carry a current
through the material.\\
In summary, the 68 naturally occurring metallic  
and semimetallic elements~\cite{sid}
can be classified as it is shown in Table 2.
%
%
%
\begin{table}
\label{tab2}
\begin{center}
Table 2. Metallic  and Semimetallic Elements
\caption{ { \bf Metallic  and Semimetallic Elements }}
\end{center}
\begin{center}
\begin{tabular}{|l|c|l|} \hline
item  &   number& elements\\
 \hline
 alkali metals & 5 & $Li$, $Na$, $K$, $Rb$, $Cs$ \\
 \hline
noble metals& 3& $Cu$, $Ag$, $Au$ \\
 \hline
polyvalent simple metals& 11& $Be$, $Mg$, $Zn$, $Cd$, $Hg$, $Al$,   
                              $Ga$, $In$, $Tl$, $Sn$, $Pb$ \\ 
 \hline
 alkali-earth metals & 4 & $Ca$, $Sr$, $Ba$, $Ra$ \\
 \hline
 semi-metals & 4 & $As$, $Sb$, $Bi$, $graphite$ \\ 
 \hline
transition metals & 23 & $Fe$, $Ni$, $Co$, $\emph{etc.}$ \\
 \hline 
rare earths & 14 &   \\
 \hline 
actinides & 4 &   \\
 \hline  
\end{tabular}
\end{center}
\end{table}
%
%
\section{Many-Particle Interacting Systems and Current operator}
%
Let us now consider a general system of $N$ interacting electrons in a volume $\Omega$ described by the Hamiltonian
\begin{eqnarray}\label{3.17}
H =  \Bigl ( \sum_{i=1}^{N} \frac{\vec{p}_{i}^{2} }{2m} + 
\sum_{i=1}^{N}U(\vec{r}_{i}) \Bigr ) + \frac{1}{2}\sum_{i\neq j} v(\vec{r}_{i} - \vec{r}_{j}) = H_{0} + H_{1}.
\end{eqnarray}
Here $U(\vec{r})$ is a one-body potential, e.g. an externally applied potential like that due to the field of the ions in  a solid, and
$v(\vec{r}_{i} - \vec{r}_{j})$  is a two-body potential like the Coulomb potential between electrons. 
It is essential that $U(\vec{r})$ and $v(\vec{r}_{i} - \vec{r}_{j})$ do not depend on the velocities of the particles.\\
It is convenient to introduce a "quantization"  in a continuous space~\cite{bb,dpetr,th1} via the operators $ { \varPsi}^{\dag}(\vec{r})$   and ${ 
\varPsi}(\vec{r}) $ which
create and destroy a particle at $\vec{r}$. In terms of $\varPsi^{\dag}$ and $\varPsi$ we have
\begin{eqnarray}\label{3.18}
H = \int d^{3}r \varPsi^{\dag}(\vec{r}) \Bigl( \frac{-\nabla^{2}}{2m} + U(\vec{r}) \Bigr) \varPsi (\vec{r})  \\ \nonumber 
 + \frac{1}{2} \int \int d^{3}r   d^{3}r' \varPsi^{\dag}(\vec{r}) \varPsi^{\dag}(\vec{r}')  v(\vec{r}  
 - \vec{r}' ) \varPsi (\vec{r}') \varPsi (\vec{r}).
\end{eqnarray}
Studies of flow problems lead to the continuity equation~\cite{enz,zub} 
\begin{equation}\label{3.35}
\frac{\partial n(\vec{r},t )}{\partial t} + \nabla \vec{j} = 0~.
\end{equation}
This equation based on the concept of conservation of certain extensive variable. In nonequilibrium thermodynamics~\cite{zub} the fundamental
flow equations are obtained using successively mass, momentum, and energy as the relevant extensive variables. The analogous equation are known from
electromagnetism. The central role plays a global conservation law of charge, $\dot{q}(t) = 0$, for it refers to the total charge in a system. Charge
is also conserved locally~\cite{kuz10}. This is described by Eq.(\ref{3.35}), where $n(\vec{r},t )$ and $\vec{j}$ are the charge and current densities, respectively.\\
In quantum mechanics there is the connection of the wavefunction $\psi(\vec{r},t )$ to the particle mass-probability current
distribution $\vec{J} $
\begin{equation}\label{3.36}
 \vec{J}(\vec{r},t ) =  \frac{\hbar}{2mi} (\psi^{*} \nabla \psi - \psi \nabla \psi^{*}),  
\end{equation}
where $\psi (\vec{r},t )$ satisfy the time-dependent Schr\"{o}dinger equation~\cite{dav,th1}
\begin{equation}\label{3.37}
i\hbar \frac{\partial}{\partial t}\psi (\vec{r},t ) = H \psi (\vec{r},t )~.
\end{equation}
Consider the motion of a particle under the action of a time-independent force determined by a real potential $V(\vec{r}).$
Equation (\ref{3.37}) becomes
\begin{equation}\label{3.39}
\left( \frac{\vec{p}^{2} }{2m} + V \right)\psi = \frac{\hbar}{2m}\nabla^{2}\psi + V \psi = i \hbar \frac{\partial}{\partial t}\psi.
\end{equation}
It can be shown that for the probability density $n(\vec{r},t ) = \psi^{*}\psi $ we have
\begin{equation}\label{3.40}
\frac{\partial n}{\partial t} + \nabla \vec{J} = 0.
\end{equation}
This is the equation of continuity and it is quite general for real potentials. The equation of continuity mathematically states the local
conservation of particle mass probability in space.\\ A thorough consideration of a current carried by a quasi-particle for a uniform gas of
fermions, containing $N$ particles in a volume $\Omega$, which was assumed to be very large, was performed within a semi-phenomenological theory of
Fermi liquid~\cite{noz}. This theory describes the macroscopic properties of a system at zero temperature and requires knowledge of the ground
state and the low-lying excited states. The current  carried by the quasi-particle $\vec{k} $ is the sum of two terms: the current which is equal 
to the velocity $v_{k}$ of the quasi-particle and the backflow of the medium~\cite{noz}. The precise definition of the current $J$ in an arbitrary
state $|\varphi \rangle $  within the Fermi liquid theory is given by
\begin{equation}\label{3.41}
 J = \langle \varphi |  \sum_{i} \frac{p_{i} }{m} |\varphi \rangle,
\end{equation}
where $p_{i}$ is the momentum of the $i$th particle and $m$ its bare mass. To measure $J$ it is necessary to use a reference frame moving
with respect to the system with the uniform velocity $\hbar q/m$. The Hamiltonian in the rest frame can be written
\begin{equation}\label{3.42}
H =  \sum_{i} \frac{p^{2}_{i} }{2m} + V.
\end{equation}
It was assumed that  $V$ depends only on the positions and the relative velocities of the particles; it is not modified by a translation.
In the moving system only the kinetic energy changes; the apparent Hamiltonian becomes
\begin{equation}\label{3.43}
H_{q} =  \sum_{i} \frac{(p_{i} - \hbar q )^{2} }{2m} + V = H - \hbar q  \sum_{i} \frac{p_{i} }{m} + N \frac{(\hbar q )^{2} }{2m}.
\end{equation}
Taking the average value of $H_{q}$ in the state $|\varphi \rangle$, and let $E_{q}$ be the energy of the system as seen from the moving reference frame,
one find in the $\lim q \rightarrow 0$
\begin{equation}\label{3.44}
\frac{\partial E_{q}}{\partial q_{\alpha}} =  - \hbar \langle \varphi |  \sum_{i} \frac{p_{i\alpha} }{m} |\varphi \rangle = - \hbar J_{\alpha},
\end{equation}
where $\alpha$ refers to one of the three coordinates. This expression gives the definition of current in the framework of the Fermi liquid theory.
For the particular case of a translationally invariant system the total current is a constant of the motion, which commutes with the interaction $V$ and
which, as a consequence, does not change when $V$ is switched on adiabatically. For the particular state containing one quasi-particle $\vec{k} $ the
total current $J_{k}$ is the same as for the ideal system
\begin{equation}\label{3.45}
J_{k} =    \frac{(  \hbar k )}{m}.  
\end{equation}
This result is a direct consequence of Galilean invariance.\\
Let us consider now the many-particle Hamiltonian (\ref{3.18})
\begin{equation}\label{3.46}
H =    H_{1} +  H_{2}. 
\end{equation}
It will also be convenient to consider density of the particles in the following form~\cite{enz} 
$$n(\vec{r}) =  \sum_{i}\delta (\vec{r}  - \vec{r}_{i} ). $$

The Fourier transform of the particle density operator becomes
\begin{equation}\label{3.47}
n(\vec{q}) = \int d^{3}r \exp (- i \vec{q} \vec{r}) \sum_{i}\delta (\vec{r}  - \vec{r}_{i} ) = \sum_{i} \exp (- i \vec{q} \vec{r}_{i}).
\end{equation}
The particle mass-probability current distribution $\vec{J} $ in this "lattice" representation will take the form
\begin{eqnarray}\label{3.48}
\vec{J}(\vec{r}) = n(\vec{r}) \vec{v} =  \frac{1}{2} \sum_{i} \{ \frac{\vec{p}_{i} }{m}  \delta (\vec{r}  - \vec{r}_{i} ) 
+ \delta (\vec{r}  - \vec{r}_{i} )\frac{\vec{p}_{i} }{m} \} = \\ \nonumber  
\frac{1}{2} \sum_{i} \{ \frac{\vec{p}_{i} }{m} \exp (- i \vec{q} \vec{r}_{i}) + \exp (- i \vec{q} \vec{r}_{i}) \frac{\vec{p}_{i} }{m} \},\\ \nonumber 
[\vec{r}_{i}, \vec{p}_{k}] = i \hbar \delta_{ik}.
\end{eqnarray}
Here $\vec{v}$ is the velocity operator.  The direct calculation shows that
\begin{eqnarray}\label{3.49}
[ n(\vec{q}), H] = 
\frac{1}{2} \sum_{i} \{ \frac{\vec{q} \vec{p}_{i} }{m} \exp (- i \vec{q} \vec{r}_{i}) + \exp (- i \vec{q} \vec{r}_{i}) \frac{\vec{q} \vec{p}_{i} }{m} \} 
= \vec{q} \vec{J}(\vec{q}).
\end{eqnarray}
Thus the equation of motion for the particle density operator becomes
\begin{eqnarray}\label{3.50}
\frac{d n(\vec{q})}{dt} = \frac{i}{\hbar}[H, n(\vec{q})] = - \frac{i}{\hbar} \vec{q} \vec{J}(\vec{q}),
\end{eqnarray}
or in another form
\begin{equation}\label{3.51}
\frac{d n(\vec{r})}{dt} = \textrm{div} \vec{J}(\vec{r}),
\end{equation}
which is the continuity equation considered above. Note, that
$$ [n(\vec{q}), H_{1}]_{-} = [n(\vec{q}), H_{2}]_{-} = 0. $$
These relations holds in general for any periodic potential and interaction potential of the electrons which depend only on the coordinates of the
electrons.\\ It is easy to check the validity of the following relation
\begin{equation}\label{3.52}
[[ n(\vec{q}), H],  n^{\dag}(\vec{q})] = [ \vec{q} \vec{J}(\vec{r}),n^{\dag}(\vec{q})] =  \frac{N q^{2}}{m}.
\end{equation}
This formulae is the known f-sum rule~\cite{noz} which is a consequence from the continuity equation (for a more 
general point of view see Ref.~\cite{klee}).\\
Now consider the second-quantized Hamiltonian (\ref{3.18}).
The particle density operator has the form~\cite{raim,bb,nego}
\begin{equation}\label{3.53}
n(\vec{r}) = e  \varPsi^{\dag}(\vec{r}) \varPsi (\vec{r}), \quad n(\vec{q}) = \int d^{3}r \exp (- i \vec{q} \vec{r}) n(\vec{r}).
\end{equation}
Then we define
\begin{equation}\label{3.54}
 \vec{j}(\vec{r}) =  \frac{e \hbar}{2mi} (\varPsi^{\dag} \nabla \varPsi - \varPsi \nabla \varPsi^{\dag}).  
\end{equation}
Here $\vec{j}$  is the probability current density, i.e. the probability flow per unit time per unit area perpendicular to $\vec{j}$. 
The continuity equation will persist for this case too. Let us consider the equation motion
\begin{eqnarray}\label{3.55}
\frac{d n(\vec{r})}{d t} = - \frac{i}{\hbar}[n(\vec{r}), H_{1} ] - \frac{i}{\hbar}[n(\vec{r}), H_{2} ] = \\ \nonumber
\frac{e \hbar}{2mi} (\varPsi^{\dag}(\vec{r}) \nabla^{2} \varPsi(\vec{r})  -   \nabla^{2} \varPsi^{\dag}(\vec{r}) \varPsi (\vec{r})).
\end{eqnarray}
Note, that $[n(\vec{r}), H_{2} ] \equiv 0$. \\ We find
\begin{equation}\label{3.56}
\frac{d n(\vec{r})}{dt} = - \nabla \vec{j}(\vec{r}).
\end{equation}
Thus the continuity equation have the same form in both the "particle" and "field" versions.
%
%
%
%
%
\section{Tight-Binding and Modified Tight-Binding Approximation}
%
Electrons and phonons are the basic elementary excitations of a metallic solid. Their mutual 
interactions~\cite{zi,call,grim,ch61,har71,kitt} 
manifest themselves in such observations as the temperature dependent resistivity and low-temperature 
superconductivity. In the quasiparticle picture, at the basis of this interaction is the individual electron-phonon
scattering event, in which an electron is deflected in the dynamically distorted lattice. 
We consider here the
scheme which is called the modified tight-binding interaction (MTBA). But firstly, we remind shortly 
the essence of the tight-binding approximation. The main purpose in using
the tight-binding method is to simplify the theory sufficiently to make workable. The tight-binding approximation
considers  solid as a giant molecule. 
%
%
%
\subsection{Tight-binding  approximation}
%
The main problem of the electron theory of solids is to calculate the energy level spectrum of electrons moving in 
an ion lattice~\cite{call,kor}. The tight binding method~\cite{call,laf,tb,calf,tb97} for energy band calculations has generally been regarded
as suitable primarily for obtaining a simple first approximation to a complex band structure. 
It was shown that  the method should 
also be quite powerful in quantitative calculations from first principles for a wide variety of materials.
An approximate treatment requires to obtain energy levels and electron wave functions for some 
suitable chosen one-particle potential (or pseudopotential), which is usually local. The standard molecular orbital
theories of band structure are founded on an independent particle model.\\
As atoms are brought together to form a crystal lattice the sharp atomic levels broaden into bands. Provided there is no 
overlap between the bands, one expects to describe the crystal state by a Bloch function of the type,
\begin{equation}\label{4.10}
 \psi_{\vec{k}}(\vec{r}) = \sum_{n} e^{i\vec{k}\vec{R}_{n}} \phi (\vec{r} - \vec{R}_{n} ),
\end{equation}
where $\phi (\vec{r})$ is a free atom single electron wave function, for example such as $1s$ and $\vec{R}_{n}$ is the
position of the atom in a rigid lattice. If the bands overlap or 
approach each other one should use instead of $\phi (\vec{r})$ a combination of the wave functions corresponding to the
levels in question, e.g. $ ( a \phi(1s) + b \phi(2p))$, \emph{etc.} In the other words, this approach, first introduced
to crystal calculation by F.Bloch, expresses  the eigenstates of an electron in a perfect crystal in a linear combination
of atomic orbitals and termed LCAO method~\cite{call,laf,tb,calf,tb97}. \\
Atomic orbitals are not the most suitable basis set due to the
nonorthogonality problem. It was shown by many authors~\cite{call,wa,bro,wko} that the very efficient basis set for the
expansion (\ref{4.10}) is the atomic-like Wannier functions $\{w (\vec{r} - \vec{R}_{n} )\}$~\cite{call,wa,bro,wko}. These 
are the Fourier transforms of the extended Bloch functions and are defined as
\begin{equation}\label{4.24}
w (\vec{r} - \vec{R}_{n})   =   \mathcal{N}^{-1/2}  \sum_{\vec{k}} e^{-i\vec{k}\vec{R}_{n}}    \psi_{\vec{k}}(\vec{r}). 
\end{equation}
Wannier functions $w (\vec{r} - \vec{R}_{n})$ form a complete set of mutually orthogonal functions localized around each lattice site $\vec{R}_{n}$
within any band or group of bands. They permit one to formulate  an effective Hamiltonian for electrons
in periodic potentials and span the space of a singly energy band. However, the real computation of Wannier functions in
terms of sums over Bloch states is a complicated task~\cite{ha,wko}.\\ To define the Wannier functions more precisely let us consider
the eigenfunctions $\psi_{\vec{k}}(\vec{r})$ belonging to a particular simple band in a lattice with the one type of atom at
a center of inversion. Let it satisfy the following equations with one-electron Hamiltonian $H$
\begin{equation}\label{4.25}
H \psi_{\vec{k}}(\vec{r}) =    E(\vec{k}) \psi_{\vec{k}}(\vec{r}), \quad  \psi_{\vec{k}}(\vec{r} + \vec{R}_{n} )  =
e^{-i\vec{k}\vec{R}_{n}}    \psi_{\vec{k}}(\vec{r}),    
\end{equation}
and the orthonormality relation $ \langle \psi_{\vec{k}}|\psi_{\vec{k'}}\rangle  = \delta_{\vec{k}\vec{k'} }$ where 
the integration is performed over the $\mathcal{N}$ unit cells in the crystal. The property of periodicity together with the property of the orthonormality
lead to the orthonormality condition of the Wannier functions
\begin{equation}\label{4.26}
\int d^{3}r w^{*} (\vec{r} - \vec{R}_{n} ) w (\vec{r} - \vec{R}_{m} )  = \delta_{nm }.
\end{equation}
The set of the Wannier functions is complete, i.e.
\begin{equation}\label{4.27}
\sum_{i} w^{*} (\vec{r}' - \vec{R}_{i} ) w (\vec{r} - \vec{R}_{i} )  = \delta (\vec{r}' - \vec{r} ).
\end{equation}
Thus it is possible to find the inversion of the Eq.(\ref{4.24}) which has the form
\begin{equation}\label{4.28}
\psi_{\vec{k}}(\vec{r})    =   \mathcal{N}^{-1/2}  \sum_{\vec{k}} e^{i\vec{k}\vec{R}_{n}} w (\vec{r} - \vec{R}_{n} ).     
\end{equation}
These conditions are not sufficient to define the functions uniquely since the Bloch states $\psi_{\vec{k}}(\vec{r}) $
are determined only within a multiplicative phase factor $\varphi(\vec{k})$ according to
\begin{equation}\label{4.29}
w (\vec{r} )   =   \mathcal{N}^{-1/2}  \sum_{\vec{k}} e^{i\varphi(\vec{k})}    u_{\vec{k}}(\vec{r}), 
\end{equation}
where $\varphi(\vec{k})$  is any real function of $\vec{k}$, and $u_{\vec{k}}(\vec{r})$ are Bloch functions~\cite{fer}. The phases
$\varphi(\vec{k})$ are usually chosen so as to localize $w (\vec{r} )$ about the origin. The usual choice of 
phase makes $\psi_{\vec{k}}(\vec{0})$ real and positive. This lead to the maximum possible value in $w(\vec{0})$
and $w(\vec{r})$ decaying exponentially away from $\vec{r} = 0$. In addition,  function  $\psi_{\vec{k}}(\vec{r})$ with this 
choice will satisfy the symmetry properties
$$ \psi_{-\vec{k}}(\vec{r}) = ( \psi_{\vec{k}}(\vec{r}))^{*} = \psi_{\vec{k}}(-\vec{r}).$$
It follows from the above consideration that the Wannier functions are real and symmetric,
$$ w(\vec{r}) = ( w(\vec{r}))^{*} = w(-\vec{r}).$$
Analytical, three dimensional Wannier functions have been constructed from Bloch states formed from a lattice 
gaussians.
Thus, in the condensed matter theory, the Wannier functions play an important role in the
theoretical description of transition metals, their compounds and disordered alloys, impurities and imperfections,
surfaces, \emph{etc.}
%
\subsection{Interacting electrons on a lattice and the Hubbard model}
%
There are big difficulties in 
description of the complicated problems of electronic and magnetic properties of a metal
with the $d$ band electrons which  are really neither "local" nor "itinerant" in a full sense. 
A better understanding of the electronic correlation effects in metallic systems can be achieved by the formulating of the
suitable flexible model that could be used to analyze major aspects of both the insulating and metallic states of solids in which
electronic correlations are important. \\ 
The Hamiltonian of the interacting electrons with pair interaction in the second-quantized form is given by Eq.(\ref{3.18}).
Consider this Hamiltonian in the Bloch representation. We have
\begin{equation}\label{4.31}
 \varPsi_{\sigma} (\vec{r}) =   \sum_{k} \varphi_{\vec{k} \sigma}(\vec{r}) a_{k \sigma}, \quad 
 \varPsi^{\dag}_{\sigma} (\vec{r}) =   \sum_{k} \varphi^{*}_{\vec{k}\sigma}(\vec{r}) a^{\dag}_{k \sigma}.  
\end{equation}
Here $\varphi_{\sigma}(\vec{k})$ is the Bloch function satisfying the equation
\begin{eqnarray}\label{4.32}
 H_{1}(r) \varphi_{\vec{k}\sigma}(\vec{r}) =  E_{\sigma}(\vec{k})\varphi_{\vec{k}\sigma}(\vec{r}), \quad E_{\sigma}(\vec{k}) = E_{\sigma}( - \vec{k}),\\ \nonumber
 \varphi_{\vec{k}}(\vec{r}) = \exp (i \vec{k} \vec{r}) u_{k} (r), \quad u_{k} (r + l) = u_{k} (r); \\ \nonumber
 \varphi_{\vec{k}\sigma}(\vec{r}) = \varphi_{- \vec{k} \sigma}(\vec{- r}), \quad \varphi^{*}_{\vec{k}\sigma}(\vec{r}) = \varphi_{- \vec{k}\sigma}(\vec{r}).
\end{eqnarray}
The functions $\{\varphi_{\vec{k}\sigma}(\vec{r}) \} $ form a complete orthonormal set of functions
\begin{eqnarray}\label{4.33}
\int d^{3}r \varphi^{*}_{\vec{k}'}(\vec{r}) \varphi_{\vec{k}}(\vec{r}) = \delta_{k k'}, \\ \nonumber 
\sum_{k}  \varphi^{*}_{\vec{k}}(\vec{r}~') \varphi_{\vec{k}}(\vec{r}) = \delta(r - r').
\end{eqnarray}
We find
\begin{eqnarray}\label{4.34}
H = \sum_{mn} \langle m |H_{1}|n \rangle a^{\dag}_{m} a_{n} + 
\frac{1}{2} \sum_{klmn} \langle kl |H_{2}|mn \rangle a^{\dag}_{k} a^{\dag}_{l} a_{m} a_{n} = \\ \nonumber 
\sum_{\vec{k}\sigma} \langle \varphi^{*}_{\vec{k},\sigma} |H_{1}|\varphi_{\vec{k},\sigma} \rangle  a^{\dag}_{k \sigma} a_{k \sigma}  + \\ \nonumber
\frac{1}{2}\sum_{\vec{k}_{4}\vec{k}_{3}\vec{k}_{2}\vec{k}_{1}}\sum_{\alpha \beta \mu \nu} 
\langle \varphi^{*}_{\vec{k}_{4},\nu} \varphi^{*}_{\vec{k}_{3},\mu}|H_{2}|\varphi_{\vec{k}_{2},\beta}\varphi_{\vec{k}_{1},\alpha} \rangle 
a^{\dag}_{\vec{k}_{4} \nu} a^{\dag}_{\vec{k}_{3} \mu}a_{\vec{k}_{2} \beta}a_{\vec{k}_{1} \alpha}.
\end{eqnarray}
Since the method of second quantization is based on the choice of suitable complete set of orthogonal normalized wave functions, we take
now the set $\{ w_{\lambda} (\vec{r} - \vec{R}_{n} )\}$  of the Wannier functions. Here $\lambda$ is the band index. The field operators
in the Wannier-function representation are given by
\begin{eqnarray}\label{4.35}
 \varPsi_{\sigma} (\vec{r}) =   \sum_{n} w_{\lambda } (\vec{r} - \vec{R}_{n} ) a_{n \lambda \sigma}, \quad 
 \varPsi^{\dag}_{\sigma} (\vec{r}) =   \sum_{n} w_{\lambda }^{*}(\vec{r} - \vec{R}_{n}) a^{\dag}_{n \lambda \sigma}.  
\end{eqnarray}
Thus we have
\begin{equation}\label{4.36}
a^{\dag}_{n \lambda \sigma} = N^{-1/2} \sum_{\vec{k}} e^{-i\vec{k}\vec{R}_{n}}a^{\dag}_{\vec{k} \lambda \sigma}, \quad 
a_{n \lambda \sigma}   =  N^{-1/2} \sum_{\vec{k}} e^{i\vec{k}\vec{R}_{n}}a_{\vec{k} \lambda \sigma}.    
\end{equation}
Many of treatment of the correlation effects are effectively restricted
to a non-degenerate band. 
The Wannier functions basis set is the background of the widely used Hubbard model.
The Hubbard model\cite{rnc,hub} is, in a
certain sense, an intermediate model (the narrow-band model) and
takes into account the specific features of transition metals and
their compounds by assuming that the $d$ electrons form a band,
but are subject to a strong Coulomb repulsion  at one lattice
site.  The single-band Hubbard   Hamiltonian is of the
form~\cite{rnc,kuz09}
\begin{equation}
\label{4.45} H =
\sum_{ij\sigma}t_{ij}a^{\dagger}_{i\sigma}a_{j\sigma} +
U/2\sum_{i\sigma}n_{i\sigma}n_{i-\sigma}.
\end{equation}
Here $a^{\dagger}_{i\sigma}$ and $a_{i\sigma}$ are the second-quantized operators
of the creation and annihilation of the electrons in the lattice state $w({\vec r} -{\vec R_{i}})$ with spin $\sigma$. 
The Hamiltonian  includes the intra-atomic Coulomb repulsion $U$ and the
one-electron hopping energy $t_{ij}$. 
The corresponding parameters of the Hubbard   Hamiltonian are given by
\begin{eqnarray}\label{4.46}
t_{ij} = \int d^{3}r w^{*}(\vec{r} - \vec{R}_{i})  H_{1}(r) w (\vec{r} - \vec{R}_{j} ), \\ 
U  =   \int \int d^{3}r  d^{3}r'
w^{*}(\vec{r} - \vec{R}_{i}) w^{*}(\vec{r'} - \vec{R}_{i})   \frac{e^{2}}{|\vec{r}  - \vec{r'}|}  w (\vec{r'}  - 
\vec{R}_{i} ) w (\vec{r} - \vec{R}_{i}).
\label{4.47}
\end{eqnarray}
The electron correlation
forces electrons to localize in the atomic-like orbitals which are
modelled here by a complete and orthogonal set of the Wannier wave
functions $\{w({\vec r} -{\vec R_{j}})\}$. On the other hand,
the kinetic energy is increased when electrons are delocalized. The
band energy of Bloch electrons $E(k)$ is defined as
follows:
\begin{equation} \label{4.48}
  t_{ij} = N^{-1}\sum_{\vec k}E(k)
 \exp[i{\vec k}({\vec R_{i}} -{\vec R_{j}}], 
\end{equation}
where $N$ is the number of lattice sites. 
The Pauli exclusion
principle which does not allow two electrons of common spin to be
at the same site, $n_{i\sigma}^{2} = n_{i\sigma}$, plays a crucial role. Note, that the standard derivation of the Hubbard model presumes the rigid
ion lattice with the rigidly fixed ion positions. We note that $s$-electrons are not explicitly taken into account in our model Hamiltonian. They
can be, however, implicitly taken into account by screening effects and effective $d$-band occupation.
%
%
\subsection{Current operator for the tight-binding electrons}
%
Let us  consider again a many-particle interacting systems on a lattice with the Hamiltonian (\ref{4.34}).
At this point, it is important to realize the fundamental difference between many-particle system which is 
uniform in space and many-particle system  on a lattice. For the many-particle systems on a lattice the proper definition of current 
operator is a  subtle problem.
It was shown above  that a physically satisfactory definition of the current operator in the quantum many-body theory is given
based upon the continuity equation. However, this point should be reconsidered carefully for the lattice fermions 
which are described by the Wannier functions.\\
Let us remind once again that the Bloch and Wannier wave functions are related to each other by the unitary transformation of the form
\begin{eqnarray} \label{4.50}
  \varphi_{k}( \vec{r} ) = N^{-1/2}\sum_{\vec R_{n}} w({\vec r} -{\vec R_{n}}) \exp[i{\vec k}{\vec R_{n}} ], \\ \nonumber
  w({\vec r} -{\vec R_{n}}) = N^{-1/2}\sum_{\vec k} \varphi_{k}( \vec{r} ) \exp[- i{\vec k}{\vec R_{n}} ].
\end{eqnarray}
The number occupation representation for a single-band case lead to
\begin{eqnarray}\label{4.51}
 \varPsi_{\sigma} (\vec{r}) =   \sum_{n} w(\vec{r} - \vec{R}_{n} ) a_{n  \sigma}, \quad 
 \varPsi^{\dag}_{\sigma} (\vec{r}) =   \sum_{n} w^{*}(\vec{r} - \vec{R}_{n}) a^{\dag}_{n \sigma}.  
\end{eqnarray}
In this representation the particle density operator and current density take the form
\begin{eqnarray}\label{4.52}
n(\vec{r}) = \sum_{ij} \sum_{\sigma} w^{*}(\vec{r}-\vec{R}_{i}) w(\vec{r}-\vec{R}_{j} )a^{\dag}_{i\sigma}a_{j\sigma}, \,\,\,\, \, \\ \nonumber
\vec{j}(\vec{r}) =  \frac{e \hbar}{2mi} \sum_{ij} \sum_{\sigma} [w^{*}(\vec{r} - \vec{R}_{i})\nabla w(\vec{r} - \vec{R}_{j} ) - 
\nabla w^{*}(\vec{r} - \vec{R}_{i}) w(\vec{r} - \vec{R}_{j} )] a^{\dag}_{i \sigma}a_{j  \sigma}.  
\end{eqnarray}
The equation of the motion for the particle density operator will consists of two contributions
\begin{eqnarray}\label{4.53}
\frac{d n(\vec{r})}{d t} = - \frac{i}{\hbar}[n(\vec{r}), H_{1} ] - \frac{i}{\hbar}[n(\vec{r}), H_{2} ]. 
\end{eqnarray}
The first contribution is
\begin{eqnarray}\label{4.54}
[n(\vec{r}), H_{1} ] = \sum_{mni} \sum_{\sigma}F_{nm}(\vec{r})( t_{mi}a^{\dag}_{n \sigma}a_{i  \sigma} -  t_{in}a^{\dag}_{i \sigma}a_{m \sigma}).   
\end{eqnarray}
Here the notation was introduced
\begin{equation}\label{4.55}
F_{nm}(\vec{r}) =  w^{*}(\vec{r} - \vec{R}_{n}) w(\vec{r} - \vec{R}_{m}).  
\end{equation}
In the Bloch representation for the particle density operator one finds
\begin{eqnarray}\label{4.56}
[n(\vec{k}), H_{1} ] = \sum_{mni} \sum_{\sigma}F_{nm}(\vec{k})( t_{mi}a^{\dag}_{n \sigma}a_{i  \sigma} -  t_{in}a^{\dag}_{i \sigma}a_{m \sigma}),   
\end{eqnarray}
where
\begin{equation}\label{4.57}
F_{nm}(\vec{k}) = \int d^{3}r  \exp[-i{\vec k}{\vec r} ]F_{nm}(\vec{r}) = \int d^{3}r  \exp[-i{\vec k}{\vec r} ]w^{*}(\vec{r} - \vec{R}_{n}) w(\vec{r} - \vec{R}_{m}).  
\end{equation}
For the second contribution $[n(\vec{r}), H_{2}]$ we find
\begin{eqnarray}\label{4.58}
[n(\vec{r}), H_{2} ] = \frac{1}{2} \sum_{mn} \sum_{fst}\sum_{\sigma \sigma'}F_{nm}(\vec{r})\cdot \\ \nonumber  \Bigl (
\langle m f|H_{2}|s t \rangle a^{\dag}_{m \sigma}a^{\dag}_{f \sigma'}a_{t \sigma'}a_{s \sigma} 
- \langle f m|H_{2}|s t \rangle a^{\dag}_{m \sigma'}a^{\dag}_{f \sigma}a_{t \sigma'}a_{s \sigma} \\ \nonumber
+  \langle f s|H_{2}|t n \rangle a^{\dag}_{f \sigma}a^{\dag}_{s \sigma'}a_{t \sigma}a_{n \sigma'}  
- \langle f s |H_{2}|n t \rangle a^{\dag}_{f \sigma}a^{\dag}_{s \sigma'}a_{t \sigma'}a_{n \sigma}  \Bigr ). \nonumber
\end{eqnarray}
For the single-band Hubbard Hamiltonian the last equation will take the form
\begin{equation}\label{4.59}
[n(\vec{r}), H_{2} ] =  U \sum_{mn} \sum_{\sigma}F_{nm}(\vec{r}) a^{\dag}_{n \sigma}a_{m \sigma} (n_{m - \sigma} - n_{n - \sigma}).
\end{equation}
The direct calculations give for the case of electrons on a lattice ($e$ is a charge of an electron)
\begin{eqnarray}\label{4.60}
\frac{d n(\vec{r})}{dt} = \,\,\, \\ \nonumber \frac{e \hbar}{2mi} \sum_{ij} \sum_{\sigma} [w^{*}(\vec{r} - \vec{R}_{i})\nabla^{2} w(\vec{r} - \vec{R}_{j} ) - 
\nabla^{2} w^{*}(\vec{r} - \vec{R}_{i}) w(\vec{r} - \vec{R}_{j} )] a^{\dag}_{i \sigma}a_{j  \sigma}  -  \\ \nonumber
- ie U \sum_{ij} \sum_{\sigma}F_{ij}(\vec{r}) a^{\dag}_{i \sigma}a_{j \sigma} (n_{j - \sigma} - n_{i - \sigma}).
\end{eqnarray}
Taking into account that
\begin{eqnarray}\label{4.61}
\textrm{div} \vec{j}(\vec{r}) = \\ \nonumber \frac{e \hbar}{2mi} \sum_{ij} \sum_{\sigma} [w^{*}(\vec{r} - \vec{R}_{i})\nabla^{2} w(\vec{r} - \vec{R}_{j} ) - 
\nabla^{2} w^{*}(\vec{r} - \vec{R}_{i}) w(\vec{r} - \vec{R}_{j} )] a^{\dag}_{i \sigma}a_{j  \sigma},  
\end{eqnarray}
we find
\begin{equation}\label{4.62}
\frac{d n(\vec{r})}{dt} = - \textrm{div} \vec{j}(\vec{r}) - 
ie U \sum_{ij} \sum_{\sigma}F_{ij}(\vec{r}) a^{\dag}_{i \sigma}a_{j \sigma} (n_{j - \sigma} - n_{i - \sigma}).
\end{equation}
This unusual result was analyzed critically by many authors. The proper definition of the
current operator for the Hubbard model has been the subject of intensive discussions~\cite{oh70,ad70,kk71,bene,baka,sada,cabi,muk,ma75,ma76,gs76}.
To clarify the situation let us consider the "total position operator" for our system of the electrons on a lattice
\begin{equation}\label{4.63}
\vec{R}  = \sum_{j=1}^{N} \vec{R}_{j}.
\end{equation}
In the "quantized" picture it has the form
\begin{eqnarray}\label{4.64}
\vec{R} = \sum_{j} \int d^{3}r   \varPsi^{\dag}(\vec{r})\vec{R}_{j} \varPsi (\vec{r})   \\ \nonumber
 =    \sum_{j} \sum_{mn}\sum_{\mu} \int d^{3}r \vec{R}_{j} w^{*}(\vec{r} - \vec{R}_{m}) w(\vec{r} - \vec{R}_{n} )a^{\dag}_{m \mu}a_{n \mu} \\ \nonumber
 = \sum_{j} \sum_{m}\sum_{\mu} \vec{R}_{j}a^{\dag}_{m \mu}a_{m \mu},
\end{eqnarray}
where we took into account the relation
\begin{equation}\label{4.65}
\int d^{3}r  w^{*}(\vec{r} - \vec{R}_{m}) w(\vec{r} - \vec{R}_{n} )  = \delta_{mn}.
\end{equation}
We find that

\begin{eqnarray}\label{4.66}
[\vec{R},a^{\dag}_{i \sigma}]_{-}  = \sum_{m}\vec{R}_{m}a^{\dag}_{i \sigma}, \\ \nonumber
[\vec{R},a_{i \sigma}]_{-}  = - \sum_{m}\vec{R}_{m}a_{i \sigma},\\ \nonumber
[\vec{R},a^{\dag}_{i \sigma}a_{i \sigma}]_{-}  = 0.
\end{eqnarray}
Let us consider the local particle density operator $n_{i \sigma} = a^{\dag}_{i \sigma}a_{i  \sigma}.$  
\begin{eqnarray}\label{4.67}
\frac{d n_{i \sigma}}{dt} = - \frac{i}{\hbar} [n_{i \sigma}, H]_{-} = 
\sum_{j}  t_{ij} ( a^{\dag}_{i \sigma}a_{j  \sigma} -  a^{\dag}_{j \sigma}a_{i \sigma}).
\end{eqnarray}
It is clear that the current operator should be defined on the basis of the equation
\begin{equation}\label{4.68}
\vec{j} = e  \Bigl ( \frac{-i}{\hbar}\Bigr ) [\vec{R}, H ]_{-}.  
\end{equation}
Defining the so-called polarization operator~\cite{oh70,kk71,sada,cabi}
\begin{equation}\label{4.69}
\mathcal{P} = e \sum_{m}\sum_{\sigma} \vec{R}_{m}n_{m \sigma}, 
\end{equation}
we find the current operator in the form
\begin{equation}\label{4.70}
\vec{j} = \mathcal{\dot{P}} = e \Bigl ( \frac{-i}{\hbar}\Bigr )\sum_{mn}\sum_{\sigma}( \vec{R}_{m} - \vec{R}_{n}) t_{mn} a^{\dag}_{m \sigma}a_{n \sigma}. 
\end{equation}
This expression of the current operator is  a suitable formulae for  studying of the transport properties of the systems of correlated electron on
a lattice~\cite{ca75,mald77,km96}. The consideration carried out in this section demonstrate explicitly the specific features of the many-particle
interacting systems on a lattice.
%
%
%
%
\subsection{Electron-lattice interaction in metals}
%
In order to understand quantitatively the electrical,
thermal and superconducting properties of metals and their alloys one
needs a proper description an electron-lattice interaction~\cite{grim}. In the physics of molecules~\cite{sla2}
the concept of an intermolecular force requires that an effective separation of the nuclear and electronic 
motion can be made. This separation is achieved in the Born-Oppenheimer approximation~\cite{sla2,baer}. Closely related to the
validity of the Born-Oppenheimer approximation is the notion of adiabaticity.  The adiabatic approximation is 
applicable if the nuclei is much slower than the electrons. The Born-Oppenheimer approximation consists of separating the nuclear 
motion and in computing only the electronic wave functions and energies for fixed position of the nuclei. In the
mathematical formulation of this approximation, the total wave function is assumed in the form of a product both of
whose factors can be computed as solutions of two separate Schr\"{o}dinger equations. In most applications the separation is valid
with sufficient accuracy, and the adiabatic approach is reasonable, especially if the electronic properties of
molecules are concerned.\\ The conventional physical picture of a metal adopts these ideas~\cite{grim,ch61} and assumes that the
electrons and ions are essentially decoupled from one another with an error which involves the small parameter
$m/M$, the ratio between the masses of the electron and the ion. The qualitative arguments for this statement are
the following estimations. The maximum lattice frequency is of the order $10^{13} sec^{-1}$ and is quite small 
compared with a typical atomic frequency. This latter frequency is of order of $ 10^{15} sec^{-1}$. If the
electrons are able to respond in times of the order of atomic times then they will effectively be following the motion
of the lattice instantaneously at all frequencies of vibration. In other words the electronic motion will be
essentially adiabatic. This means that the wave functions of the electrons adjusting instantaneously to the motion
of the ions. It is intuitively clear that the electrons would try to follow the motion of the ions in such a way as
to keep the system locally electrically neutral. In other words, it is expected that the electrons will try to
respond to the motion of the ions in such a way as to screen out the local charge fluctuations.\\
The construction of an electron-phonon interaction requires the separation of the Hamiltonian describing mutually
interacting electrons and ions into terms representing electronic quasiparticles, phonons, and a residual 
interaction~\cite{zi,ha,call,grim,ch61,har71}. For the simple metals the
interaction between the electrons and the ions can be described within the pseudopotential method or the muffin-tin
approximation. These methods could not handled well the $d$ bands in the transition metals. They are too narrow to be
approximated as free-electron-like bands but too broad to be described as core ion states.
The electron-phonon interaction in solid is usually described by the Fr\"{o}hlich Hamiltonian~\cite{grim,fr}. We consider below
the main ideas and approximations concerning to the derivation of the explicit form of the electron-phonon interaction
operator. \\ Consider the total
Hamiltonian for the electrons with coordinates $\vec{r_{i}}$ and the ions with coordinates $\vec{R_{m}}$, with the electron cores which 
can be regarded as tightly bound to the nuclei.
The Hamiltonian of the $N$ ions is 
\begin{eqnarray}\label{4.71}
H = - \frac{\hbar^{2}}{2M}\sum_{m=1}^{N} \nabla^{2}_{\vec{R_{m}}} 
- \frac{\hbar^{2}}{2m}\sum_{i=1}^{ZN} \nabla^{2}_{\vec{r_{i}}}  
+ \frac{1}{2} \sum_{i,j=1}^{ZN} \frac{e^{2}}{|\vec{r_{i}} - \vec{r_{j}}| }  + \\ \nonumber  
\sum_{n>m} V_{i}(\vec{R_{m}} - \vec{R_{n}}) + 
\sum_{m=1}^{N}U_{ie}(\vec{r_{i}};\vec{R_{m}}).
\end{eqnarray}
Each ion is assumed to contribute $Z$ conduction electrons with coordinates
$\vec{r_{i}}$ $ ( i = 1, \ldots, ZN)$. The first two terms in Eq.(\ref{4.71}) are the kinetic energies of the electrons and the ions.
The third term is the direct electron-electron Coulomb interaction between the conduction electrons. The next two terms
 are short for the potential energy for direct ion-ion interaction and the potential energy of 
the $ZN$ conduction electrons moving in the field from the nuclei and the ion core electrons, when the ions take
instantaneous position $\vec{R_{m}}$ $ ( m = 1, \ldots, N)$.
The term $V_{i}(\vec{R_{m}} - \vec{R_{n}})$ is the interaction potential of the ions with each other, 
while $U_{ie}(\vec{r_{i}};\vec{R_{m}})$ represents the interaction between an electron at $\vec{r_{i}}$ and an ion at
$\vec{R_{m}}$. Thus the total Hamiltonian of the system can be represented as the sum of an electronic and ionic part.
\begin{equation}\label{4.72}
H = H_{e} + H_{i},  
\end{equation}
where
\begin{equation}\label{4.73}
H_{e} =  - \frac{\hbar^{2}}{2m}\sum_{i=1}^{ZN} \nabla^{2}_{\vec{r_{i}}}  
+ \frac{1}{2} \sum_{i,j=1}^{ZN} \frac{e^{2}}{|\vec{r_{i}} - \vec{r_{j}}| } +
\sum_{m=1}^{N}U_{ie}(\vec{r_{i}};\vec{R_{m}}),
\end{equation}
and
\begin{equation}\label{4.74}
H_{i}  = - \frac{\hbar^{2}}{2M}\sum_{m=1}^{N} \nabla^{2}_{\vec{R_{m}}} + \sum_{n>m} V_{i}(\vec{R_{m}} - \vec{R_{n}}).
\end{equation}
The Schr\"{o}dinger equation for the electrons in the presence of fixed ions is
\begin{equation}\label{4.75}
H_{e}\Psi( \vec{K},\vec{R},\vec{r}) = E ( \vec{K},\vec{R}) \Psi( \vec{K},\vec{R},\vec{r}),
\end{equation}
in which $\vec{K}$ is the total wave vector of the system, $\vec{R}$ and $\vec{r}$ denote the set of all electronic and
ionic coordinates. It is seen that the energy of the electronic system and the wave function of the electronic state
depend on the ionic positions. The total wave function for the entire system of electrons plus ions 
$\Phi( \vec{Q},\vec{R},\vec{r})$  can be expanded, in principle, with respect to the $\Psi$ as basis functions
\begin{equation}\label{4.76}
\Phi( \vec{Q},\vec{R},\vec{r}) = \sum_{\vec{K}} L( \vec{Q},\vec{K},\vec{R})\Psi( \vec{K},\vec{R},\vec{r}).  
\end{equation}
We start with the approach which uses a fixed set of basis states. Let us suppose that the ions of the crystal lattice
vibrate around their equilibrium positions $\vec{R}_{m}^{0}$ with a small amplitude, namely 
$\vec{R}_{m} = \vec{R}_{m}^{0} + \vec{u}_{m}, $ where $\vec{u}_{m}$ is the deviation from
the equilibrium position $\vec{R}_{m}^{0}$. Let us consider an idealized system in which the ions are fixed in these
positions. Suppose that the energy bands $E_{n}( \vec{k})$ and wave functions $\psi_{n}( \vec{k},\vec{r})$ are known.
As a result of the oscillations of the ions, the actual crystal potential differs from that of the rigid lattice. This
difference is possible to treat as a perturbation. This is the Bloch formulation of the electron-phonon interaction.\\
To proceed we must expand the potential energy $V(\vec{r} - \vec{R})$ of an electron at $\vec{r}$ 
in the field of an ion at $\vec{R}_{m}$ in the atomic displacement $\vec{u}_{m}$
\begin{equation}\label{4.77}
V(\vec{r} - \vec{R}_{m})  \simeq V(\vec{r} - \vec{R}_{m}^{0}) - \vec{u}_{m} \nabla V(\vec{r} - \vec{R}_{m}^{0}) + \ldots
\end{equation}
The perturbation potential, including all atoms in the crystal, is
\begin{equation}\label{4.78}
\widetilde{V} = - \sum_{m} \vec{u}_{m} \nabla V(\vec{r} - \vec{R}_{m}^{0}).
\end{equation} 
This perturbation will produce transitions between one-electron states with the corresponding matrix element
of the form
\begin{equation}\label{4.79}
M_{m k, n q} = \int \psi^{*}_{m}( \vec{k},\vec{r}) \widetilde{V} \psi_{n}( \vec{q},\vec{r}) d^{3}r.
\end{equation} 
To describe properly the lattice subsystem let us remind that the normal coordinate $Q_{\vec{q},\lambda}$ is defined by the 
relation~\cite{call,grim}
\begin{equation}\label{4.80}
(\vec{R}_{m} - \vec{R_{m}^{0}} ) = \vec{u}_{m} = ( \hbar / 2 \mathcal{N}M)^{1/2}  \sum_{\vec{q},\nu} Q_{\vec{q},\nu}
\vec{e}_{\nu}(\vec{q}) \exp (i \vec{q}\vec{R_{m}^{0}}),
\end{equation} 
where $\mathcal{N}$ is the number of unit cells per unit volume and $\vec{e}_{\nu}(\vec{q})$ is the polarization vector 
of the phonon. The Hamiltonian of the phonon subsystem in terms of normal coordinates is written as~\cite{call,grim}
\begin{equation}\label{4.81}
H_{i} = \sum_{\mu,\vec{q}}^{BZ} \left ( \frac{1}{2} P^{\dagger}_{\vec{q},\mu}P_{\vec{q},\mu} + 
\frac{1}{2} \Omega^{2}_{\vec{q},\mu} Q^{\dagger}_{\vec{q},\mu}Q_{\vec{q},\mu} \right ),
\end{equation} 
where $\mu$ denote polarization direction and the $\vec{q}$ summation is restricted to the Brillouin zone 
denoted as $BZ$. It is convenient to express $ u_{m}$ in terms of the second quantized phonon operators
\begin{equation}\label{4.82}
\vec{u}_{m}  = ( \hbar / 2 \mathcal{N}M)^{1/2}  \sum_{\vec{q},\nu} [( \omega^{1/2}_{\nu} ( \vec{q})]^{-1}
\vec{e}_{\nu}(\vec{q}) [\exp (i \vec{q}\vec{R_{m}^{0}}) b_{\vec{q},\nu} + 
\exp (-i \vec{q}\vec{R_{m}^{0}}) b^{\dagger}_{\vec{q},\nu} ],
\end{equation} 
in which $\nu$ denotes a branch of the phonon spectrum, $\vec{e}_{\nu}(\vec{q})$ is the eigenvector for a 
vibrational state of wave vector $\vec{q}$ and branch $\nu$, 
and $b^{\dagger}_{\vec{q},\nu} (b_{\vec{q},\nu})$ is a phonon creation (annihilation) operator. The matrix
element $M_{m k, n q}$ becomes
\begin{equation}\label{4.83}
M_{m k, n q} = - ( \hbar / 2 \mathcal{N}M)^{1/2}  \sum_{\vec{q},\nu} \Bigl ( \vec{e}_{\nu} (\vec{k} - \vec{q})
A_{mn}(\vec{k},\vec{q}) [\omega_{\nu} (\vec{k} - \vec{q})]^{-1/2}
 (b_{\vec{k} - \vec{q},\nu} +  b^{\dagger}_{\vec{q} - \vec{k},\nu}) \Bigr ).
\end{equation} 
Here the quantity $A_{mn}$ is given by
\begin{equation}\label{4.84}
A_{mn}(\vec{k},\vec{q}) = \mathcal{N} \int \psi^{*}_{m}( \vec{k},\vec{r}) \nabla V(\vec{r}) \psi_{n}( \vec{q},\vec{r}) d^{3}r.
\end{equation} 
It is well known~\cite{call,grim} that there is the distinction between normal processes in which 
vector $(\vec{k} - \vec{q})$ is inside the Brillouin zone and \emph{Umklapp} processes in which 
vector $(\vec{k} - \vec{q})$ must be brought back into the zone by addition of a reciprocal lattice vector $\vec{G}$.\\
The standard simplification in the theory of metals consists of replacement of the Bloch 
functions $\psi_{n}( \vec{q},\vec{r})$ by the plane waves
$$ \psi_{n}( \vec{q},\vec{r}) = \mathcal{V}^{-1/2}\exp (i \vec{q}\vec{r}),$$
in which $\mathcal{V}$ is the volume of the system. With this simplification we get
\begin{equation}\label{4.85}
A_{mn}(\vec{k},\vec{q}) = i (\vec{k} - \vec{q}) V((\vec{k} - \vec{q})). 
\end{equation} 
Introducing the field operators $\psi(\vec{r}),  \psi^{\dagger}(\vec{r})$  and the fermion second quantized creation
and annihilation operators $ a_{n\vec{k}}^{\dagger}, \, a_{n\vec{k}} $    for an electron of wave vector $\vec{k}$ in band 
$n$ in the plane wave basis
$$\psi(\vec{r}) = \sum_{\vec{q} n} \psi_{n}( \vec{q},\vec{r})a_{n\vec{k}}$$
and the set of quantities 
$$\Gamma_{mn,\nu}(\vec{k},\vec{q}) =
 - \left( \hbar / 2 M \omega_{\nu} ( \vec{k}-\vec{q}) \right)^{1/2} \vec{e}_{\nu} (\vec{k} - \vec{q}) A_{mn}(\vec{k},\vec{q}),$$
we can write an interaction Hamiltonian for the electron-phonon system in the form
\begin{eqnarray}\label{4.86}
H_{ei} = \mathcal{N}^{1/2} \sum_{nl \nu}\sum_{\vec{k}\vec{q} }\Gamma_{mn,\nu}(\vec{k},\vec{q})  
\left( a_{n\vec{k}}^{\dagger}a_{l\vec{q}}  b_{\vec{k} - \vec{q},\nu} + 
a_{n\vec{k}}^{\dagger}a_{l\vec{q}}  b^{\dagger}_{\vec{q} - \vec{k},\nu} \right).
\end{eqnarray} 
This Hamiltonian describes the processes of  phonon absorption or emission by an electron in the lattice, which
were first considered by Bloch. Thus the electron-phonon interaction is essentially dynamical and affects the
physical properties of metals in a characteristic way.
\\ It is possible to show~\cite{grim} that in the Bloch momentum representation
the Hamiltonian of a system of conduction electrons in metal interacting with phonons will have the form
\begin{equation}\label{4.87}
H = H_{e} + H_{i} + H_{ei},
\end{equation} 
where
\begin{equation}\label{4.88}
 H_{e} = \sum_{\vec{p} } E(\vec{p}) a^{\dagger}_{\vec{p}}a_{\vec{p}},
\end{equation} 
\begin{equation}\label{4.89}
H_{i} = \frac{1}{2} \sum_{\vec{q},\nu}^{|\vec{q}|< q_{m} } \omega_{\nu} ( \vec{q})( b^{\dagger}_{\vec{q}}b_{\vec{q}} +
b^{\dagger}_{-\vec{q}}b_{-\vec{q}}),
\end{equation} 
\begin{equation}\label{4.90}
 H_{ei} = \sum_{\nu} \sum_{\vec{p'}=\vec{p}+\vec{q} + \vec{G} } \Gamma_{\vec{q} \nu}(\vec{p}-\vec{p'})a^{\dagger}_{\vec{p'}}a_{\vec{p}}
 (b_{\vec{q}\nu} + b^{\dagger}_{-\vec{q}\nu}).
\end{equation} 
The Fr\"{o}hlich model ignores the \emph{Umklapp} processes $(\vec{G} \neq 0)$ and transverse phonons and takes  the
unperturbed electron and phonon energies as
$$E(\vec{p}) = \frac{\hbar^{2}p^{2}}{2m} - E_{F}, \quad \omega( \vec{q}) = v^{0}_{s}q, 
\quad (q < q_{m}). $$
Here $v_{s}$ is the sound velocity of the free phonon. The other notation are:
$$|\vec{q}|= q,\quad |\vec{p}|= p, \quad q_{m} = (6\pi^{2}n_{i})^{1/3}.$$
Thus we obtain
\begin{equation}\label{4.91}
 H_{ei} = \sum_{p,q} v(q)a^{\dagger}_{p+q}a_{p}
 (b_{q} + b^{\dagger}_{-q}),
\end{equation} 
where $v(q)$ is the Fourier component of the interaction potential
$$ v(q) = g (\omega(q)/2 )^{1/2}, \quad g = [2E_{F}/3 M n_{i}v^{2}_{s}]^{1/2}.$$
Here $n_{i}$ is the ionic density.
The point we should like to emphasize in the present context is that the derivation of this Hamiltonian is
based essentially on the plane wave representation for the electron wave function.
%
\subsection{Modified tight-binding approximation}
%
%
%
Particular properties of the transition metals, their alloys and compounds follow, to a great extent,
from the dominant role of $d$-electrons.
The natural approach to description of electron-lattice effects in such type of materials is
the modified tight-binding approximation (MTBA).
The electron-phonon matrix element in the Bloch picture is taken between electronic states of the undeformed
lattice.  For transition metals it is not easy task to estimate the electron-lattice interaction matrix 
element due to the anisotropy and other factors~\cite{mor,wbut,but,var}. 
There is an alternative description, introduced by Fr\"{o}hlich~\cite{mitr,mit,mi1} and which was termed the modified tight-binding 
approximation (MTBA). In this approach  the electrons  are moving adiabatically with the ions. Moreover, the coupling
of the electron to the displacement of the ion nearest to it, is already included in zero order of approximation. 
This is the basis of modified tight-binding calculations of the electron-phonon interaction which purports to remove certain difficulties
of the conventional Bloch tight-binding approximation for electrons in narrow band. The standard Hubbard Hamiltonian should be rederived
in this approach in terms of the new basis wave functions for the vibrating lattice. This was carried 
out by Barisic, Labbe and Friedel~\cite{blf}. They derived a model Hamiltonian which
is a generalization of the single-band Hubbard model~\cite{hub} including the lattice vibrations. 
The hopping integral $t_{ij}$ of the single-band Hubbard model (\ref{4.45}) is given by 
\begin{equation}\label{4.92}
t_{ij} = \int d^{3}r w^{*}({\vec r} -{\vec R_{j}})\left( \frac{\hbar^{2}p^{2}}{2m} + \sum_{l} V_{sf}(\vec{r} - \vec{R}_{l})\right)
w({\vec r} -{\vec R_{i}}).
\end{equation} 
Here we assumed that $V_{sf}$ is a short-range, self-consistent potential of the lattice suitable screened by outer
electrons. Considering small vibrations of ions we replace in Eq.(\ref{4.92}) the ion position ${\vec R}_{i}$ by
$({\vec R}^{0}_{i} + {\vec u}_{i})$ , i.e. its equilibrium position plus displacement. The unperturbed electronic 
wave functions must be written as a Bloch sum of displaced and suitable (approximately) orthonormalized atomic-like
functions
\begin{equation}\label{4.93}
 \int d^{3}r w^{*}({\vec r} -{\vec R^{0}_{j}} - {\vec u}_{j})
w({\vec r} -{\vec R^{0}_{i}} - {\vec u}_{i}) \approx \delta_{ij}.  
\end{equation} 
As it follows from Eq.(\ref{4.93}), the creation and annihilation operators $a^{+}_{k\sigma}, a_{k\sigma}$ may be introduced
in the deformed lattice so as to take partly into account the adiabatic follow up of the electron upon the 
vibration of the lattice. The Hubbard Hamiltonian Eq.(\ref{4.45}) can be rewritten in the form~\cite{khp,kz}
\begin{equation}
\label{4.94} H =  t_{0}\sum_{i\sigma}h_{i\sigma} + \sum_{i\neq j \sigma}t({\vec R^{0}_{j}} + {\vec u}_{j}  - {\vec R}^{0}_{i} - {\vec u}_{i})a^{\dagger}_{i\sigma}a_{j\sigma} +
U/2\sum_{i\sigma}n_{i\sigma}n_{i-\sigma}.
\end{equation}
For small displacements ${\vec u}_{i}$, we may expand $t({\vec R})$ as
\begin{equation}
\label{4.95} 
t({\vec R^{0}_{j}} + {\vec u}_{j}  - {\vec R}^{0}_{i} - {\vec u}_{i}) \approx
t({\vec R^{0}_{j}}  - {\vec R}^{0}_{i} )  +  \frac{\partial t({\vec R})}{\partial {\vec R}}|_{{\vec R} = 
{\vec R^{0}_{j}}  - {\vec R}^{0}_{i}}  ( {\vec u}_{j} - {\vec u}_{i})
+ \ldots
\end{equation}
Using the character of the exponential decrease of the Slater and Wannier functions the following approximation
may be used~\cite{blf,khp,kz}
\begin{equation}
\label{4.96}
\frac{\partial t({\vec R})}{\partial {\vec R}} \simeq - q_{0} \frac{{\vec R}}{|{\vec R}|} t({\vec R}).
\end{equation}
Here $q_{0}$ is the Slater coefficient~\cite{sla3} originated in the
exponential decrease of the wave functions of $d$-electrons; $q^{-1}_{0}$ related to the range of the $d$ function and is
of the order of the interatomic distance. The Slater coefficients for various metals are tabulated~\cite{blf}. The typical values are given in Table 3.
%
%
%
%
\begin{table}
\label{tab3}
\begin{center}
\caption{ { \bf Slater coefficients }}
Table 3. Slater coefficients
\end{center}
\begin{center}
\begin{tabular}{|l||l|l|c|l|l|l|} \hline
$q_{0} \quad ( A^{-1})$ &   element& element& element& element& element& element\\
 \hline
$q_{0} = 0.93 $ & $Ti$ & $V$& $Cr$& $Mn$& $Fe$& $Co$ \\
 \hline
$q_{0} = 0.91$& $Zr$ & $Nb$& $Mo$& $Tc$& $Ru$ & $Rh$\\
 \hline
$q_{0} = 0.87$& $Hf$& $Ta$& $W$& $Re$& $Os$& $Ir$\\   
\hline  
\end{tabular}
\end{center}
\end{table}
%
%
%
\\ It is of use to rewrite the total model Hamiltonian of transition
metal  $H = H_{e} + H_{i} + H_{ei}$ in the quasi-momentum representation. We have
\begin{equation}\label{4.97}
H_{e} =  \sum_{k\sigma}E(k)a^{\dagger}_{k\sigma}a_{k\sigma} +
U/2N \sum_{k_{1}k_{2}k_{3}k_{4}G}a^{\dagger}_{k_{1}\uparrow}a_{k_{2}\uparrow}a^{\dagger}_{k_{3}\downarrow}
a_{k_{4}\downarrow} \delta (\vec {k}_{1} - \vec {k}_{2} + \vec {k}_{3} - \vec {k}_{4} + \vec {G}).
\end{equation} 
For the tight-binding electrons in crystals we 
use $E(\vec {k}) = 2 \sum_{\alpha}t(a_{\alpha}) \cos (k_{\alpha}a_{\alpha})$, where $t(\vec {a})$ is the
hopping integral between nearest neighbours, and $a_{\alpha} ( \alpha = x, y. z)$ denotes the lattice vectors
in a simple lattice with an inversion center.\\ The electron-phonon interaction is rewritten as
\begin{equation}\label{4.98}
H_{ei} =  \sum_{kk_{1}} \sum_{qG} \sum_{\nu \sigma}g^{\nu}_{kk_{1}}a^{\dagger}_{k_{1}\sigma}a_{k\sigma}
(b^{\dagger}_{q\nu} + b_{-q\nu}) \delta (\vec {k}_{1} - \vec {k} + \vec {q} + \vec {G}),
\end{equation} 
where
\begin{equation}\label{4.99}
g^{\nu}_{kk_{1}} = \Bigl (   \frac{1}{( \mathcal{N} M \omega_{\nu}(k))}\Bigr )^{1/2} I^{\nu}_{kk_{1}},
\end{equation} 
\begin{equation}\label{4.100}
I^{\nu}_{kk_{1}} = 2i q_{0} \sum_{\alpha}
t(\vec {a}_{\alpha})   \frac{\vec {a}_{\alpha} \vec{e}_{\nu}(\vec {k}_{1})}{|\vec {a}_{\alpha}|}
\left( \sin (\vec {a}_{\alpha} \vec k)
- \sin (\vec (a_{\alpha} \vec {k}_{1})\right),
\end{equation}
where $\mathcal{N}$ is the
number of unit cells in the crystal and M is the ion mass. The $\vec
e_{\nu}(\vec q)$ are the polarization vectors of the phonon modes. 
Operators $b^{\dagger}_{q\nu}$ and $  b_{q\nu}$ are the creation and annihilation phonon operators and
$\omega_{\nu}(k)$ are the acoustical phonon frequencies. 
Thus we can  describe~\cite{dee,dee1,khp,kz} the transition metal by
the one-band model which takes into consideration the electron-electron and electron-lattice interaction in the framework
of the MTBA. It is possible to rewrite (\ref{4.98}) in the following form~\cite{khp,kz}  
\begin{equation}\label{4.101}
H_{ei} = \sum_{\nu\sigma}\sum_{kq}
V^{\nu}(\vec k, \vec k + \vec q)Q_{\vec q\nu}a^{+}_{k+q\sigma}
a_{k\sigma},
\end{equation}
where
\begin{equation}\label{4.102}
V^{\nu}(\vec k, \vec k + \vec q) =
\frac{2iq_{0}}{( \mathcal{N} M )^{1/2}}\sum_{\alpha}
t(\vec a_{\alpha})e^{\alpha}_{\nu}(\vec q)\left( \sin \vec a_{\alpha} \vec k
- \sin \vec a_{\alpha} (\vec k - \vec q)\right).
\end{equation}
The  one-electron hopping $t(\vec a_{\alpha})$ is the overlap integral between a given site $\vec R_{m}$ and one
of the two nearby sites lying on the lattice axis $\vec a_{\alpha}$.
For the ion subsystem we have
\begin{equation}\label{4.103}
H_{i} =
\frac{1}{2} \sum_{q\nu}
\left( P^{+}_{q\nu}P_{q\nu} +
\omega^{2}_{\nu}(q)\right) Q^{+}_{q\nu}Q_{q\nu}  = \sum_{q\nu} \omega_{\nu}(q) ( b^{\dagger}_{q\nu}b_{q\nu}  + 1/2),
\end{equation}
where
$P_{q\nu}$ and $Q_{q\nu}$ are the normal coordinates. 
Thus, as in the
Hubbard model~\cite{hub}, the $d$- and $s(p)$-bands are replaced by
one effective band in our  model. However, the $s$-electrons give
rise to screening effects and are taken into effects by choosing proper
value of $U$ and the acoustical phonon frequencies. It was shown by Ashkenazi, Dacorogna and Peter~\cite{adp,asda}
that the MTBA approach for calculating electron-phonon coupling constant based on wave functions moving with
the vibrating atoms lead to same physical results as the Bloch approach within the harmonic approximation.
For transition metals and narrow band compounds the MTBA approach seems to be yielding more accurate results,
especially in predicting anisotropic properties.
%

%
\section{Charge and Heat Transport}
%
We now tackle the transport problem in a qualitative fashion. This crude picture has many 
obvious shortcomings. Nevertheless, the qualitative description of conductivity is instructive. Guided
by this instruction the results of the more advanced and careful calculations of the transport coefficients will be
reviewed below in the next sections.
\subsection{Electrical resistivity and Ohm law}
%
Ohm law is one of the equations used in the analysis of electrical circuits.
When a steady current flow through a metallic wire, Ohm law tell us that an electric field exists in the circuit, that like the current
this field is directed along the uniform wire, and that its magnitude is $\mathbf{J}/ \sigma,$ where $\mathbf{J}$ is the current density and
$\sigma$ the conductivity of the conducting material.
Ohm law  states that, in an electrical circuit, the current passing through most materials is directly proportional to the potential difference 
applied across them. A voltage source,  $V$, drives an electric current, $I$ , through resistor, $R$, the three quantities obeying Ohm  law: 
$V = IR$. \\ 
In other terms, this is written often as: $I = V/R,$ where $I$ is the current, $V$ is the potential difference, and $R$ is a proportionality 
constant called the \emph{resistance}. The potential difference is also known as the voltage drop, and is sometimes denoted by $E$ or $U$ instead of $V$. 
The SI unit of current is the \emph{ampere}; that of potential difference is the \emph{volt}; and that of resistance 
is the \emph{ohm}, equal to one \emph{volt} per \emph{ampere}. 
The law is named after the physicist Georg Ohm, who formulated it in 1826 .  
The continuum form of Ohm's law is often of use
\begin{equation}\label{5.1}
 \mathbf{J} = \sigma \cdot \mathbf{E},
\end{equation}
where $\mathbf{J}$ is the current density (current per unit area), $\sigma$ is the conductivity (which can be a tensor in anisotropic materials) and $E$ is 
the electric field . The common form $V = I \cdot R$ used in circuit design is the macroscopic, averaged-out version. The continuum form of the 
equation is only valid in the reference frame of the conducting material. \\ A conductor may be defined as a material within which there are free
charges, that is, charges that are free to move when a force is exerted on them by an electric field. Many conducting materials, mainly the
metals, show a linear dependence of $I$ on $V.$ The essence of Ohm law is this \textbf{linear}  relationship.
The important problem is the applicability of Ohm law. The relation $ R \cdot I = W$ is the generalized form of Ohm law for the current 
flowing through the system from terminal $A$ to terminal $B.$ Here $I$ is a steady \emph{dc} current, which is zero if the work $W$ done per unit
charge is zero, while $I \neq 0$ or $W \neq 0.$ If the current in not too large, the current $I$ must be simply proportional to $W.$ Hence 
one can write $ R \cdot I = W$, where the proportionality constant is called the \emph{resistance} of the two-terminal system. The basic equations
are:
\begin{equation}\label{5.2}  
 \vec{\nabla }\times  \vec{E}  =  4 \pi n,
\end{equation}
Gauss law, and
\begin{equation}\label{5.3}  
 \frac{\partial n}{\partial t}  +  \vec{\nabla }\times   \mathbf{J}  =  0,
\end{equation}
charge conservation law. Here $n$ is the number density of charge carriers in the system.  Equations (\ref{5.2})   and (\ref{5.3}) are fundamental. The Ohm law is not. However, in the absence of nonlocal effects,
Eq.(\ref{5.1}) is still valid.  In an electric conductor with finite cross section
it must be possible a surface conditions on the current density $\mathbf{J}.$ Ohm law does not permit this and cannot, therefore, be quite correct. It has to be
supplemented by terms describing a viscous flow. Ohm law is a statement of the behavior of many, but not all conducting bodies, and in this
sense should be looked upon as describing a special property of certain materials and not a general property of all matter.
%
\subsection{Drude-Lorentz model }
%
The phenomenological picture described above  requires the microscopic justification.
We are concerned in this paper with the transport of electric charge and heat by the
electrons in a solid. When our sample is in uniform thermal equilibrium the distribution of electrons over the eigenstates
available to them in each region of the sample is described by the Fermi-Dirac distribution function and the electric
and heat current densities both vanish everywhere. Non-vanishing macroscopic current densities arise whenever the
equilibrium is made non-uniform by varying either the electrochemical potential or the temperature 
from point to point in the sample. The electron distribution in each region of the crystal is then perturbed because 
electrons move from filled states to adjacent empty states.\\ 
The electrical conductivity of a material is determined by the mobile carriers and is
proportional to the number density of charge carriers in the system, denoted by $n$, and their
mobility, $\mu$, according to
\begin{equation}\label{5.4}
 \sigma \simeq n e\mu.
\end{equation}
Only in metallic systems the number density of charge carriers is large enough to make the electrical 
conductivity sufficiently large. 
The precise conditions under which one substance has a large
conductivity and another substance has low ones are determined by the microscopic physical
properties of the system such as energy band structure, carrier effective mass,  carrier mobility, lattice
properties, and the presence of impurities and imperfections.\\
Theoretical considerations of the electric conductivity were started by P. Drude within the  classical picture 
about hundred years ago~\cite{slat,drud}. He put 
forward a free electron model that assumes a relaxation of the independent charge carriers due to driving forces 
(frictional force and the electric field). The current density was written as
\begin{equation}\label{5.5}
 J  = \frac{n e^{2}}{m}  E \tau.  
\end{equation}
Here $\tau$ is the average time between collisions, $E$ is the electric field, $m$ and $e$ are the mass and the charge
of the electron.
The electric conductivity in the Drude model~\cite{slat} is given by 
\begin{equation}\label{5.6}
 \sigma =  \frac{n e^{2}\tau}{m}. 
\end{equation}
The time $\tau$ is called the mean lifetime or electron relaxation time. 
Then the Ohm's law can be expressed as the linear relation between current density $\vec{J}$ and
electric field $\vec{E}$
\begin{equation}\label{5.7}
 \mathbf{J} = \sigma \mathbf{E}.  
\end{equation}
The electrical resistivity  $R$  of the material is equal to
\begin{equation}\label{5.8}
R  = \frac{E}{J}.       
\end{equation}
The free-electron model of Drude is the limiting case of the total delocalization of the outer atomic electrons in 
a metal. The former valence electrons became conduction electrons. They move independently through the entire body of the metal;
the ion cores are totaly ignored. The theory of Drude was refined by Lorentz. Drude-Lorentz theory assumed that the
free conduction electrons formed an electron gas and were impeded in their motion through crystal by collisions with the
ions of the lattice. In this approach, the number of free electrons $n$ and the collision time $\tau$, related to the
mean free path $r_{l} = 2\tau v$ and the mean velocity $v$, are still adjustable parameters.\\
Contrary to this, in the Bloch model for the electronic structure of a crystal, 
though each valence electron is treated as an independent particle, it is recognized that the presence of the 
ion cores and the other valence electrons modifies the motion of that valence electron.\\ 
In spite of its simplicity, Drude model
contains some delicate points. Each electron changes its direction of propagation with an average period of $2\tau$.
This change of propagation direction  is mainly due to a collision of an electron with an impurity or defect and the 
interaction of electron with a lattice vibration. In an essence, $\tau$ is the average time of the electron motion 
to the first collision. Moreover, it is assumed that the  electron forgets its history on each collision, \emph{etc.}
To clarify these points let us consider the notion of the electron drift velocity. The electrons which contribute to
the conductivity have large velocities, that is large compared to the {\em drift velocity} which is due to the electric
field, because they are at the top of the Fermi surface and very energetic.
The drift velocity of the
carriers $v_{d}$ is intimately connected with the collision time  $\tau$
$$v_{d} = \alpha \tau,$$
where $\vec{\alpha}$ is a constant acceleration between collision of the charge carriers. In general, the mean
drift velocity of a particle over $N$ free path is
$$v_{d} \sim \frac{1}{2}  \alpha [\tau  + (\Delta t)^{2}/\tau]~.  $$
This expression shows that the drift velocity depends not only on the average value $\tau$ but also on the standard
deviation $(\Delta t) $ of the distribution of times between collisions. An analysis shows that the times between
collisions have an exponential probability distribution. For such a distribution, $\Delta t = \tau $ and one obtains
$v_{d} = \alpha \tau$ and $ J  =  n e^{2}/m   E \tau   $.  Assuming that the time between collisions always has the same value $\tau$ we find that
$(\Delta t) = 0$ and $v_{d} = \frac{1}{2}\alpha \tau$ and $ J  = n e^{2}/2m   E \tau   $. \\
The equations (\ref{5.7}) and (\ref{5.8}) are the most fundamental formulas in the physics of electron conduction.
Note, that  resistivity is not zero even at absolute zero, but is equal to the so called "residual resistivity".
For most typical cases it reasonably to assume that scattering by impurities or defects and scattering by lattice vibrations
are independent events. As a result, the  relation  (\ref{5.6}) will take place.
There is a big variety (and irregularity) of the resistivity values for the elements not speaking on the huge variety
of substances and materials~\cite{crc05,hand05,ldres,ldres2}.\\
In a metal with  spherical Fermi surface in the presence of an electric field $\vec{E}$, the Fermi surface would affect a $\Delta \vec{k}$ displacement,
$\Delta \vec{k} = \vec{k} - \vec{k}_{0}$. The simplest approximation is to suppose a rigid displacement 
of the Fermi sphere with a single relaxation time $\tau$,
\begin{equation}\label{5.9}
 \hbar \frac{d \vec{k}}{dt} + \hbar \frac{( \vec{k} - \vec{k_{0}})}{\tau} =   e\vec{E}.  
\end{equation}
Thus we will have at equilibrium
\begin{equation}\label{5.10}
\Delta \vec{k} = \frac{e\tau}{\hbar} \vec{E}.  
\end{equation}
The corresponding current density will take the form
\begin{equation}\label{5.11}
\vec{J} =  \frac{2}{(2\pi)^{3}} \int_{\Omega_{k}} e\vec{v}   d\Omega_{k} =   
\frac{2}{(2\pi)^{3}} \int_{S_{k}} e\vec{v} \Delta \vec{k}  \delta \vec{S}_{k_0}.
\end{equation}
We get from Eq.(\ref{5.7})
\begin{equation}\label{5.12}
\sigma =  \frac{2}{(2\pi)^{3}} \frac{e^{2}\tau}{\hbar}    
\int_{S_{k}}  \vec{v}   d \vec{S}_{k_0}.
\end{equation}
Let us consider briefly the frequency dependence of $\sigma.$ 
Consider a gas of noninteracting electrons of number density $n$ and collision time $\tau.$
At low frequencies collisions occur so frequently that
the charge carriers are moving as if within a viscous medium, whereas at high frequencies the charge carriers behave as if
they were free. These two frequency regimes are well-known in the transverse electromagnetic response 
of metals~\cite{mott,zi,ol,mea,slat,ros,hum}. The electromagnetic energy given to the electrons is lost in collisions
with the lattice, which is the "viscous medium". The relevant frequencies in this case satisfy the condition $\omega \tau \ll 1.$ Thus in a
phenomenological description~\cite{froh56} one should introduce a conductivity $\sigma$ and viscosity $\eta$ by
\begin{equation}\label{5.13}
\sigma =  \frac{e^{2}\bar{\tau}_{c}n }{m}, \quad \eta =    \frac{1}{2 } m \overline{v^{2}}n \bar{\tau}_{c}.
\end{equation}
On the other hand, for $\omega \tau \gg 1$ viscous effects are negligible, and the electrons behave as the nearly free particles. For 
optical frequencies they can move quickly enough to screen out the applied field. Thus, two different physical mechanisms are suitable 
in the different regimes defined by $\omega \tau \ll 1$ and $\omega \tau \gg 1.$\\
In a metal impurity atoms and phonons determine the scattering processes of the conduction electrons. The electrical force on the electrons
is $e \mathbf{E}.$ The "viscous" drag force is given by $- m \vec{v}/ \tau.$ Then one can write the equation
\begin{equation}\label{5.14}
m \dot{\vec{v}} = e \mathbf{E} - \frac{m \vec{v}}{\tau}.  
\end{equation}
For $\mathbf{E} \sim \exp (- i \omega t),$ the oscillating component of the current is given by
\begin{equation}\label{5.15}
\mathbf{J}(\omega)  = n e \vec{v}(\omega)  = \sigma(\omega) \mathbf{E}(\omega),  
\end{equation}
where 
\begin{equation}\label{5.16}
 \sigma(\omega) = \frac{\sigma_{0}}{1 - i \omega \tau},  \quad \sigma_{0} = \frac{e^{2}n \tau}{m}.
\end{equation}
For low frequencies we may approximate Eq.(\ref{5.14}) as $ \vec{v} \sim (e \tau / m)\mathbf{E}.$ For high frequencies we may neglect the
collision term, so $ \vec{v} \sim (e / m) \mathbf{E}.$ Thus the behavior of the conductivity as a function of frequency can be described
on the basis of the formula Eq.(\ref{5.16}).\\
Let us remark  on a residual resistivity, i.e. the resistivity at absolute zero. Since real crystals
always contain impurities and defects the resistivity is not equal zero even at absolute zero. If one assume that the scattering of a wave caused by impurities (or defects)
and by lattice vibrations are independent events, then the total probability for scattering will be the sum of the two
individual probabilities. The scattering probability is proportional to $1/\tau$, where $\tau$ is the mean lifetime  
or relaxation time of the electron motion. Denoting by $1/\tau_{1}$ the scattering probability due to impurities and defects
and by $1/\tau_{2}$ the scattering probability due to lattice vibrations we obtain for total probability the equality
\begin{equation}\label{5.17}
1/\tau =  1/\tau_{1} +  1/\tau_{2}; \quad   1/\sigma =  1/\sigma_{1} +  1/\sigma_{2}. 
\end{equation}
This relation is called Matthiessen rule. In practice, this relation is not fulfilled well (see Refs.~\cite{mat1,mat2}).
The main reason for the violation of the Matthiessen rule are the interference effects between phonon and impurity
contributions to the resistivity. Refs.~\cite{mat1,mat2}  give a comprehensive review of the subject of deviation from
Matthiessen rule and detailed critical evaluation of both theory and experimental data.
%
%
%
\subsection{The low- and high-temperature dependence of  conductivity}
%
One of the most informative and fundamental properties of a metal is the behavior of its electrical resistivity
as a function of temperature. The temperature dependence of the resistivity is a good indicator of important
scattering mechanisms for the conduction electrons. It can also suggests in a general way what the solid-state electronic
structure is like. There are two limiting cases, namely, the low temperature dependence of the resistivity for the case 
when  $ T \leq \theta_{D}$, where $\theta_{D}$ is effective Debye temperature, and the high temperature dependence 
of the resistivity, when $ T \geq \theta_{D}$. \\
The electrical resistivity of metals is due to two mechanisms, namely, (i) scattering of electrons on impurities
(static imperfections in the lattice), and (ii) scattering of electrons by phonons. Simplified treatment assumes that
one scattering process is not influenced by the other (Matthiessen rule). The first process is usually temperature
independent.
For a typical metal the electrical resistivity $R(T)$, as a function of the absolute temperature $T$, can be written as
\begin{equation}\label{5.18}
 R(T) =  R_{0} +  R_{i}(T),
\end{equation}
where $R_{0}$ is the residual electrical resistivity independent of $T$, and $R_{i}(T)$ is the temperature-dependent
intrinsic resistivity. The quantity $R_{0}$ is due to the scattering of electrons from chemical and structural 
imperfections. The term $R_{i}(T)$ is assumed to result from the interaction of electrons with other degrees of freedom 
of a crystal. In general, for the temperature dependence of the 
resistivity three scattering mechanisms are essential, (i) electron-phonon scattering, (ii) electron-magnon scattering 
and (iii) electron-electron scattering. The first one gives $T^{5}$ or $T^{3}$ dependence at low temperatures~\cite{zi}.
The second one, the magnon scattering is essential for the transition metals because some of them show ferromagnetic and
antiferromagnetic properties~\cite{mea}. This mechanism can give different temperature dependence due to the
complicated (anisotropic) dispersion of the magnons in various structures. The third mechanism, the electron-electron
scattering is responsible for the $R \sim T^{2}$ dependence of resistivity.\\ Usually, the temperature-dependent electrical resistivity
is tried to fit to an expression of the form
\begin{equation}\label{5.19}
 R(T) =  R_{0} +  R_{i}(T) = R_{0} + A T^{5} + B T^{2} + (C T^{3} ) + \ldots
\end{equation}
This dependence corresponds to Mathiessen rule, where the different terms are produced by different scattering
mechanisms. The early approach for  studying of
the temperature variation of the conductivity~\cite{mott,zi,som} was carried out by Sommerfeld, Bloch and Houston. Houston explained the
temperature variation of  conductivity applying the wave mechanics and assuming that  the wave-lengths of 
the electrons were in most cases long compared with the interatomic distance. He then solves the Boltzmann equation, 
using for the collision term an expression taken from the work of Debye and Waller on the thermal scattering of
x-rays. He obtained an expression for the conductivity as a function of a mean free path, which can be determined in
terms of the scattering of the electrons by the thermal vibrations of the lattice. Houston found a resistance proportional
to the temperature at high temperatures and to the square of temperature at low temperatures. The model used by Houston for the
electrons in a metal was that of Sommerfeld - an ideal gas in a structureless potential well. Bloch improved  this 
approach by taking the periodic structure of the lattice into account. For the resistance law at low temperatures both
Houston and Bloch results were incorrect. Houston realized that the various treatments of the mean free path would give
different variations of resistance with temperature. In his later work~\cite{hou} he also realized that the Debye 
theory of scattering was inadequate at low temperatures. He applied the Brillouin theory of scattering and arrived
at $T^{5}$ law for the resistivity at low temperatures and $T$ at high temperatures. 
Later on, it was shown by many authors~\cite{zi} that the distribution function obtained in the steady state under the action of an electric field 
and the phonon collisions does indeed lead to $R \sim T^{5}$. The calculations of the electron-phonon scattering
contribution to the resistivity by Bloch~\cite{fblo} and  Gruneisen~\cite{grun} lead to the following expression
\begin{equation}\label{5.20}
 R(T) \sim   \frac{T^{5}}{\theta^{6}} \int^{\theta/T}_{0} dz \frac{z^{5}}{(e^{z} - 1)(1 - e^{-z})}~,  
\end{equation}
which is known as the Bloch-Gruneisen law.\\
A lot of efforts has been devoted to the theory of transport processes in simple metals~\cite{mea,kav,bas}, such as the alkali
metals. The Fermi surface of these metals is nearly spherical, so that band-structure effects can be either neglected or
treated in some simple approximation. The  effect of the electron-electron interaction in these systems is not very
substantial. Most of the scattering is due to impurities and phonons. It is expected that the characteristic $T^{2}$
dependence of electron-electron interaction effects can only be seen at very low temperatures $T$, where phonon scattering
contributes a negligible $T^{5}$ term. In the non-simple metallic conductors, and in  transition metals, the
Fermi surfaces are usually far from being isotropic. Moreover, it can be viewed as the two-component systems~\cite{mald} where one carrier is an electron and the 
other is an inequivalent electron  (as in $s-d$ scattering) or a hole. It was shown that anisotropy such as that arising
from a nonspherical Fermi surface or from anisotropic scattering can yield a $T^{2}$ term in the resistivity at low
temperatures, due to the deviations from Mathiessen rule. This term disappears at sufficiently high $T$. The electron-electron
\emph{Umklapp} scattering contributes a $T^{2}$ term  even at high $T$. It was conjectured (see Ref.~\cite{md}) that the effective electron-electron interaction due to the exchange of
phonons should contribute to the electrical resistivity in exactly the same way as the direct Coulomb interaction, 
namely, giving rise to a $T^{2}$ term in the resistivity at low temperatures. The estimations of this contribution
show~\cite{ds} that it can alter substantially the coefficient of the $T^{2}$ term in the resistivity of simple and polyvalent
metals. The role of electron-electron scattering in transition metals was discussed in Refs.~\cite{rice,black,pot}  
A calculation of the electrical and thermal resistivity of $Nb$ and $Pd$ due to electron-phonon scattering was
discussed in Ref.~\cite{but1}. A detailed investigation~\cite{abut} of the temperature dependence of the 
resistivity  of $Nb$ and $Pd$ showed
that a simple power law fit cannot reconcile the experimentally observed behavior of the transition metals. 
Matthiessen rule breaks down and simple Bloch-Gruneisen theory is inadequate to account for the experimental data. In
particular, in Ref.~\cite{abut}   it has been shown that the resistivity of $Pd$ can be expressed by a $T^{2}$ function where,
on the other hand, the temperature dependence of the resistivity of $Nb$ should be represented by a function of $T$
more complicated than the $T^{3}$. It seems to be plausible that  the low-temperature behavior of  the 
resistivity  of transition metals may be described by a rational function of $(A T^{5} + B T^{2})$. This conjecture
will be  justified in section  \ref{nfermi}.\\
For real metallic systems the precise measurements show a quite complicated picture in which the term
$R_{i}(T)$ will not necessarily be proportional to $T^{5}$ for every metal (for detailed review see 
Refs.~\cite{mea,kav,bas}).
The purity of the samples and size-effect contributions and other experimental limitations can lead
to the deviations from the $T^{5}$ law. 
There are a lot of other reasons for such a deviation.  First, the electronic
structures of various pure metals differ very considerably. For example, the Fermi surface of sodium is nearly close to
the spherical one, but those of transition and rare-earth metals are much more complicated, having groups of  
electrons of very different velocities. The phonon spectra are also different for different metals. It is possible to
formulate that the $T^{5}$ law can be justified for a metal of a spherical Fermi surface and for a Debye phonon spectrum.
Moreover, the additional assumptions are  an
assumption that the electron and phonon systems are separately in equilibrium so that only one phonon is 
annihilated or created in an electron-phonon collision, that the \emph{Umklapp} processes can be neglected, and
an assumption of a constant volume at any temperature. Whenever these conditions are not satisfied in principle, 
deviations from $T^{5}$ law can be expected. This takes place, for example, in transition metals as a result of the
$s-d$ transitions~\cite{mott,mottm} due to the scattering of $s$ electrons by phonons. This process can be approximately
described as being proportional to $T^{\gamma}$ with $\gamma$ somewhere between 5 and 3. The $s-d$ model of electronic
transport in transition metals was developed by Mott~\cite{mott,mottm, mo36}. In this model, the motion of the electrons
is assumed to take place in the nearly-free-electron-like $s$-band conduction states. These electrons are then assumed
to be scattered into the localized $d$ states. Owing to the large differences in the effective masses of the $s$ and 
$d$ bands, large resistivity result.\\
In Ref.~\cite{webb}   the temperature of the normal-state electrical resistivity of
very pure niobium was reported. The measurements were carried out in the temperature range from the superconducting
transition ($T_{c} = 9.25 K$) to $300 K$ in zero magnetic field. The resistance-versus-temperature data were analyzed
in terms of the possible scattering mechanisms likely to occur in niobium. To fit the data a single-band model was
assumed. The best fit can be expressed as
\begin{eqnarray}\label{5.21}
 R(T)  = (4.98 \pm 0.7) 10^{-5} + (0.077 \pm 3.0)10^{-7}T^{2}  \,\,\,\,\,\,\,\,\,\,\,\, \\
+ (3.10 \pm 0.23)10^{-7}[ T^{3}J_{3}(\theta_{D} / T) /7.212] +
(1.84 \pm 0.26)10^{-10}[ T^{5}J_{5}(\theta_{D} / T) /124.4], \nonumber
\end{eqnarray}
where $J_{3}$ and $J_{5}$ are integrals occurring in the Wilson and Bloch theories~\cite{zi} and the best 
value for $\theta_{D}$, the effective Debye temperature, is $(270 \pm 10) K$. Over most of the temperature range
below 300 K, the $T^{3}$ Wilson term dominates. Thus it was concluded that inter-band scattering is quite important in
niobium. Because of the large magnitude of inter-band scattering, it was  difficult to determine the precise amount of
$T^{2}$  dependence in the resistivity.  Measurements of the electrical resistivity of the high purity specimens of 
niobium were carried out in Refs.~\cite{fr77,sh79,sh80,sh81}  It was shown that Mott theory is obeyed at high temperature in
niobium. In particular, the resistivity curve reflects the variation of the density of states at the Fermi surface when the
temperature is raised, thus demonstrating the predominance of $s-d$ transitions. In addition, it was found impossible
to fit a Bloch-Gruneisen or Wilson relation to the experimental curve. Several arguments were presented to indicate
that even a rough approximation of the Debye temperature has no physical significance and that it is necessary to take the
\emph{Umklapp} processes into account. Measurements of low-temperature electrical and thermal resistivity of 
tungsten~\cite{wag,sto}
and vanadium~\cite{gau} showed the effects of the electron-electron scattering between different branches of the Fermi surface in
tungsten and vanadium, thus concluding that electron-electron scattering does contribute measurable to electrical resistivity 
of these substances at low temperature.\\ 
In transition metal compounds, e.g. $MnP$ the
electron-electron scattering is attributed~\cite{akta} to be dominant at low temperatures, and furthermore the $3d$
electrons are thought to carry electric current. It is remarkable that the coefficient of the $T^{2}$ resistivity is  very 
large, about a hundred times those of $Ni$ and $Pd$ in which $s$ electrons coexist with $d$ electrons and electric current
is mostly carried by the $s$ electrons. This fact suggests strongly that in $MnP$ $s$ electrons do not exist at the
Fermi level and current is carried by the $3d$ electrons. This is consistent with the picture~\cite{lmf} that in
transition metal compounds the $s$ electrons are shifted up by the effect of antibonding with the valence electrons
due to a larger mixing matrix, compared with the $3d$ electrons, caused by their larger orbital extension.\\
It should be noted that the temperature coefficients of resistance  can be positive and negative in different materials.
A semiconductor material exhibits the temperature dependence of the resistivity quite different than in metal. A qualitative
explanation of this different behavior follows from considering the number of free charge carriers per unit volume, $n$,
and their mobility, $\mu$. In metals $n$ is essentially constant, but $\mu$ decreases with increasing temperature, owing to
increased lattice vibrations which lead to a reduction in the mean free path of the charge carriers. This decrease in mobility also occurs
in semiconductors, but the effect is usually masked by a rapid increase in $n$ as more charge carriers are set free and made
available for conduction. Thus, intrinsic semiconductors exhibit a negative temperature coefficient of resistivity. The
situation is different in the case of extrinsic semiconductors in which ionization of impurities in the crystal lattice
is responsible for the increase in $n$. In these, there may exist a range of temperatures over which essentially all the
impurities are ionized, that is, a range over which $n$ remains approximately constant. The change in resistivity is
than almost entirely due to the change in $\mu$, leading to a positive temperature coefficient.\\  
It is believed that the electrical resistivity of a solid at high temperatures is  primarily due to the
scattering of electrons by phonons and by impurities~\cite{zi}. It is usually assumed, in accordance with Matthiessen 
rule that the effect of these two contributions to the resistance are simply additive.  At high temperature (not 
lower than Debye temperature) lattice vibrations can be well represented by the Einstein model. In this case,
$1/\tau_{2} \sim T$, so that $1/\sigma_{2} \sim T$. If the properties and concentration of the lattice defects are 
independent of temperature, then $1/\sigma_{1}$ is also independent of temperature and we obtain
\begin{equation}\label{5.22}
 1/\sigma \simeq  a +  bT,
\end{equation}
where $a$ and $b$ are constants. 
However, this additivity is
true only if the effect of both impurity and phonon scattering can be represented by means of single relaxation times
whose ratio is independent of velocity~\cite{fri61}. It was shown~\cite{fri61} that the addition of impurities will
always decrease the conductivity. Investigations of the deviations from Matthiessen  rule at high temperatures in
relation to the electron-phonon interaction were carried out in Refs.~\cite{fr77,sh79,sh80,sh81}  It was shown~\cite{sh81}, in
particular, that changes in the electron-phonon interaction parameter $\lambda$, due to dilute impurities were caused
predominantly by interference between electron-phonon and electron-impurity scattering.\\
The electronic band structures of transition metals are extremely complicated and make calculations of the electrical 
resistivity due to structural disorder and phonon scattering very difficult. In addition the nature of the electron-phonon matrix
elements is not well understood~\cite{mor73}. The analysis of the matrix elements for scattering between states was
performed in Ref.~\cite{mor73} It was concluded that even in those metals where a fairly spherical Fermi surface exists,
it is more appropriate to think of the electrons as tightly bound in character rather than free electron-like. In addition,
the "single site" approximations are not likely to be appropriate for the calculation of the transport properties of
structurally-disordered transition metals. 
%
\subsection{Conductivity of alloys }
%
The theory of metallic conduction can be applied for explaining the conductivity of alloys~\cite{viss,moi,rep02,rep03,vysrep05,rep05}.
According to the Bloch-Gruneisen theory, the contribution of the electron-phonon interaction to the $dc$ electrical resistivity
of a metal at high temperatures is essentially governed by two factors, the absolute square of the electron-phonon coupling constant, and
the thermally excited mean square lattice displacement. Since the thermally excited mean lattice displacement is proportional to the
number of phonons, the high temperature resistivity $R$ is linearly proportional to the absolute temperature $T$, and the slope $dR/dT$ 
reflects the magnitude of the electron-phonon coupling constant. However, in many high resistivity  metallic alloys, the resistivity
variation $dR/dT$ is found to be far smaller than that of the constituent materials. In some cases $dR/dT$ is not even always positive.
There are two types of alloy, one of which the atoms of the different metals are distributed at random over the lattice points, another
in which the atoms of the components are regularly arranged. Anomalous behavior in electrical resistivity was observed 
in many amorphous and disordered substances~\cite{moi,ohka}. At low temperatures, the resistivity increases in $T^{2}$ instead
of the usual $T^{5}$ dependence. Since $T^{2}$ dependence is usually observed in alloys which include a large 
fraction of transition metals, it has been considered to be due to spins. In some metals, $T^{2}$ dependence might be
caused by spins. However, it can be caused by disorder itself. The calculation of transport coefficients in disordered transition 
metal alloys become a complicated task if the 
random fluctuations of the potential  are too large. It can be shown that strong potential fluctuations force the
electrons into localized states. Another anomalous behavior occurs in highly resistive metallic systems~\cite{moi,ohka} 
which is characterized by small
temperature coefficient of the electrical resistivity, or by even negative temperature coefficient. \\
According to  Matthiessen rule~\cite{mat1,mat2}, the electrical resistance of a dilute alloy is separable into a
temperature-dependent part, which is characteristic of the pure metal, and a residual part due to impurities. The
variation with temperature of the impurity resistance was calculated by Taylor~\cite{tay}. The total resistance
is composed of two parts, one due to elastic scattering processes, the other to inelastic ones. At the zero of 
temperature the resistance is entirely due to elastic scattering, and is smaller by an amount $\gamma_{0}$ than the
resistance that would be found if the impurity atom were infinitely massive. The factor $\gamma_{0}$ is typically
of the order of $10^{-2}$. As the temperature is raised the amount of inelastic scattering increases, while the
amount of elastic scattering decreases. However, as this happens the ordinary lattice resistance, which varies
as $T^{5}$, starts to become appreciable. For a highly impure specimen for which the lattice resistance at room
temperature, $R_{\theta}$, is equal to the residual resistance,  $R_{0}$, the total resistance at low temperatures
will have the form
\begin{equation}\label{5.23}
 R(T) \approx   10^{-2} (\frac{T}{\theta})^{2} + 500(\frac{T}{\theta})^{5} + R_{0}. 
\end{equation}
The first term arises from incoherent scattering and the second from coherent scattering, according to the
usual Bloch-Gruneisen theory. It is possible to see from this expression that $T^{2}$ term would be hidden by the lattice
resistance except at temperatures below $\theta/40$. This represents a resistance change of less than $10^{-5}R_{0}$,
and is not generally really observable. \\ In disordered metals the Debye-Waller factor in electron scattering  
by phonons may be an origin for negative temperature coefficient of the resistivity. The residual resistivity
may decreases as $T^{2}$ with increasing temperature because of the influence of the Debye-Waller factor. But resulting
resistivity increases as $T^{2}$ with increasing temperature at low temperatures even if the Debye-Waller factor is
taken into account.  It is worthy to note that the deviation from Matthiessen rule in electrical resistivity is large
in the transition metal alloys~\cite{yama,willi} and dilute alloys~\cite{grim1,berg}. In certain cases the temperature
dependence of the electrical resistivity of transition metal alloys at high temperatures can be connected with change
electronic density of states~\cite{is76}. The electronic density of states for $V-Cr$, $Nb-Mo$ and $Ta-W$ alloys have been
calculated in the coherent potential approximation. From these calculated results, temperature dependence of the electrical
resistivity $R$ at high temperature have been estimated. It was shown that the concentration variation of the temperature
dependence in $R/T$ is strongly dependent on the shape of the density of states near the Fermi level.\\
Many amorphous metals and disordered alloys exhibit a constant or negative temperature coefficient of the electrical 
resistivity~\cite{moi,ohka} in contrast to the
positive temperature coefficient of the electrical resistivity of normal metals. Any theoretical models of this phenomenon must include both the
scattering (or collision) caused by the topological or compositional disorder, and also the modifications to this collision induced by the
temperature or by electron-phonon scattering. If one assumes that the contributions to
the resistivity from scattering mechanisms other than the electron-phonon interaction are either independent of $T$, like impurity
scattering, or are saturated at high $T$, like magnetic scattering, the correlation between the quenched temperature dependence and high
resistivity leads one to ask whether the electron-phonon coupling constant is affected by the collisions of the electrons.\\
The effect of collisions on charge redistributions is the principal contributor to the electron-phonon
interaction in metals. It is studied as a mechanism which could explain the observed lack of temperature dependence of the electrical resistivity
of many concentrated alloys. The collision time-dependent free electron deformation potential can be derived from a self-consistent 
linearized Boltzmann equation. The results indicates that the collision effects are not very important for real systems. It can be understood assuming
that the charge redistribution produces only a negligible correction to the transverse phonon-electron interaction. In addition, although the
charge shift is the dominant contribution to the longitudinal phonon-electron interaction, this deformation potential is not affected 
by collisions until the root mean square electron diffusion distance in a phonon period is less than the Thomas-Fermi screening length.
This longitudinal phonon-electron interaction reduction requires collision times of the order of $10^{-19} sec$ in typical metals before it is effective.
Thus, it is highly probable that it is never important in real metals. Hence, this collision effect does not account for the observed, quenched
temperature-dependence of the resistivity of these alloys. However, these circumstances suggest that the validity of the adiabatic approximation,
i.e., the Born-Oppenheimer approximation, should be relaxed far beyond the previously suggested criteria. All these factors make the proper
microscopic formulation of the theory of the electron-phonon interaction in strongly disordered alloys a very complicated problem. A consistent
microscopic theory of the electron-phonon interaction in substitutionally disordered crystalline transition metal alloys was formulated by
Wysokinski and Kuzemsky~\cite{wk82} within the MTBA. This approach combines the Barisic, Labbe and Friedel model~\cite{blf} with the more complex details
of the CPA (coherent potential approximation).\\
The low-temperature resistivity of many disordered paramagnetic materials often shows a $T^{3/2}$ rather than a $T^{2}$
dependence due to spin-fluctuation-scattering resistivity. The coefficient of the $T^{3/2}$ term often
correlates with the magnitude of the residual resistivity as the amount of disorder is varied. A model
calculations that exhibits such behavior were carried our in Ref.~\cite{ris}  In the absence of disorder the
spin-fluctuations drag suppresses the spin-fluctuation  $T^{2}$ term in the resistivity. Disorder produces a finite
residual resistivity and also produces a finite spin-fluctuation-scattering rate.
%
\subsection{Magnetoresistance and the Hall effect}
%
The Hall effect and the magnetoresistance~\cite{hurd,hall,dre,arg60,sim,wan,pip68} are the manifestations of the Lorentz force on a subsystem of charge carries in
a conductor constrained to move in a given direction and subjected to a transverse magnetic field. Let us consider
a confined stream of a carriers, each having a charge $e$ and a steady-state velocity $v_{x}$ due to the applied
electric field $E_{x}$. A magnetic field $H$ in the $z$ direction produces a force $F_{y}$ which has the following form
\begin{equation}\label{5.24}
\vec{F}  = e \left( \vec{E} + (1/c) \vec{v} \times \vec{H}\right ).
\end{equation}
The boundary conditions lead to the equalities
\begin{equation}\label{5.25}
F_{y} = 0 = E_{y} -  (1/c) v_{x} H_{z}.                
\end{equation}
The transverse field $E_{y}$ is termed the Hall field $E_{y} \equiv E^{H}$ and is given by
\begin{equation}\label{5.26}
 E^{H}  =  (1/c) v_{x} H_{z}  =\frac{J_{x} H_{z} }{n e c} ; \quad   J_{x} = n e v_{x},          
\end{equation}
where $J_{x}$ is the current density and $n$ is the charge carrier concentration. The Hall field can be related
to the current density by means of the Hall coefficient $R_{H}$
\begin{equation}\label{5.27}
E^{H} = R_{H} J_{x} H_{z}; \quad  R_{H} = \frac{1}{nec}~.               
\end{equation}
The essence of the Hall effect is that Hall constant is inversely proportional to the charge carrier density $n$, and
that is negative for electron conduction and positive for hole conduction. A useful notion is the so-called Hall
angle which is defined by the relation
\begin{equation}\label{5.28}
\theta = \tan^{-1} (E_{y}/E_{x})~.              
\end{equation}
Thus the Hall effect may be regarded as the rotation of the electric field vector in the sample as a result of the
applied magnetic field. The Hall effect is an effective practical tool for studying the electronic characteristics 
of solids. The above consideration helps to understand how thermomagnetic effects~\cite{zi,gold10,bloh} can arise in the framework of 
simple free-electron model. The Lorentz force acts as a velocity selector. In other words, due to this force the slow
electrons will be deflected less than the more energetic ones. This effect  will lead to a temperature gradient in
the transverse direction. This temperature difference will result in a transverse potential difference due to the
Seebeck coefficient of the material. This phenomenon is called the Nernst-Ettingshausen effect~\cite{zi,gold10}.\\ It should be
noted that the simple expression for the Hall coefficient $R_{H}$ is the starting point only for the studies of the
Hall effect in metals and alloys~\cite{hurd,hall}. It implies $R_{H}$ is temperature independent and that $E^{H}$ varies
linearly with applied field strength. Experimentally the dependence $R_{H} = 1/n e c$ do not fit well the situation in
any solid metal. Thus there are necessity to explain these discrepancies. One way is to consider an effective carrier
density $n^{*}(n)$ which depends on $n$, where $n$ is now the mean density of electrons calculated from the valency. This
interrelation is much more complicated for the alloys where $n^{*}(n)$ is the function of the concentration of solute too.
It was shown that the high-field Hall effect reflects global properties of the Fermi surface such as its connectivity, 
the volume of occupied phase space, \emph{etc.} The low-field Hall effect depends instead on microscopic details of the dominant 
scattering process.A quantum-mechanical theory of transport of 
charge for an electron gas in a magnetic field which takes account of the quantization of the electron orbits has been given by Argyres~\cite{arg60}. \\
Magnetoresistance~\cite{pi,wan,lu,bud,wa2} is an important galvanomagnetic effect which is observed in a wide range of substances and under a variety
of experimental conditions~\cite{yang,font,many}. The transverse magnetoresistance is defined by
\begin{equation}\label{5.29}
\varrho_{MR}(H) = \frac{R(H) - R}{R} \equiv \frac{\Delta R}{R},            
\end{equation}
where $R(H)$ is the electrical resistivity measured in the direction perpendicular to the magnetic field $H,$ and 
$R$ is the resistivity corresponding to the zero magnetic field. The zero-field resistivity $R$ is the inverse of the zero-field
conductivity and is given approximately by
\begin{equation}\label{5.30}
R \sim \frac{m^{*} \langle v \rangle}{n e l},             
\end{equation}
according to the simple kinetic theory applied to a single-carrier system. Here $e$, $m^{*}$, $n$, $\langle v \rangle$ and $l$ are
respectively, charge, effective mass, density, average speed, and mean free path of the carrier. In this simplified picture the four
characteristics, $e$, $m^{*}$, $n$, and $\langle v \rangle$, are unlikely to change substantially when a weak magnetic field is applied. The change
in the mean free path  $l$ should then approximately determine the behavior of the magnetoresistance $\Delta R/R$ at low fields.\\
The magnetoresistance  practically of all conducting pure single-crystals has been experimentally found to be positive and a strong argument for
this were given on the basis of nonequilibrium statistical mechanics~\cite{pi}. In some substances, e.g. carbon, $CdSe$, $Eu_{2}CuSi_{3}$, \emph{etc.}, 
magnetoresistance is negative while in $CdMnSe$ is positive and much stronger than in $CdSe$~\cite{saw,jar,maju}. 
A qualitative interpretation of the magnetoresistance  suggests that those physical processes which make the mean free path larger for 
greater values  of $H$ should contribute to the negative magnetoresistance.  Magnetic scattering leads to negative magnetoresistance~\cite{yama1}
characteristic for ferro- or paramagnetic case, which comes from the suppression of fluctuation of the localized spins by the magnetic field. 
A comprehensive derivation of the quantum transport equation for electric and magnetic fields  was carried out by Mahan~\cite{gmah}. More detailed discussion of the various aspects of theoretical
calculation of the magnetoresistance in concrete substances are given in Refs.~\cite{pip68,yama1,toy,fuche,kuk} 
%
\subsection{Thermal conduction in solids}
%
Electric and thermal conductivities are intimately connected since the thermal energy also is mainly transported by the
conduction electrons.
The thermal conductivity~\cite{tri04,car} of a variety of substances, metals and nonmetals, depends on temperature region and varies
with temperature substantially~\cite{ldtc} (see Fig.3). 
\begin{figure}[bt]
\centerline{ \includegraphics[width=3.65in]{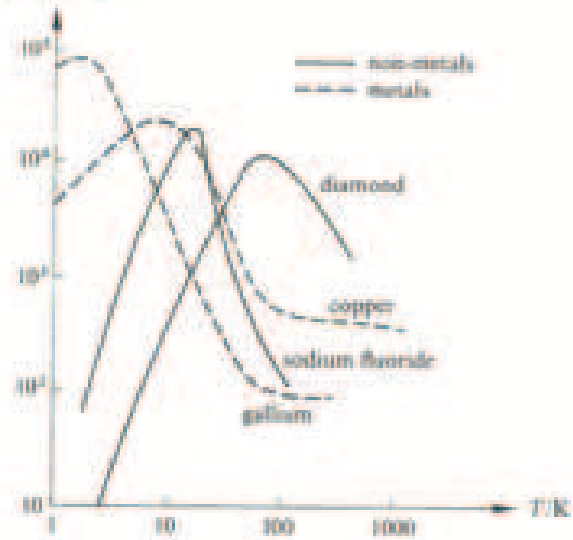}}
\vspace*{8pt}
\caption{Schematic form of the thermal conductivity of various materials} 
\label{f.3}
\end{figure}
Despite a rough similarity in the form of the curves for metallic and
nonmetallic materials there is a fundamental difference in the mechanism whereby heat is transported in these two types
of materials. In metals~\cite{tri04,pow} heat is conducted by electrons; in nonmetals~\cite{tri04,sla} it is conducted through coupled vibrations of the
atoms. The empirical data~\cite{ldtc} show that the better the electrical conduction of a metal, the better its thermal conduction.
Let us consider a sample with a temperature gradient $dT/dx$ along the $x$ direction. Suppose that the electron located
at each point $x$ has thermal energy $E(T)$ corresponding to the temperature $T$ at the point $x$. It is possible to
estimate the net thermal energy carried by each electron as
\begin{equation}\label{5.31}
E(T) - E(T + \frac{dT}{dx}\tau v \cos \theta) = - \frac{d E(T)}{dT} \frac{dT}{dx}\tau v \cos \theta.             
\end{equation}
Here we denote by $\theta$ the angle between the propagation direction of an electron and the $x$ direction and by $v$
the average speed of the electron. Then the average distance travelled in the $x$ direction by an electron until it
scatters is $\tau v \cos \theta $. The thermal current density $J_{q}$ can be estimated as
\begin{equation}\label{5.32}
J_{q} = - n \frac{d E(T)}{dT} \frac{dT}{dx}\tau v^{2} \langle \cos^{2} \theta  \rangle,            
\end{equation}
where $n$ is the number of electrons per unit volume. If the propagation direction of the electron is random, then
$\langle \cos^{2} \theta  \rangle  = 1/3$ and the thermal current density is given by
\begin{equation}\label{5.33}
J_{q}  = - \frac{1}{3} n \frac{d E(T)}{dT} \tau v^{2} \frac{dT}{dx}.              
\end{equation}
Here $n d E(T)/dT $ is the electronic heat capacity $C_{e}$ per unit volume. We obtain for the thermal 
conductivity $\kappa$ the following expression
\begin{equation}\label{5.34}
\kappa = \frac{1}{3}C_{e}\tau v^{2}; \quad  Q = - \kappa \frac{dT}{dx}.              
\end{equation}
The estimation of $\kappa$ for a degenerate Fermi distribution can be given by
\begin{equation}\label{5.35}
\kappa = \frac{1}{3} \frac{\pi^{2}k^{2}_{B} T}{2 \zeta_{0}} n \frac{6 \zeta_{0}}{5m}\tau = \frac{\pi^{2}k^{2}_{B} T}{5m}n,              
\end{equation}
where $ 1/2 mv^{2} = 3/5\zeta_{0} $ and $k_{B}$ is the Boltzmann constant. It is possible to eliminate $n\tau$ with the aid of 
equality $\tau = m \sigma /n e^{2}$. Thus we obtain~\cite{ber}
\begin{equation}\label{5.36}
\frac{\kappa}{\sigma} \cong \frac{\pi^{2}}{5}\frac{k^{2}_{B}}{e^{2}}T.                
\end{equation}
This relation is called by the Wiedemann-Franz law. The more precise calculation gives a more accurate factor
value for the quantity $\pi^{2}/5$ as $\pi^{2}/3$. The most essential conclusion to be drawn from the
Wiedemann-Franz law is that $\kappa/\sigma$ is proportional to $T$ and the proportionality constant is independent
of the type of metal. In other words, a metal having high electrical conductivity has a high thermal conductivity at a
given temperature. The coefficient $\kappa/\sigma T$ is called the Lorentz number. At sufficiently high temperatures, where
$\sigma$ is proportional to $1/T$, $\kappa$ is independent of temperature.
Qualitatively, the Wiedemann-Franz law is based upon the fact that the heat and electrical transport both involve the free 
electrons in the metal. 
The thermal conductivity increases with the average particle velocity since that increases the forward transport of energy. 
However, the electrical conductivity decreases with particle velocity increases because the collisions divert the electrons 
from forward transport of charge. This means that the ratio of thermal to electrical conductivity depends upon the average velocity squared, which is proportional to the 
kinetic temperature. \\ 
Thus, there are relationships between the transport 
coefficients of a metal in a strong magnetic field and a very low temperatures. 
Examples of  such relations are the Wiedemann-Franz law for the heat conductivity $\kappa$, which we rewrite in a more general form
\begin{equation}
\label{5.37}
   \kappa  = L T   \sigma, 
\end{equation}
 and the Mott rule~\cite{mot36} for the thermopower $S$
\begin{equation}
\label{5.38}
 S = e L T   \sigma^{-1}\frac{d \sigma}{d \mu}.
\end{equation}
Here T is the temperature, $\mu$ denotes the chemical potential. The Lorentz number $L = 1/3 (\pi k_{B} / e)^{2}$, where $k_{B}$  is the Boltzmann constant, 
is universal for all metals. Useful analysis of the Wiedemann-Franz law  and the Mott rule was carried out  by Nagaev~\cite{nag}.
%
%

\section{Linear Macroscopic Transport Equations }
%
We give here a brief refresher of the standard formulation of the macroscopic transport equations from the
most general point of view~\cite{gro}. One of the main problems of electron transport theory is the finding of the perturbed
electron distribution which determines the magnitudes of the macroscopic current densities. Under the standard
conditions it is reasonably to assume that the gradients of the electrochemical potential and the temperature are both very
small. The macroscopic current densities are then  linearly related to those gradients and the ultimate objective of the
theory of transport processes in solids (see Table 4). 
%
%
%
\begin{table}
\label{tab4}
\begin{center}
\caption{ { \bf Fluxes and Generalized Forces }}
Table 4. Fluxes and Generalized Forces
\end{center}
\begin{center}
\begin{tabular}{|l|l|l|l|} \hline
Process  & Flux & Generalized force &  Tensor character\\
 \hline
 Electrical conduction & $\vec{J}_{e}$ & $\nabla \phi$ & vector\\
 \hline
Heat conduction& $\vec{J}_{q}$ & $\nabla (1/T)$& vector \\
 \hline
Diffusion & Diffusion flux $\vec{J}_{p}$ & $-(1/T)[\nabla n] $&  vector\\ 
 \hline
Viscous flow  & Pressure tensor ${\bf P}$ & $-(1/T)\nabla \vec{u} $& Second-rank tensor\\
 \hline
 Chemical reaction& Reaction rate $W_{r}$& Affinity $a_{r}/T$ & Scalar \\ 
 \hline
\end{tabular}
\end{center}
\end{table}
%
%
Let $\eta$ and $T$ denote respectively the electrochemical potential and temperature
of the electrons. We suppose that both the quantities vary from point to point with small gradients $\nabla \eta$ and
$\nabla T$. Then, at each point in the crystal, electric and heat current densities $ \vec{J}_{e}$  and $ \vec{J}_{q}$
will exist which are linearly related to the electromotive force $\mathcal{\vec{E}} = 1/e \nabla \eta$ and
$\nabla T$ by the basic transport equations
\begin{equation}\label{6.19}
\vec{J}_{e} =  L_{11} \mathcal{\vec{E}} + L_{12} \nabla T,
\end{equation}
\begin{equation}\label{6.20}
\vec{J}_{q} =  L_{21}  \mathcal{\vec{E}}    +  L_{22} \nabla T.
\end{equation}
The coefficients $L_{11} $, $L_{12}$, $L_{21}$ and $L_{22}$ in these equations are the
transport coefficients which describe irreversible processes in linear approximation. We note that in a homogeneous
isothermal crystal, $\mathcal{\vec{E}} $ is equal to the applied electric field $ \vec{E}$. The basic transport equations
in the form (\ref{6.19}) and (\ref{6.20}) describe responses $\vec{J}_{e}$ and $\vec{J}_{q}$ under the influence 
of $\mathcal{\vec{E}}$ and $\nabla T$.
The coefficient $L_{11} = \sigma$ is the electrical conductivity. The other three 
coefficients, $L_{12}$, $L_{21}$ and $L_{22}$ have no generally accepted nomenclature because these quantities are
hardly ever measured directly. From the experimental point of view it is usually more convenient to fix $\vec{J}_{e}$ and
$\nabla T$ and then measure $\mathcal{\vec{E}} $ and $\vec{J}_{q}$. To fit the experimental situation the equations
(\ref{6.19}) and (\ref{6.20}) must be rewritten in the form 
\begin{equation}\label{6.21}
\mathcal{\vec{E}}  = R \vec{J}_{e}  + S \nabla T,
\end{equation}
\begin{equation}\label{6.22}
\vec{J}_{q} =  \Pi  \vec{J}_{e}     -   \kappa \nabla T,
\end{equation}
where
\begin{equation}\label{6.23}
 R = \sigma^{-1},
\end{equation}
\begin{equation}\label{6.24}
 S = - \sigma^{-1} L_{12},
\end{equation}
\begin{equation}\label{6.25}
 \Pi = L_{21} \sigma^{-1},
\end{equation}
\begin{equation}\label{6.26}
 \kappa = L_{21} \sigma^{-1}L_{12} = L_{22},
\end{equation}
which are known respectively as the resistivity, thermoelectric power, Peltier coefficient and thermal conductivity.
These are the quantities which are measured directly in experiments.\\
All the coefficients in the above equations are tensors of rank 2 and they depend on the magnetic induction 
field $ \vec{B}$ applied to the crystal. By considering crystals with full cubic symmetry, when $\vec{B} = 0$, one
reduces to a minimum the geometrical complications associated with the tensor character of the coefficients. In this case
all the transport coefficients must be invariant under all the operations in the point group $m3m$~\cite{woo}.
This high degree of symmetry implies that the coefficients must reduce to scalar multiples of the unit tensor and
must therefore be replaced by scalars. When $\vec{B} \neq 0$ the general form of transport tensors is complicated
even in cubic crystals~\cite{bloh,woo}. In the case, when an expansion to second order in $ \vec{B}$ is
sufficient the conductivity tensor takes the form
\begin{equation}\label{6.27}
 \sigma_{\alpha\beta} (\vec{B}) = \sigma_{\alpha\beta} (0) +  
 \sum_{\mu}\frac{\partial \sigma_{\alpha\beta} (0)}{\partial B_{\mu}} B_{\mu} + 
 \frac{1}{2} \sum_{\mu \nu} \frac{\partial^{2} \sigma_{\alpha\beta} (0)}{\partial B_{\mu} B_{\nu}} B_{\mu} B_{\nu} + \ldots
\end{equation}
Here the $(\alpha\beta)-$th element of $\sigma$ referred to the cubic axes $(0xyz)$. For the case when it is possible
to confine ourselves by the proper rotations in $m3m$ only, we obtain that
\begin{equation}\label{6.28}
\sigma_{\alpha\beta} (0) = \sigma_{0} \delta_{\alpha\beta},
\end{equation}
where $\sigma_{0}$ is the scalar conductivity when $\vec{B} = 0$. We also find that
$$ \frac{\partial \sigma_{\alpha\beta} (0)}{\partial B_{\mu}} = \varsigma \epsilon_{\alpha\beta\gamma},$$
$$\frac{\partial^{2} \sigma_{\alpha\beta} (0)}{\partial B_{\mu} B_{\nu}} = 
2 \xi \delta_{\alpha\beta} \delta_{\mu\nu} + \eta [\delta_{\alpha\mu} \delta_{\beta\nu} + 
\delta_{\alpha\nu} \delta_{\beta\mu}] + 2 \zeta \delta_{\alpha\beta} \delta_{\alpha\mu} \delta_{\alpha\nu},$$
where $\varsigma, \xi, \eta, \zeta$ are all scalar and $\epsilon_{\alpha\beta\gamma}$ is the three-dimensional 
alternating symbol~\cite{woo}.
Thus we obtain a relation between $ \vec{j} $ and $\mathcal{\vec{E}}$ (with $\nabla T  = 0$)
\begin{equation}\label{6.29}
 \vec{j}  = \sigma_{0} \mathcal{\vec{E}} + \varsigma \mathcal{\vec{E}}\times \vec{B} + 
 \xi \vec{B}^{2} \mathcal{\vec{E}} + \eta \vec{B} (\vec{B}\mathcal{\vec{E}} ) + \zeta \Phi \mathcal{\vec{E}},
\end{equation}
where $\Phi$ is a diagonal tensor with $\Phi_{\alpha \alpha} = B^{2}_{\alpha}$ (see Refs.~\cite{sm,bla}). The most
interesting transport phenomena is the electrical conductivity under homogeneous isothermal conditions. In general,
the calculation of the scalar transport coefficients $\sigma_{0}, \varsigma, \xi, \eta, \zeta$ is complicated task.
As was mentioned above, these coefficients are not usually measured directly. In practice, one measures the
corresponding terms in the expression for $\mathcal{\vec{E}}$ in terms of $\vec{j}$ up to terms of second order in
$\vec{B}$. To show this clearly, let us iterate the equation (\ref{6.29}). We then find that
\begin{equation}\label{6.30}
\mathcal{\vec{E}} = R_{0}\vec{j}  - R_{H} \vec{j}\times \vec{B} + R_{0} \bigl ( b \vec{B}^{2}\vec{j} +
c \vec{B} (\vec{B}\vec{j} ) + d \vec{\Phi} \vec{j} \bigr ),
\end{equation}
where 
\begin{equation}\label{6.31}
R_{0} = \sigma_{0}^{-1}, \quad R_{H} = \sigma_{0}^{-2} \varsigma,
\end{equation}
are respectively the low-field resistivity and Hall constant~\cite{pi,hall,dre,wan} and
\begin{equation}\label{6.32}
b = - R_{0}(\xi + R_{0}\varsigma^{2}); \quad c = - R_{0}(\eta - R_{0}\varsigma^{2}); \quad 
d = - \varrho_\zeta
\end{equation}
are the magnetoresistance coefficients~\cite{pi}. These are the quantities which are directly measured.
%
%
%
\section{Statistical Mechanics and Transport Coefficients}
%
%
The central problem of nonequilibrium statistical mechanics is to derive a set of equations which
describe irreversible processes from the reversible equations of motion~\cite{zub,zub81,kuz07,lib,vliet08,bb,zub61,zub70,toda92,kubo91,macl89,ichi94,zw,luz00,luz02}.
The consistent calculation of transport coefficients
is of particular interest because  one can get information on the microscopic
structure of the condensed matter. There exist a lot of theoretical methods for  calculation
of transport coefficients as a rule having a fairly restricted range of validity and
applicability~\cite{kuz07,zw,luz00,luz02,lvr,howl,fu,hl00,hml00}.  The most extensively developed theory of transport processes is that based on the
Boltzmann equation~\cite{kuz07,lib,vliet08,pe,petr09,fu}. However, this approach has strong restrictions and can reasonably be applied 
to a strongly rarefied gas of point particles. For systems in the state of statistical equilibrium, there
is the Gibbs distribution by means of which it is possible to calculate an average value 
of any dynamical quantity. No such universal distribution has been formulated for irreversible
processes. Thus, to proceed to the solution of problems of statistical mechanics of
nonequilibrium systems, it is necessary to resort to various approximate methods.
During the last decades, a number of schemes have been 
concerned with a more general and consistent approach to transport 
theory~\cite{zub,zub81,kuz07,zub61,zub70,toda92,kubo91,macl89,zw}. This field is very active and
there are many aspects to the problem~\cite{kuz07,lvr,howl,hl00,hml00}. 

%
\subsection{Variational principles for transport coefficients}
%
The variational principles for transport coefficients are the special techniques for bounding transport
coefficients, originally developed by Kohler, Sondheimer, Ziman, \emph{etc}. (see Refs.~\cite{zi,ichi94}). This approach
is equally applicable for  both the electronic and thermal transport. It starts from a Boltzmann-like
transport equation for the space-and-time-dependent distribution function $f_{\vec{q}}$ or the occupation number 
$n_{q}(\vec{r},t)$
of a single quasiparticle state specified by indices $q$ (e.g., wave vector for electrons or wave vector and polarization,
for phonons). Then it is necessary to find or fit a functional $F [f_{\vec{q}}, n_{q}( \vec{r},t )]$ which has a stationary point 
at the distribution $f_{\vec{q}}$,   $n_{q}(\vec{r},t)$ satisfying the transport equation, and whose 
stationary value is the suitable transport coefficient. By evaluating $F$ for a distribution only approximately 
satisfying the transport equation, one then obtains an upper or lower bound on the transport coefficients.
 Let us mention briefly  the phonon-limited electrical 
resistivity in metals~\cite{zi,grim}. With the neglect of phonon drag, the electrical resistivity can be written
\begin{equation}\label{7.9}
 R \leq  \frac{1}{2k_{B}T}   \frac{\int\int (\Phi_{\vec{k}} - \Phi_{\vec{k'}})^{2}  W(\vec{k},\vec{k'}) d\vec{k} d\vec{k'}}
 { | \int e \vec{v}_{\vec{k}} \Phi_{\vec{k}} \Bigl ( \partial f^{0}_{\vec{k}}/\partial \epsilon(\vec{k})\Bigr ) d\vec{k}|^{2} }.
\end{equation} 
Here $W(\vec{k},\vec{k'})$ is the transition probability from an electron state $\vec{k}$ to a state $\vec{k}'$,
 $\vec{v}_{\vec{k}}$ is the electron velocity and $f^{0}$ is the equilibrium Fermi-Dirac statistical factor. The variational
principle~\cite{zi} tell us that the smallest possible value of the right hand side obtained for any 
function $\Phi_{\vec{k}}$ is also the actual resistivity. In general we do not know the form of the function
$\Phi_{\vec{k}}$ that will give the right hand side its minimal value. For an isotropic system the correct choice is
$\Phi_{\vec{k}} = \vec{u}\vec{k},$ 
where $\vec{u}$ is a unit vector along the direction of the applied field. Because of its simple form, this function
is in general used also in calculations for real systems. The resistivity will be overestimated but it can still be a
reasonable approximation. This line of reasoning leads to the Ziman formula for the electrical resistivity
\begin{equation}\label{7.10}
 R \leq  \frac{3\pi}{e^{2} k_{B}T S^{2} \bar{k}^{2}_{F}}
 \sum_{\nu}\int\int \frac{d^{2}\vec{k}}{v} \frac{d^{2}\vec{k}'}{v'}  
 \frac{ (\vec{k} - \vec{k'})^{2} \omega_{\nu}( \vec{q}) A^{2}_{\nu} (\vec{k},\vec{k'})}
 { \bigl ( \exp(\hbar \omega_{\nu} ( \vec{q}) / k_{B}T  - 1 \bigr ) 
 \bigl (1 -  \exp(- \hbar \omega_{\nu} ( \vec{q}) / k_{B}T \bigr ) }. 
\end{equation} 
Here $\bar{k}_{F}$ is the average magnitude of the Fermi wave vector and $S$ is the free area of the Fermi surface.
It was shown in Ref.~\cite{nw}    that the formula for the electrical resistivity should not contains any electron-phonon
enhancements in the electron density of states. The electron velocities in Eq.(\ref{7.10}) are therefore the same as
those in Eq.(\ref{7.9}). 
%
\subsection{Transport theory  and electrical conductivity}
%
%
Let us summarize the results of the preceding sections.
It was shown above that in zero magnetic field, the quantities of main interest are the conductivity $\sigma$ (or the
electrical resistivity, $R = 1/\sigma$) and the thermopower $S$. When a magnetic field $\vec{B}$ is applied, the quantity of
interest is a magnetoresistance
\begin{equation}\label{7.11}
 \varrho_{MR} = \frac{R(\vec{B}) - R( B=0)}{R( B=0)}
\end{equation} 
and the Hall coefficient $R_{H}$. The third of the (generally) independent transport coefficients is the thermal conductivity $\kappa$.
The important relation which relates $\kappa$ to $R$ at low and high temperatures is the Wiedemann-Franz law~\cite{zi,langer}. In simple
metals and similar metallic systems, which have well-defined Fermi surface it is possible to interpret all the transport coefficients
mentioned above, the conductivity (or resistivity), thermopower, magnetoresistance, and Hall coefficient in terms of the rate of scattering of conduction
electrons from initial to final states on the Fermi surface. Useful tool to describe this in an approximate way is the Boltzmann transport
equation, which, moreover usually simplified further by introducing a concept of a relaxation time. In the approach of this kind we are interested in
low-rank velocity moments of the distribution function such as the current
\begin{equation}\label{7.12}
 \vec{j} = e \int \vec{v} f(\vec{v})d^{3}v. 
\end{equation} 
In the limit of weak fields one expects to find Ohm law $\vec{j} = \sigma \vec{E}. $ The validity of such a formulation of the Ohm law was 
analyzed by Bakshi and Gross~\cite{bagr}.
This approach was generalized and developed by many authors. The most popular  kind of consideration starts from the 
linearized Boltzmann equation which can be derived assuming weak scattering processes. 
For example, for the scattering of electrons
by "defect" (substituted atom or vacancy) with the scattering potential $V^{d}(\vec{r} )$ the perturbation theory gives
\begin{equation}\label{7.13}
 R \sim  \frac{3 \pi^{2}m^{2}\Omega_{0}N}{4\hbar^{3}e^{2} k_{F}^{2}}  
 \int^{2k_{F}}_{0} |\langle k+q | V^{d}(\vec{r} )|k \rangle |^{2} q^{3} dq, 
\end{equation} 
where $N$ and $\Omega_{0}$ are the number and volume of unit cells, $k_{F}$ the magnitude of the Fermi wavevector, the integrand 
$\langle k+q | V^{d}(\vec{r} )|k \rangle$ represents the matrix elements of the total scattering potential $V^{d}(\vec{r} )$, and
the integration is over the magnitude of the scattering wavevector $\vec{q} $ defined by $\vec{q} = \vec{k} -  \vec{k}'$.\\
Lax have analyzed in detail the general theory of carriers mobility
in solids~\cite{lax}. 
Luttinger and Kohn~\cite{lat,lat1},  Greenwood~\cite{gre},  Chester and Thellung~\cite{che}, and Fujiita~\cite{fu}
developed  approaches to the calculation of the electrical conductivity on the basis of the generalized 
quantum kinetic equations. The basic theory of transport for the case of scattering by static impurities has been given in the works of
Kohn and Luttinger~\cite{lat,lat1} and Greenwood~\cite{gre} (see also Ref.~\cite{ram98}). In these works the usual Boltzmann transport equation and its
generalizations were used to write down the equations for the occupation probability in the case of a weak, uniform
and static electric field. It was shown that in the case of static impurities the exclusion principle for the electrons
has no effect at all on the scattering term of the transport equation. In the case of scattering by phonons, where
the electrons scatter inelastically, the exclusion principle plays a very important role and the transport problem is more
involved. 
On the other hand
transport coefficients can be calculated by means of theory of the linear response  such as the  
Kubo formulae for the electrical conductivity.
New consideration of the transport processes in solids which involve weak assumptions 
and easily generalizable methods are of interest because they increase our understanding of 
the validity of the equations and approximations used~\cite{kuz07,vliet08,ram98}. Moreover, it permits one to consider
more general situations and apply the equations derived to a variety of physical systems.
%
%
\section{The Method of Time Correlation Functions}
%
The method of time correlation functions~\cite{zub,zub81,kuz07,kuz09,howl,hl00,hml00,lee} is an attempt to base a linear macroscopic transport equation theory directly
on the Liouville equation. In this approach one starts with complete $N$-particle distribution function which contains
all the information about the system. In the method of time correlation functions it is assumed that the 
$N$-particle distribution function can be written as a local equilibrium $N$-particle distribution function plus
correction terms. The local equilibrium function depends upon the local macroscopic variables, temperature, density and
mean velocity and upon the position and momenta of the $N$ particles in the system. The corrections to this distribution
functions is  determined on the basis of the Liouville equation. The main assumption is that at some initial time
the system was in local equilibrium (quasi-equilibrium)  but at later time is tending towards complete equilibrium.
 It was shown by many authors (for comprehensive review see Refs.~\cite{zub,zub81,kuz07}) that the suitable solutions to the Liouville equation
can be constructed and an expression for the corrections to local equilibrium  in powers of the gradients of the 
local variables can be found as well. The generalized linear macroscopic transport equations can be derived by
retaining the first term in the gradient expansion only. In principle, the expressions obtained in this way should depend
upon the dynamics of all $N$ particles in the system and apply to any system, regardless of its density.
%
%
\subsection{Linear response theory}
%
The linear response theory was anticipated in many works (see Refs.~\cite{zub,zub81,kuz07,mi95,hnak}   for details) on the theory of transport 
phenomena and nonequilibrium statistical mechanics. The important contributions have been made by many authors. 
By solving the Liouville equation to the first order in the 
external electric field, Kubo~\cite{kb56,k57,rk57,rk59,kb59} formulated an expression for the electric conductivity in microscopic terms.\\
He used linear response theory to give exact expressions for transport coefficients in terms of correlation
functions for the equilibrium system. To evaluate such correlation functions for any particular system, approximations have
to be made.\\
In this section we shall formulate briefly some general expressions for the conductivity tensor within the linear response theory.
Consider a many-particle system with the Hamiltonian of a system denoted by $H$.  This includes everything in the absence of the field; the
interaction of the system with the applied electric field is denoted by $H_{ext}$. The total Hamiltonian is
\begin{equation}\label{8.1}
 \mathcal{H} = H  + H_{ext}.
\end{equation} 
The conductivity tensor for an oscillating electric field  will be expressed in the form~\cite{kb56}
\begin{equation}\label{8.2}
\sigma_{\mu\nu} = \int_{0}^{\beta} \int_{0}^{\infty} \textrm{Tr} \rho_{0}  j_{\nu}(0)  j_{\mu}(t + i \hbar \lambda) e^{-i \omega t} dt d \lambda,
\end{equation} 
where $\rho_{0}$ is the density matrix representing the equilibrium distribution of the system in absence of the electric field 
\begin{equation}\label{8.3}
 \rho_{0} = e^{-\beta H}/ [\textrm{Tr} e^{-\beta H}],
\end{equation} 
$\beta$ being equal to $1/k_{B}T$. Here $j_{\mu}, j_{\nu}$ are the current operators of the whole system in the $\mu, \nu$ directions respectively, and
$j_{\mu}$ represents the evolution of the current as determined by the Hamiltonian $H$
\begin{equation}\label{8.4}
 j_{\mu}(t) = e^{i H t/\hbar}j_{\mu}e^{- i H t/\hbar}.
\end{equation} 
Kubo derived his expression Eq.(\ref{8.2}) by a simple perturbation calculation. He assumed that at $t = - \infty$ the system was in the
equilibrium represented by $\rho_{0}.$ A sinusoidal electric field was switched on at $t = - \infty,$ which however was assumed to be sufficiently weak.
Then he considered the equation of motion of the form
\begin{equation}\label{8.5}
i \hbar \frac{\partial}{\partial t} \rho = [H + H_{ext}(t), \rho].
\end{equation} 
The change of $\rho$ to the first order of $H_{ext}$ is given by
\begin{equation}\label{8.6}
 \rho - \rho_{0} =     \frac{1}{i \hbar}\int^{t}_{- \infty} e^{(-  H t'/i \hbar)} [H_{ext}(t'), \rho_{0}] e^{( H t'/i \hbar)} + O(H_{ext}).
\end{equation} 
Therefore the averaged current will be written as
\begin{equation}\label{8.7}
 \langle j_{\mu}(t) \rangle =  \frac{1}{i \hbar}\int^{t}_{- \infty} \textrm{Tr} [H_{ext}(t'), \rho_{0}] j_{\mu}(- t') dt' , 
\end{equation} 
where $H_{ext}(t')$ will be replaced by $ - e_{d}E(t')$, $e_{d}$ being the total dipole moment of the system. Using the relation
\begin{equation}\label{8.8}
 [A, e^{-\beta H}] =    \frac{\hbar}{i}      e^{-\beta H} \int_{0}^{\beta}   e^{\lambda H }[A, \rho]e^{- \lambda H} d \lambda,
\end{equation} 
the expression for the current can be transformed into Eq.(\ref{8.7}). The conductivity can be also written in terms of the correlation
function $ \langle j_{\nu}(0)  j_{\mu}(t) \rangle_{0}.$ The average sign $ \langle \ldots \rangle_{0}$ means the average over the density
matrix $\rho_{0}.$\\
The correlation of the spontaneous currents may be described by the correlation function~\cite{kb56}
\begin{equation}\label{8.9}
 \Xi_{\mu \nu}(t) = \langle j_{\nu}(0) j_{\mu}(t) \rangle_{0} =  \langle j_{\nu}(\tau) j_{\mu}(t+\tau) \rangle_{0}.
\end{equation} 
The conductivity can be also written in terms of these correlation functions. For the symmetric $("s")$ part of the conductivity 
tensor Kubo~\cite{kb56} derived a relation of the form
\begin{equation}\label{8.10}
Re \sigma^{s}_{\mu\nu}(\omega) =  \frac{1}{\varepsilon_{\beta}(\omega)}   \int_{0}^{\infty} \Xi_{\mu \nu}(t)\cos \omega t dt, 
\end{equation} 
where  $\varepsilon_{\beta}(\omega)$ is the average energy of an oscillator with the frequency $\omega$ at the temperature $T = 1/k_{B} \beta.$ This
equation represents the so-called fluctuation-dissipation theorem, a particular case of which is the Nyquist theorem for
the thermal noise in a resistive circuit. The fluctuation-dissipation theorems were established~\cite{ku66,case} for systems in thermal equilibrium.
It relates the conventionally defined noise power spectrum of the dynamical variables of a system to the corresponding admittances which describe the
linear response of the system to external perturbations.\\
The linear response theory is very general and effective tool for the calculation of transport coefficients of the systems which are rather close
to a thermal equilibrium. Therefore, the two approaches, the linear response theory and the traditional kinetic equation theory share a domain in which
they give identical results. A general formulation of the linear response theory was given by Kubo ~\cite{k57,rk57,rk59,kb59} for the case of
mechanical disturbances of the system with an external source in terms of an additional Hamiltonian.\\
A mechanical disturbance is represented by a force $F(t)$ acting on the system which may be given function of time. The interaction energy 
of the system may then be written as
\begin{equation}\label{8.11}
H_{ext}(t) = - A F(t),
\end{equation} 
where $A$ is the quantity conjugate to the force $F.$ The deviation of the system from equilibrium is observed through measurements of certain physical 
quantities. If $\Delta \bar{B}(t) $ is the observed deviation of a physical quantity $B$ at the time $t$, we may assume, if only the force $F$ is
weak enough, a linear relationship between $\Delta \bar{B}(t) $ and the force $F(t)$, namely
\begin{equation}\label{8.12}
\Delta \bar{B}(t) = \int_{- \infty}^{t} \phi_{BA}(t,t') F(t') dt',
\end{equation} 
where the assumption that the system was in equilibrium at $t = - \infty$,  when the force had been switched on, was introduced. This assumption
was formulated mathematically by the asymptotic condition,
\begin{equation}\label{8.13}
F(t) \sim e^{\varepsilon t} \quad {\rm as }\quad  t \rightarrow - \infty \quad (\varepsilon > 0).
\end{equation} 
Eq.(\ref{8.12}) assumes the causality and linearity. Within this limitation it is quite general. Kubo called the function $\phi_{BA}$ of response function
of $B$ to $F$, because it represents the effect of a delta-type disturbance of $F$ at the time $t'$ shown up in the quantity $B$ at a later time $t$.
Moreover, as it was claimed by Kubo, the linear relationship (\ref{8.12}) itself not in fact restricted by the assumption of small deviations from
equilibrium. In principle, it should be true even if the system is far from equilibrium as far as only differentials of the forces and responses are
considered. For instance, a system may be driven by some time-dependent force and superposed on it a small disturbance may be exerted; the
response function then will depend both on $t$ and $t'$ separately. If, however, we confine ourselves only to small deviations from equilibrium,
the system is basically stationary and so the response functions depend only on the difference of the time of pulse and measurement, $t$ and $t'$,
namely
\begin{equation}\label{8.14}
\phi_{BA}(t,t') = \phi_{BA}(t - t').
\end{equation} 
In particular, when the force is periodic in time
\begin{equation}\label{8.15}
F(t) = \textrm{Re} F e^{i \omega t},
\end{equation} 
the response of B will have the form
\begin{equation}\label{8.16}
\Delta \bar{B}(t) = \textrm{Re} \chi_{BA}(\omega)F e^{i \omega t},
\end{equation} 
where $\chi_{BA}(\omega)$ is the admittance
\begin{equation}\label{8.17}
\chi_{BA}(\omega) = \int_{- \infty}^{t} \phi_{BA}(t) e^{- i \omega t} dt.
\end{equation} 
More precisely~\cite{suz}, the response $\Delta \bar{B}(t)$ to an external periodic force $F(t) = F \cos (\omega t)$ conjugate to a physical quantity $A$
is given by Eq.(\ref{8.16}), where the admittance $\chi_{BA}(\omega)$ is defined as
\begin{equation}\label{8.18}
\chi_{BA}(\omega) = \lim_{\varepsilon \rightarrow +0}\int_{0}^{\infty} \phi_{BA}(t) e^{- (i \omega + \varepsilon)t} dt.
\end{equation} 
The response function $\phi_{BA}(\omega)$ is expressed as
\begin{eqnarray}\label{8.19}
\phi_{BA}(t) =  i \textrm{Tr} [A, \rho]B(t) = - i \textrm{Tr}  \rho [A,B(t)] \\ \nonumber
 = \int_{0}^{\beta} \textrm{Tr} \rho \dot{A}(-i \lambda) B(t)dt = -   \int_{0}^{\beta} \textrm{Tr} \rho A (-i \lambda) \dot{B}d \lambda,
\end{eqnarray} 
where $\rho$ is the canonical density matrix
\begin{eqnarray}\label{8.20}
\rho = \exp( - \beta (H - \Omega)), \quad \exp( - \beta \Omega) = \textrm{Tr} \exp( - \beta H).
\end{eqnarray} 
In  certain problems it is convenient to use the relaxation function defined by
\begin{eqnarray}\label{8.21}
\Phi_{BA}(t) = \lim_{\varepsilon \rightarrow +0}\int_{t}^{\infty} \phi_{BA}(t') e^{- \varepsilon t'} dt' \\ \nonumber
= i \int_{t}^{\infty} \langle [B(t'),A] \rangle dt' \\ \nonumber
= - \int_{t}^{\infty} dt'  \int_{0}^{\beta}  d \lambda \langle A(i \lambda) \dot{B}(t') \rangle \\ \nonumber
= \int_{0}^{\beta} d \lambda \Bigl ( \langle A (-i \lambda)B(t) \rangle - \lim_{t \rightarrow \infty}\langle A (-i \lambda)B(t) \rangle \Bigr ) \\ \nonumber
= \int_{0}^{\beta} d \lambda \langle A (-i \lambda)B(t) \rangle - \beta \lim_{t \rightarrow \infty}\langle A B(t) \rangle \\ \nonumber
= \int_{0}^{\beta} d \lambda \langle A (-i \lambda)B(t) \rangle - \beta \langle A^{0} B^{0} \rangle.
\end{eqnarray} 
It is of use to represent the last term in terms of the matrix elements
\begin{eqnarray}\label{8.22}
\int_{0}^{\beta} d \lambda \langle A (-i \lambda)B(t) \rangle - \beta \langle A^{0} B^{0} \rangle \\ \nonumber
= \Bigl ( 1/ \sum_{i} \exp( - \beta E_{i})\Bigr ) \sum_{n,m} \langle n|A|m \rangle \langle m|B|n \rangle e^{-i t (E_{n} - E_{m})}.
\frac{e^{ - \beta E_{n}} - e^{-  \beta E_{m}}}{E_{m} - E_{n}}.
\end{eqnarray} 
Here $|m \rangle$ denotes an eigenstate of the Hamiltonian with an eigenvalue $E_{m}$ and $A^{0}$  and $ B^{0}$ are the diagonal parts of $A$ and $B$
with respect to $H$.
The response function $\chi_{BA}(\omega)$ can be rewritten in terms of the relaxation function. We have
\begin{eqnarray}\label{8.23}
\chi_{BA}(\omega) = - \lim_{\varepsilon \rightarrow +0}\int_{0}^{\infty} \dot{\phi}_{BA}(t) e^{- (i \omega + \varepsilon)t} dt \\ \nonumber
= \phi_{BA}(0) - i \omega \lim_{\varepsilon \rightarrow +0}\int_{0}^{\infty} \phi_{BA}(t) e^{- (i \omega + \varepsilon)t} dt \\ \nonumber
= - \lim_{\varepsilon \rightarrow +0}\int_{0}^{\infty} dt \int_{0}^{\beta} d \lambda e^{- (i \omega + \varepsilon)t} 
\frac{d}{dt} \langle A B(t + i \lambda) \rangle \\ \nonumber
= i \lim_{\varepsilon \rightarrow +0}\int_{0}^{\infty} dt e^{- (i \omega + \varepsilon)t}\Bigl ( \langle A B(t + i \beta) \rangle - \langle A B(t) \rangle \Bigr ) \\ \nonumber
= i \lim_{\varepsilon \rightarrow +0}\int_{0}^{\infty} dt e^{- (i \omega + \varepsilon)t} \langle [ B(t),A ] \rangle \\ \nonumber
= \sum_{n,m} \frac{A_{mn}B_{nm}}{\omega + \omega_{mn} + i \varepsilon} (e^{ - \beta E_{n}} - e^{-  \beta E_{m}}),
\end{eqnarray} 
where
$$ A_{mn} = \frac{\langle m|A|n \rangle}{(\sum_{n}e^{ - \beta E_{n}} )^{1/2}}~.$$
In particular, the static response $\chi_{BA}(0)$ is given by
\begin{eqnarray}\label{8.24}
\chi_{BA}(0) = \phi_{BA}(0) \\ \nonumber
= \int_{0}^{\beta} d \lambda \Bigl (\langle A B( i \lambda) \rangle -  \lim_{t \rightarrow \infty} \langle A B(t + i \lambda) \rangle \Bigr ) \\ \nonumber
= i \lim_{\varepsilon \rightarrow +0}\int_{0}^{\infty} e^{- \varepsilon t} dt \Bigl ( \langle A B( t + i \beta) \rangle  - 
\langle A B( t) \rangle \Bigr ) \\ \nonumber
= i \lim_{\varepsilon \rightarrow +0}\int_{0}^{\infty} e^{- \varepsilon t} dt \langle [B( t), A] \rangle.
\end{eqnarray} 
This expression can be compared with the isothermal response defined by
\begin{equation}\label{8.25}
\chi_{BA}^{T} = \int_{0}^{\beta} \Bigl (\langle A B( i \lambda) \rangle - \langle A  \rangle \langle B \rangle \Bigr ) d \lambda.
\end{equation} 
The difference of the two response functions is given by
\begin{eqnarray}\label{8.26}
\chi_{BA}^{T} - \chi_{BA}(0) = \lim_{t \rightarrow \infty} \int_{0}^{\beta} d \lambda \langle A B(t + i \lambda) \rangle  - 
\beta \langle A  \rangle \langle B \rangle \\ \nonumber
= \beta \Bigl ( \lim_{t \rightarrow \infty} \langle A B( t) \rangle  - \langle A  \rangle \langle B \rangle  \Bigr ).
\end{eqnarray} 
The last expression suggests that it is possible to think that  the two response functions are equivalent for the systems which satisfy the condition 
\begin{equation}\label{8.27}
\lim_{t \rightarrow \infty} \langle A B( t) \rangle = \langle A  \rangle \langle B \rangle.
\end{equation} 
It is possible to speak about these systems in terms of ergodic (or quasi-ergodic) behavior, however, with a certain 
reservation (for a  recent analysis of the ergodic  behavior of many-body systems see Refs.~\cite{lee01,lee06,lee07a,lee07b,lee07,lee07ab,lee08,lee08a,lee09,lee10}).\\
It may be of use to remind a few useful properties of the relaxation function.\\
If $A$ and $B$ are both Hermitian, then
\begin{equation}\label{8.28}
\Phi_{BA}(t) = {\rm real}, \quad \int_{0}^{\infty} \Phi_{AA}(t) \geq 0.
\end{equation} 
The matrix-element representation of the relaxation function have the form
\begin{eqnarray}\label{8.29}
\Phi_{BA}(t) = \sum_{m,n} \Bigl ( R^{1}_{mn} \cos (\omega_{mn}) - R^{2}_{mn} \sin (\omega_{mn})\Bigr ) \\ \nonumber
+ i \sum_{m,n} \Bigl (R^{1}_{mn} \sin (\omega_{mn}) - R^{2}_{mn} \cos (\omega_{mn})\Bigr ),
\end{eqnarray} 
where
$$ R^{1}_{mn} = \frac{1}{2} (A_{nm}B_{mn} + A_{mn}B_{nm})R^{3}_{mn},$$
$$ R^{2}_{mn} = \frac{1}{2i} (A_{nm}B_{mn} - A_{mn}B_{nm})R^{3}_{mn},$$
$$ A_{mn} = \frac{\langle m|A|n \rangle}{(\sum_{n}e^{ - \beta E_{n}} )^{1/2}},$$
$$ R^{3}_{mn} = \frac{e^{- \beta E_{n}} - e^{- \beta E_{m}}}{\omega_{mn}}, \quad \omega_{mn} = E_{m} - E_{n}.$$
This matrix-element representation is very useful and informative. It can be shown that the relaxation function has the property 
\begin{equation}\label{8.30}
{\rm Im}\Phi_{BA}(t) = 0 ,
\end{equation} 
which follows from the odd symmetry of the matrix-element representation. The time integral of the relaxation function is given by
\begin{equation}\label{8.31}
\int_{0}^{\infty} \Phi_{BA}(t)dt = \frac{\pi \beta}{2} \sum_{m,n} \Bigl (A_{nm}B_{mn} + B_{nm}A_{mn}\Bigr )e^{- \beta E_{n}} \delta (\omega_{mn}).
\end{equation} 
In particular, for $A = B,$ we obtain
\begin{equation}\label{8.32}
\int_{0}^{\infty} \Phi_{AA}(t)dt = \pi \beta \sum_{m,n} |A_{nm}|^{2}e^{- \beta E_{n}} \delta (\omega_{mn}) \geq 0.
\end{equation} 
It can be shown also~\cite{suz} that if $A$ and $B$ are both bounded, then we obtain
\begin{eqnarray}\label{8.33}
\int_{0}^{\infty} \Phi_{\dot{B}\dot{A}}(t)dt = i \int_{0}^{\infty} dt \int_{t}^{\infty} \langle [\dot{B}( t'), \dot{A}] \rangle dt'\\ \nonumber
= - i \int_{0}^{\infty} dt \langle [B( t), \dot{A}] \rangle = i \langle [A, B] \rangle.
\end{eqnarray} 
For $A = B,$ we have
\begin{equation}\label{8.34}
\int_{0}^{\infty} \Phi_{\dot{A}\dot{A}}(t)dt =  0, \quad \int_{0}^{\infty} dt \int_{0}^{\beta} d \lambda \langle \dot{A}( - i \lambda), \dot{A}( t) \rangle = 0
\end{equation} 
and
\begin{equation}\label{8.35}
\lim_{t \rightarrow \infty} \langle \dot{A} \dot{B}( t) \rangle = 0, 
\end{equation} 
if $A$ and $B$ are both bounded.\\
Application of this analysis may not be limited to admittance functions~\cite{rk72}. For example, if one write a frequency dependent mobility
function $\mu(\omega)$ as
\begin{equation}\label{8.36}
\mu(\omega) = [i\omega + \gamma(\omega)]^{-1}, 
\end{equation} 
the frequency-dependent friction $\gamma(\omega)$ is also related to a function $\phi(t),$ which is in fact the correlation 
function of a random force~\cite{rk72}. An advanced analysis and generalization of the Kubo linear response theory was carried out in series of papers 
by Van Vliet and co-authors~\cite{vliet08}. Fluctuations and response in nonequilibrium steady state were considered within the
nonlinear Langevin equation approach by Ohta and Ohkuma~\cite{ohta08}. It was shown that the steady probability current plays an important role for the
response and time-correlation relation and violation of the time reversal symmetry. 
%
%
\subsection{Green functions in the theory of irreversible processes}
%
Green functions are not only applied to the case of statistical equilibrium~\cite{zub,bt,dnz60,tyab,bb,rnc,howl}. They are a convenient means of studying 
processes where the deviation from the state of statistical equilibrium is small. The use of the Green functions permits one to evaluate the
transport coefficients of these processes . Moreover, the transport coefficients are written in terms of Green functions evaluated for the
unperturbed equilibrium state without explicitly having recourse to setting up a transport equation.
The linear response theory can be reformulated in terms of double-time temperature-dependent (retarded and advanced) Green functions~\cite{bt,dnz60,tyab}.
We shall give a brief account of this reformulation~\cite{zub,dnz60} and its simplest applications to the theory of irreversible processes.\\
The retarded two-time thermal Green functions arise naturally within the linear response formalism, as it was shown by Zubarev~\cite{zub,dnz60}.
To show this we consider the reaction of a quantum-mechanical system with a time-independent Hamiltonian H when an external perturbation
\begin{equation}\label{8.37}
H_{ext}(t) = - A F(t),
\end{equation} 
is switched on. The total Hamiltonian is equal to
\begin{equation}\label{8.38}
 \mathcal{H} = H  + H_{ext},
\end{equation} 
where we assume that there is no external perturbation at $\lim t \rightarrow - \infty$
\begin{equation}\label{8.39}
H_{ext}(t)|_{\lim t \rightarrow - \infty} = 0.
\end{equation} 
The last condition means that
\begin{equation}\label{8.40}
 \lim_{ t \rightarrow - \infty} \rho (t) = \rho_{0} = e^{-\beta H}/ [\textrm{Tr} e^{-\beta H}],
\end{equation} 
where $\rho (t) $ is a statistical operator which satisfies the equation of motion
\begin{equation}\label{8.41}
i \hbar \frac{\partial}{\partial t} \rho (t) = [H + H_{ext}(t), \rho],
\end{equation} 
This equation of motion together with the initial condition (\ref{8.40}) suggests to look for a solution of Eq.(\ref{8.41}) of the form
\begin{equation}\label{8.42}
\rho (t) = \rho  + \Delta \rho(t).
\end{equation} 
Let us rewrite Eq.(\ref{8.42}), taking into account that $[H, \rho] = 0,$ in the following form
\begin{eqnarray}\label{8.43}
i \hbar \frac{\partial}{\partial t}(\rho + \Delta \rho (t)) = i \hbar \frac{\partial}{\partial t} \Delta \rho (t) = \\ \nonumber
 [H + H_{ext}(t), \rho + \Delta \rho (t)] = [H, \Delta \rho (t)] + [H_{ext}(t), \rho ] + [H_{ext}(t), \Delta \rho (t)]. 
\end{eqnarray} 
Neglecting terms $H_{ext}(t)\Delta \rho,$ since we have assumed that the system is only little removed from a state of statistical
equilibrium, we get then
\begin{eqnarray}\label{8.44}
i \hbar \frac{\partial}{\partial t} \Delta \rho (t)  = 
  [H, \Delta \rho (t)] + [H_{ext}(t), \rho ], 
\end{eqnarray} 
where
\begin{equation}\label{8.45}
\Delta \rho(t)|_{\lim t \rightarrow - \infty} = 0.
\end{equation} 
Processes for which we can restrict ourselves in Eq.(\ref{8.44}) to terms linear in the perturbation are called linear dissipative processes.
For a discussion of higher-order terms it is convenient to introduce a transformation
\begin{eqnarray}\label{8.46}
\Delta \rho (t)  = e^{- i H t/\hbar}\varrho (t) e^{ i H t/\hbar},
\end{eqnarray} 
Then we have
\begin{eqnarray}\label{8.47}
i \hbar \frac{\partial}{\partial t} \Delta \rho (t)  = 
  [H, \Delta \rho (t)] + e^{- i H t/\hbar} \left ( i \hbar \frac{\partial}{\partial t} \varrho (t) \right ) e^{ i H t/\hbar}.
\end{eqnarray} 
This equation can be transformed to the following form
\begin{eqnarray}\label{8.48}
i \hbar \frac{\partial}{\partial t} \varrho (t) = [e^{ i H t/\hbar} H_{ext}(t) e^{- i H t/\hbar}, \rho ] + 
[e^{ i H t/\hbar} H_{ext}(t) e^{- i H t/\hbar}, \varrho (t) ],
\end{eqnarray} 
where
\begin{equation}\label{8.49}
\varrho (t)|_{\lim t \rightarrow - \infty} = 0.
\end{equation} 
In the equivalent integral form the the above equation reads
\begin{eqnarray}\label{8.50}
\varrho (t) = \frac{1}{i \hbar} \int_{- \infty}^{t} d \lambda [e^{ i H \lambda/\hbar} H_{ext}(\lambda) e^{- i H \lambda/\hbar}, \rho ] \\ \nonumber
+ \frac{1}{i \hbar} \int_{- \infty}^{t} d \lambda [e^{ i H \lambda/\hbar} H_{ext}(\lambda) e^{- i H \lambda/\hbar}, \varrho (\lambda) ].
\end{eqnarray} 
This integral form is convenient for the iteration procedure which can be written as
\begin{align}\label{8.51}
\varrho (t) = \frac{1}{i \hbar} \int_{- \infty}^{t} d \lambda [e^{ i H \lambda/\hbar} H_{ext}(\lambda) e^{- i H \lambda/\hbar}, \rho ] \\ \nonumber 
+ \Bigl (\frac{1}{i \hbar}\Bigr )^{2}\int_{- \infty}^{t} d \lambda \int_{- \infty}^{\lambda} d \lambda' 
\left [ e^{ i H \lambda/\hbar} H_{ext}(\lambda) e^{- i H \lambda/\hbar}, [e^{ i H \lambda'/\hbar} H_{ext}(\lambda') e^{- i H \lambda'/\hbar}, \rho ] \right ]
+ \ldots
\end{align} 
In the theory of the linear reaction of the system on the external perturbation usually the only first term is retained
\begin{equation}\label{8.52}
\Delta \rho (t) = \frac{1}{i \hbar} \int_{- \infty}^{t} d \tau e^{- i H (t - \tau)/\hbar} [H_{ext}(\tau), \rho ] e^{ i H (t - \tau)/\hbar}.
\end{equation} 
The average value of observable $A$  is
\begin{equation}\label{8.53}
\langle A \rangle_{t} = \textrm{Tr} ( A \rho (t)) =  \textrm{Tr} ( A \rho_{0}) + \textrm{Tr} ( A \Delta \rho (t)) = \langle A \rangle +  \Delta\langle A \rangle_{t}.
\end{equation} 
From this we find
\begin{eqnarray}\label{8.54}
\Delta \langle A \rangle_{t} = \frac{1}{i \hbar} \int_{- \infty}^{t} d\tau \cdot \nonumber  \\  
\textrm{Tr} \left ( e^{ i H (t - \tau)/\hbar} A e^{- i H (t - \tau)/\hbar} H_{ext}(\tau) \rho  
- H_{ext}(\tau)e^{ i H (t - \tau)/\hbar} A e^{- i H (t - \tau)/\hbar} \rho \right ) + \ldots  \nonumber \\ 
= \frac{1}{i \hbar} \int_{- \infty}^{t} d\tau \langle [A(t - \tau), H_{ext}(\tau)]_{-} \rangle + \ldots \quad \quad
\end{eqnarray} 
Introducing under the integral the sign function
\begin{eqnarray}\label{8.55}
\theta (t - \tau) = 
\begin{cases}
1 & {\rm if}  \quad  \tau < t ,\cr
 0 &  {\rm if}  \quad  \tau > t,
\end{cases}
\end{eqnarray}
and extending the limit of integration to $ - \infty < \tau < + \infty$ we finally find
\begin{equation}\label{8.56}
\Delta\langle A \rangle_{t} =  \int_{- \infty}^{\infty} d\tau \frac{1}{i \hbar} \theta (\tau) \langle [A(\tau),H_{ext}(t - \tau)]_{-} \rangle + \ldots
 \end{equation} 
Let us consider an adiabatic switching on a periodic perturbation of the form
$$H_{ext}(t) = B \exp {\frac{1}{i \hbar}(E + i\varepsilon)t}.$$
The presence in the exponential function of the infinitesimal factor $\varepsilon > 0, \varepsilon \rightarrow 0$ make for the adiabatic
switching of the perturbation. Then we obtain
\begin{equation}\label{8.57}
\Delta\langle A \rangle_{t} =  \exp {\frac{1}{i \hbar}(E + i\varepsilon)t} \int_{- \infty}^{\infty} d\tau 
\exp {\frac{- 1}{i \hbar}(E + i\varepsilon)\tau} \, \frac{1}{i \hbar} \theta (\tau) \langle [A(\tau),B]_{-} \rangle.  
 \end{equation} 
It is clear that the last expression can be rewritten as
\begin{eqnarray}\label{8.58}
\Delta\langle A \rangle_{t} =  \exp {\frac{1}{i \hbar}(E + i\varepsilon)t} \int_{- \infty}^{\infty} d\tau 
\exp {\frac{- 1}{i \hbar}(E + i\varepsilon)\tau} \, G^{ret}(A,B;\tau) = \\ \nonumber
\exp {\frac{1}{i \hbar}(E + i\varepsilon)t}\, G^{ret}(A,B;E) = \exp {\frac{1}{i \hbar}(E + i\varepsilon)t} \,  \langle \langle A | B \rangle \rangle_{E + i\varepsilon}~.
\end{eqnarray} 
Here $E = \hbar \omega$ and $\langle \langle A | B \rangle \rangle_{E + i\varepsilon}$ is the Fourier component of the \emph{retarded} Green function
$\langle \langle A(t) ; B(\tau) \rangle \rangle.$ \\ The change in the average value of an operator when a periodic perturbation is switched on
adiabatically can thus be expressed in terms of the Fourier components of the retarded Green functions which connect the perturbation operator 
and the observed quantity.\\ In the case of an instantaneous switching on of the interaction
\begin{eqnarray}\label{8.59}
H_{ext}(t) = 
\begin{cases}
0 & {\rm if}  \quad  t < t_{0} ,\cr
 \sum_{\Omega} \exp \Bigl (\Omega t/i \hbar  \Bigr ) V_{\Omega} &  {\rm if}  \quad  t > t_{0},
\end{cases}
\end{eqnarray}
where $V_{\Omega}$ is an operator which does not explicitly depend on the time, we get
\begin{eqnarray}\label{8.60}
\Delta\langle A \rangle_{t} =  \sum_{\Omega} \int_{t_{0}}^{\infty} d\tau 
\langle \langle A(t) ; V_{\Omega}(\tau) \rangle \rangle \exp {\frac{1}{i \hbar}(\Omega + i\varepsilon)\tau},
\end{eqnarray} 
i.e., the reaction of the system can also be expressed in terms of the retarded Green functions.\\ Now we can define the
generalized susceptibility of a system on a perturbation $H_{ext}(t)$ as
\begin{equation}\label{8.61}
\chi (A, B; E) =  \chi (A, z B; E) = \lim_{z \rightarrow 0} \frac{1}{z} \Delta\langle A \rangle_{t} \exp {\frac{- 1}{i \hbar}(E + i\varepsilon)t} =
\langle \langle A | B \rangle \rangle_{E + i\varepsilon}.
\end{equation} 
In the time representation the above expression reads
\begin{equation}\label{8.62}
\chi (A, B; E) =  \frac{ 1}{i \hbar} \int_{- \infty}^{\infty} dt  \exp {\frac{- 1}{i \hbar}(E + i\varepsilon)t} \,\,
\theta (t) \langle [A(t),B]_{-} \rangle.
\end{equation} 
This expression is an alternative form of the  fluctuation-dissipation theorem, which show explicitly the connection of the relaxation 
processes in the system with  the dispersion of the physical quantities.\\ The particular case where the external perturbation is 
periodic in time and contains only one harmonic frequency $\omega$ is of interest. Putting in that case $\Omega = \pm \hbar \omega$ in
Eq.(\ref{8.60}), since
\begin{eqnarray}\label{8.63}
H_{ext}(t ) = -h_{0} \cos \omega t e^{\varepsilon t}B,
\end{eqnarray}
where $h_{0},$ the amplitude of the periodic force, is a $c-$number and where $B$ is the operator part of the perturbation, we get
\begin{eqnarray}\label{8.64}
\Delta\langle A \rangle_{t} =  -h_{0} \exp \left( {\frac{1}{i \hbar}\omega t + \varepsilon t} \right)
 \langle \langle A | B \rangle \rangle_{E = \hbar \omega} \\ \nonumber
- h_{0} \exp \left(
{\frac{- 1}{i \hbar}\omega t + \varepsilon t} \right)
 \langle \langle A | B \rangle \rangle_{E = - \hbar \omega}~.
\end{eqnarray} 
Taking into account that $\langle A \rangle_{t}$ is a real quantity we can write it as follows
\begin{equation}\label{8.65}
\Delta\langle A \rangle_{t} = {\rm Re}\left ( \chi (E) h_{0} e^{\frac{1}{i \hbar}E t + \varepsilon t}\right ).
\end{equation} 
Here $\chi (E)$ is the complex admittance, equal to
\begin{equation}\label{8.66}
\chi (E) = - 2 \pi \langle \langle A | B \rangle \rangle_{E = \hbar \omega}~.
\end{equation} 
These equations elucidate the physical meaning of the Fourier components of the Green 
function $\langle \langle A | B \rangle \rangle_{E = \hbar \omega}$ as being the complex admittance that describes the influence of the 
periodic perturbation on the average value of the quantity $A.$
%
%
%
%
\subsection{The electrical conductivity tensor}
%
When a uniform electric field of strength $\mathcal{\vec{E}}$ is switched on then the perturbation acting upon the system of charged
particles assumes the form $H_{ext} = - \mathcal{\vec{E}}\cdot \vec{d}(t)$, where $\vec{d}(t)$ is the total dipole moment of the system.
In this case the average operator $A(t)$ is the current density operator $\vec{j}$ and the function $\chi $ is the complex electrical conductivity
tensor denoted by $ \sigma_{\alpha \beta}(\omega).$ If the volume of the system is taken to be equal to unity, then we have
\begin{equation}\label{8.67}
\frac{d}{d t} d_{\alpha}(t) = j_{\alpha}(t).
\end{equation} 
The Kubo formula (\ref{8.65})  relates the linear response of a system to its equilibrium correlation functions. Here we consider the
connection between the electrical conductivity tensor and Green functions~\cite{zub,dnz60}. Let us start with a simplified treatment when 
there be switched on adiabatically an
electrical field $\mathcal{E}(t),$ uniform in space and changing periodically in time with a frequency $\omega$
\begin{equation}\label{8.68}
\mathcal{\vec{E}}(t) = \mathcal{\vec{E}}\cos \omega t.
\end{equation} 
The corresponding perturbation operator is equal to
\begin{equation}\label{8.69}
H_{ext}(t)  = - e \sum_{j} (\mathcal{\vec{E}} \vec{r}_{j})\cos \omega t e^{\varepsilon t}.
\end{equation} 
Here $e$ is the charge of an electron, and the summation is over all particle coordinates $\vec{r}_{j}.$ Under the influence of the perturbation
there arises in the system an electrical current
\begin{equation}\label{8.70}
j_{\alpha}(t)
 = \int_{- \infty}^{\infty} d \tau    \langle \langle j_{\alpha}(t) ; H_{ext}(\tau) \rangle \rangle, 
\end{equation} 
where
\begin{eqnarray}\label{8.71}
H_{ext}(\tau) = H_{\tau}^{1}(\tau) \cos \omega \tau e^{\varepsilon \tau}, \\ \nonumber
H_{\tau}^{1}(\tau) = - e \sum_{j \alpha} \mathcal{E}_{\alpha} r_{j}^{\alpha}(\tau), \quad  
j_{\alpha}(t) = e \sum_{j} \dot{r}_{j}^{\alpha}(t).
\end{eqnarray} 
Here $j_{\alpha}$ is the current density operator, if the volume of the system is taken to be unity. The equation (\ref{8.70}) can be transformed 
to the following form
\begin{eqnarray}\label{8.72}
j_{\alpha}(t) = - {\rm Re} \{ \int_{- \infty}^{\infty} d \tau \langle \langle j_{\alpha}(t); H_{\tau}^{1}(\tau) \rangle \rangle
\frac{ e^{\frac{i}{ \hbar}\omega t + \varepsilon \tau}}{i \omega + \varepsilon} +  \\ \nonumber \langle [j_{\alpha}(0),H_{\tau}^{1}(0)]_{-} \rangle e^{\frac{i}{ \hbar}\omega t + \varepsilon t} \frac{1}{\omega - i \varepsilon} \}.
\end{eqnarray} 
Noting that
\begin{equation}\label{8.73}
\dot{H}_{\tau}^{1}(\tau) = - (\mathcal{\vec{E}} \vec{j}(\tau)), \quad [r_{i}^{\alpha},r_{j}^{\beta}] = \frac{1}{i \hbar m}\delta_{\alpha \beta}\delta_{ij},
\end{equation} 
we get from this equation
\begin{equation}\label{8.74}
j_{\alpha}(t) = {\rm Re} \{ \sigma_{\alpha \beta}(\omega)\mathcal{E}_{\beta}\exp (i \omega t + \varepsilon t) \},
\end{equation} 
where
\begin{equation}\label{8.75}
 \sigma_{\alpha \beta}(\omega) = - \frac{i e^{2}n }{m \omega} \delta_{\alpha \beta}  + 
 \int_{- \infty}^{\infty} d \tau \langle \langle j_{\alpha}(0); j_{\beta}(\tau) \rangle \rangle  \frac{\exp (i \omega t + \varepsilon t)}{i \omega + \varepsilon}
\end{equation} 
is the conductivity tensor, and $n$ the number of electrons per unit volume. The first term in Eq.(\ref{8.75}) corresponds to the electrical
conductivity of a system of free charges and is not connected with the inter-particle interaction. As $\omega \rightarrow \infty$ the second term
decreases more strongly than the first one
$$ \lim_{\omega \rightarrow \infty}{\rm Im} \, \omega \sigma_{\alpha \beta}(\omega) = - \frac{ e^{2}n }{m \omega} \delta_{\alpha \beta},$$
and the system behaves as a collection of free charges.\\
Let us discuss the derivation of the conductivity tensor  in general form. Consider a system of charged particles in  electrical 
field $\mathcal{\vec{E}},$ which is directed along the axis $\beta \quad ( \beta = x, y, z).$ The corresponding electrostatic potential $\varphi^{(\beta)}$

$$\mathcal{\vec{E}}^{(\beta)} = - \nabla \varphi^{(\beta)}$$
has the following form
\begin{equation}\label{8.76}
\varphi^{(\beta)} (\vec{r}, t) = \sum_{\Omega}  \frac{1}{V}\sum_{q} \varphi^{(\beta)}
(\vec{q}, \Omega)\exp \Bigl ( (\Omega t + i\varepsilon t)/i \hbar - i\vec{q}\vec{r} \Bigr ).
\end{equation} 
In the momentum representation the above expression reads
\begin{equation}\label{8.77}
\mathcal{E}^{(\beta)}_{\alpha}(\vec{q}, \Omega) = 
\frac{i}{ \hbar}q_{\alpha}\varphi^{(\beta)}(\vec{q}, \Omega) = \delta_{\alpha \beta}\mathcal{E}_{\beta}(\vec{q}, \Omega).
\end{equation} 
Consider the case when the perturbation $H^{ext}$ has the form
\begin{equation}\label{8.78}
H^{ext}_{\vec{q}\Omega}  = \frac{1}{V} e \varphi^{(\beta)} (\vec{q}, \Omega) \eta^{\dag}_{q}  \exp \Bigl ( (\Omega t + i\varepsilon t)/i \hbar  \Bigr ).
\end{equation} 
Here
$$\eta_{q} = \sum_{p} a^{\dag}_{p+q}a_{p}, \quad \eta^{\dag}_{q} = \eta_{-q}.$$
is the particle density operator.\\
A reaction of the system on the perturbation is given by
\begin{eqnarray}\label{8.79}
I^{(\beta)}_{\alpha}(\vec{q}, \Omega) = \Delta \langle e j_{\alpha}(\vec{q})\rangle = 
\frac{1}{V}  \langle \langle e j_{\alpha}(\vec{q}) | e \varphi^{(\beta)} (\vec{q}, \Omega) \eta^{\dag}_{q} \rangle \rangle_{\Omega  + i\varepsilon} \\ \nonumber
= e^{2}\varphi^{(\beta)} (\vec{q}, \Omega) \frac{1}{V} \int_{- \infty}^{\infty} d \omega J(\eta^{\dag}_{q},j_{\alpha}(\vec{q}); \omega  )
\frac{e^{\omega / \theta} - 1}{\Omega - \omega + i\varepsilon},
\end{eqnarray} 
where $I^{(\beta)}_{\alpha}(\vec{q}, \Omega)$ denotes the component of the density of the current in the $\alpha$-direction when external electric
field directed along the $\beta$-axis.
%
%
%
\subsection{Linear response theory: pro et contra}
%
%
%
It was shown in the previous sections that the formulation of the linear response theory can be generalized so as to be applied to a rather wide
class of the problems.
It is worth to note that the "exact" linear expression for electrical conductivity for an arbitrary system was derived originally by Kubo~\cite{k57} in
a slightly different form than Eq.(\ref{8.75})
\begin{equation}\label{8.80}
\sigma_{\mu \nu} = \lim_{\varepsilon \rightarrow 0} \frac{1}{\varepsilon}\left (\phi_{\mu \nu}(0)  + 
\int_{- \infty}^{\infty} dt  e^{- \varepsilon t} \dot{\phi}_{\mu \nu}(t)  \right),
\end{equation}
where
\begin{equation}\label{8.81}
\phi_{\mu \nu}(t) = \frac{1}{i \hbar}{\rm Tr}\Bigl ( n, \sum_{i} e_{i} x_{i\nu}\Bigr ) \sum_{i} e_{i}\dot{x}_{i\nu}(t)
\end{equation}
is the current response in the $\mu-$direction when a pulse of electric field ${\mathcal E}(t)$  is applied in the $\nu-$direction
at $t = 0;$ $e_{i}$ is the charge of the $i$th particle with position vector $\vec{r}_{i}$ and $n$ is a density operator. In the one-electron
approximation Eq.(\ref{8.80}) reduces to (cf. Ref.~\cite{gre})
\begin{equation}\label{8.82}
\sigma_{i j} = 2 \pi e^{2} h \sum_{nm}\langle m|v_{i}|n \rangle \langle n|v_{j}|m \rangle \left (\frac{\partial f}{\partial E} \right)_{n}.
\end{equation}
Here $v_{i}$ is the velocity operator and $f$ is the Fermi function.\\
To clarify the general consideration of the above sections it is of interest to consider here a simplified derivation of this formula, using only the lowest order of time-dependent
perturbation theory~\cite{luk}. This approach is rooted in the method of derivation of Callen and Welton~\cite{ca}, 
but takes explicitly into account degeneracy of the states. 
The linear response theory formulated by Kubo~\cite{k57} was motivated in part~\cite{mi95,hnak} by the prior 
work of Callen and Welton~\cite{ca}, who proposed a
quantum mechanical perturbative calculation with the external forces exerted on a dissipative system. They pointed out a 
general relationship
between the power dissipation induced by the perturbation and the average of a squared fluctuation of the current of the system in thermal
equilibrium.\\
Let us discuss first the role of dissipation. A system may be called to be dissipative if it absorbs
energy when subjected to a time-periodic perturbation, and linear if the dissipation (rate of absorption of energy) is quadratic in the perturbation.
For a linear system an impedance may be defined and the proportionality constant between the power and the square of the perturbation amplitude
is simply related to the impedance. In the case of electrical current in a material  one can write down that
$$ \overline{W} = \frac{1}{2} \frac{R}{({\rm Impedance})^{2}} = V^{2},$$
where $\overline{W}$ is the average power and $V$ is the voltage. If we calculate the power microscopically in some way and find it quadratic in the
applied force (voltage), then comparison with this equation will give the conductivity of the substance.\\ Consider a situation when an electron of charge
$e$ is situated a distance $x$ from the end of a resistor of length $L$ and then a voltage $V = V_{0} \sin \omega t$ is applied in the $x$ direction~\cite{luk}.
The perturbation term in the Hamiltonian will be of the form
\begin{equation}\label{8.83}
H^{ext}_{\omega}  = V_{0}e  \frac{x}{L}  \sin \omega t.
\end{equation} 
The Hamiltonian of a system in the absence of the field (but including all other interactions) is denoted by $H_{0}$ with corresponding
wave function $\psi_{n}$ such that
\begin{equation}\label{8.84}
H_{0}\psi_{n} = E_{n}\psi_{n}.
\end{equation} 
The total wave function may be expanded in terms of the $\psi_{n}$
\begin{equation}\label{8.85}
\Psi_{n} = \sum_{n} a_{n}(t)\psi_{n}, 
\end{equation} 
where the coefficients $a_{n}(t)$ may be approximately determined by first-order perturbation theory~\cite{dav}. The rate of transition is
then given by
\begin{eqnarray}\label{8.86}
\frac{d p_{n}}{d t} =  \frac{1}{2}  \frac{\pi e^{2}V_{0}^{2} }{\hbar} \sum_{m n} |\langle m|x|n \rangle |^{2}
[\delta (E_{m} - (E_{n} + \hbar \omega)) + \delta (E_{m} - (E_{n} - \hbar \omega))].\quad
\end{eqnarray} 
The first term corresponds to a transition to a state $E_{m} = E_{n} + \hbar \omega$ in which energy $\hbar \omega$ is absorbed, whereas
the second term corresponds to a transition to a state $E_{m} = E_{n} - \hbar \omega$ in which energy is emitted. Hence, the net rate of absorption
of energy is given by
\begin{eqnarray}\label{8.87}
\frac{d E_{n}}{d t} =  \frac{1}{2}  \frac{\pi e^{2}V_{0}^{2} }{\hbar} \hbar \omega \sum_{m n} |\langle m|x|n \rangle |^{2}
[\delta (E_{m} - E_{n} - \hbar \omega) + \delta (E_{m} - E_{n} + \hbar \omega)],\quad
\end{eqnarray} 
which is quadratic in $V_{0}.$\\ This equation gives the absorption rate for a single, isolated electrons, but in a real system we are dealing
with an ensemble of these, which we shall represent by the Fermi function. One must therefore find~\cite{luk} the average absorption by averaging over all 
initial states $|n \rangle$ and taking the Pauli exclusion principle as well as the two spin directions into account. The result is~\cite{luk}
\begin{eqnarray}\label{8.88}
\langle \frac{d E_{n}}{d t}\rangle = \pi e^{2}V_{0}^{2}  \omega \sum_{m n} |\langle m|x|n \rangle |^{2} \\ \nonumber
\{ f(E_{n})\delta (E_{m} - E_{n} - \hbar \omega) (1 - f(E_{m})) \\ \nonumber -  f(E_{n})\delta (E_{m} - E_{n} + \hbar \omega) (1 - f(E_{m}))\}\\ \nonumber
= \pi e^{2}V_{0}^{2}  \omega \{   \sum_{n}|\langle n + \hbar \omega |x|n \rangle |^{2}[f(E_{n}) - f(E_{n})f(E_{n} + \hbar \omega)] -  \\ \nonumber
\sum_{n'}|\langle n'|x|n'  + \hbar \omega  \rangle |^{2}[f(E_{n'} + \hbar \omega ) - f(E_{n'} + \hbar \omega)f(E_{n'})]\},
\end{eqnarray} 
where we have put $n' = n - \hbar \omega.$ The above formula can be transformed to the form
\begin{eqnarray}\label{8.89}
\langle \frac{d E_{n}}{d t}\rangle = \pi e^{2}V_{0}^{2}  \omega \sum_{m n} |\langle m|x|n \rangle |^{2} \cdot \\ \nonumber
[f(E_{n}) - f(E_{m})]\delta (E_{m} - E_{n} - \hbar \omega).
\end{eqnarray}
This expression may be simplified by introducing the matrix element of the velocity operator
\begin{equation}\label{8.90}
\frac{d x}{d t} = \frac{i}{\hbar}[H, x].
\end{equation} 
In principle, on the right-hand side of Eq.(\ref{8.90}) the total Hamiltonian
\begin{equation}\label{8.91}
 \mathcal{H} = H  + H_{ext}
\end{equation} 
should be written. In this case, however, the terms of higher than quadratic order in $V_{0}$ appear. For a linear system one can neglect these
and use Eq.(\ref{8.90}). Thus we have
\begin{equation}\label{8.92}
 |\langle m|\dot{x}|n \rangle | = \frac{i}{\hbar} \langle m|[H, x]|n \rangle = 
 \frac{i}{\hbar}(E_{m}\langle m|x|n \rangle  -  E_{n}\langle m|x|n \rangle) 
\end{equation} 
and
\begin{equation}\label{8.93}
 |\langle m|\dot{x}|n \rangle |^{2} = \frac{-1}{\hbar^{2}}  (E_{m} - E_{n})^{2}|\langle m|x|n \rangle |^{2} = 
 \omega^{2}|\langle m|x|n \rangle |^{2}.
\end{equation} 
Therefore
\begin{eqnarray}\label{8.94}
\langle \frac{d E_{n}}{d t}\rangle = - \frac{\pi e^{2}}{\omega} V_{0}^{2}  \sum_{m n} |\langle m|\dot{x}|n \rangle |^{2} 
[f(E_{n}) - f(E_{n} + \hbar \omega)]\delta (E_{m} - E_{n} - \hbar \omega).
\end{eqnarray}
If we now assume the current to be in phase with the applied voltage, the average energy dissipation becomes
\begin{equation}\label{8.95}
 {\rm Power} = \frac{1}{2} \frac{V_{0}^{2}}{R(\omega)}
\end{equation} 
as the resistance $R(\omega)$ is now equal to the impedance $Z(\omega).$ Referring everything to unit volume, and noting that the resistance per 
unit volume is the resistivity, we get for the conductivity $\sigma$
\begin{eqnarray}\label{8.96}
\sigma_{xx}(\omega) = - \frac{2 \pi e^{2}}{\omega}   \sum_{m n} | \langle m|v|n \rangle \langle n|v|m \rangle | \\ \nonumber
[f(E_{n}) - f(E_{n} + \hbar \omega)]\delta (E_{m} - E_{n} - \hbar \omega).
\end{eqnarray}
A straightforward generalization of this procedure, using a perturbation
\begin{equation}\label{8.97}
 H_{ext} = \sum_{i} V_{0} e x_{i}/L \, \sin (\omega t), 
\end{equation} 
leads to the definition of an impedance matrix and a conductivity tensor
\begin{equation}\label{8.98}
\sigma_{i j} = \frac{2 \pi e^{2}}{\omega}   
\sum_{nm} |\langle m|v_{i}|n \rangle \langle n|v_{j}|m \rangle |[f(E_{n}) - f(E_{n} + \hbar \omega)]\delta (E_{m} - E_{n} - \hbar \omega),
\end{equation}
which is the Kubo-Greenwood equation~\cite{gre,k57}. The derivation~\cite{luk} presented here confirms that fact that the linear response theory
is based on the fluctuation-dissipation theorem, i.e. that the responses to an external perturbation are essentially determined by fluctuations of 
relevant physical quantities realized in the absence of the perturbation. Thus the linear response theory has a special appeal since it deals
directly with the quantum mechanical motion of a process.\\
The linear response theory (or its equivalent) become soon a very popular tool of the transport theory~\cite{swen63,rzw64}. As was expressed by
Langer~\cite{lan62}, the Kubo formula " probably provides the most rigorous possible point of departure for transport theory. Despite its extremely
formal appearance, it has in fact proved amenable to direct evaluation for some simple models". Edwards~\cite{ed58} and Chester and Thellung~\cite{che}
have used the Kubo formula to calculate the impurity resistance of a system of independent electrons, and have recovered the usual solution of the
linearized Boltzmann equation. Verboven~\cite{ver60} has extended this work to higher orders in the concentration of impurities and has found corrections to the
conductivity not originally derived via Boltzmann techniques. It was concluded that the Kubo formula might be most fruitfully applied in the full many-body
problem, where it is not clear that any Boltzmann formulation is valid. However Izuyama~\cite{izu} casted doubt  on the Kubo formula for electrical conductivity. He claimed
that it is not, in fact, an exact formula for electrical conductivity, but is rather a coefficient relating current to an "external" field, which
coefficient is equal to the conductivity only in special case.
A correlation-function formula for electrical conductivity was derived by Magan~\cite{mag} by
a formalism which gives prominence to the total electric field, including fields which may arise from the charged particles which are part of the
system being studied.\\
Langer~\cite{lan62} have evaluated the impurity resistance of an interacting electron gas on the basis of the Kubo formula at low but finite
temperatures. The calculations are exact to all orders in the electron-electron interactions and to lowest order in the concentration of impurities.
In the previous papers~\cite{lan60,lan61}, the impurity resistance of this gas was computed at absolute zero temperature. It was shown~\cite{lan62}
that the zero-temperature limit of this calculation yields the previous result. In Ref.~\cite{langer}   Kubo formula for thermal conductivity was evaluated
for the case of an interacting electron gas and random, fixed, impurities. The heat flux was examined in some detail and it was shown that in a normal
system where the many-body correlations are sufficiently weak, the Wiedemann-Franz law remains valid. The relationships between the transport 
coefficients of a metal in a strong magnetic field and at very low temperatures were discussed by Smrcka and Streda~\cite{ss77}. Formulae 
describing the electron coefficients as functions of the conductivity were derived on the basis of the linear response 
theory. As was mentioned earlier,
examples of  such relations are the Wiedemann-Franz law for the heat conductivity $\kappa$ and the Mott rule~\cite{mot36} for the thermopower $S.$
It was shown that that the Wiedemann-Franz law and the Mott rule are obeyed even in the presence of a quantized magnetic field $\omega \tau > 1$  if the 
scattering of electrons is elastic and if $\hbar \omega \gg k T.$\\
A theoretical analysis, based on Kubo formalism, was made for the ferromagnetic Hall effect by Leribaux~\cite{leri} in the case of transport limited
by electron-phonon scattering. The antisymmetric, off-diagonal conductivity was, to first order in magnetization, found to be of order zero in the
electron-phonon interaction (assumed to be weak) and, to this order, was equivalent to Karplus and Luttinger results~\cite{kl54}.
Tanaka, Moorjani and Morita~\cite{tmm}
expressed the nonlinear transport coefficients in terms of many-time Green functions and made an attempt to calculate the higher-order transport coefficients.
They applied their theory to the calculation of the nonlinear susceptibility of a Heisenberg ferromagnet and nonlinear polarizability problem. 
Schotte~\cite{kds}reconsidered the linear response theory to show the closeness of it and kinetic equations.\\ At the same time, several 
authors~\cite{coh,ps,prs} raised an important question as to the general validity of the correlation formulae for transport coefficients. The relation
for the electrical conductivity was, in principle, not questioned as in this case (as was shown above) one may obtain it by a straightforward 
application  of basic statistical mechanical principles and perturbation techniques. One simply calculates the response of the system to an electric field.
What has been questioned, however, was the validity of the correlation relations for the transport coefficients of a fluid - the diffusion constant, the
thermal conductivity, and the viscosity - where no such straightforward procedure as was used for the electrical conductivity was available~\cite{lut64}.\\
However, Jackson and Mazur~\cite{jm64} presented a derivation of the correlation formula for the viscosity which is similar in spirit, and as free of
additional assumptions, as that for the electrical conductivity. The correlation formula for the viscosity was obtained by calculating statistically
the first order response of a fluid, initially in equilibrium, to an external shearing force. It was shown that on the basis of this derivation,
the correlation formula for the viscosity, the exactness of which had been questioned, was placed on as firm a theoretical basis as the Kubo
relation for the electrical conductivity. In addition, Resibois~\cite{re64,res64} demonstrated the complete equivalence between the kinetic
approach developed by Prigogine and coworkers~\cite{prig62} and the correlation function formalism for the calculation of linear thermal
transport coefficients. It was shown that in both cases these transport coefficients are determined by the solution of an inhomogeneous
integral equation for a one-particle distribution function which is the generalization to strongly coupled system of the Chapman-Enskog first
approximation of the Boltzmann equation~\cite{kuz07,chest}. 
Interesting  remarks concerning the comparison of the linear response theory  and Boltzmann equation approach  were formulated 
by R. Peierls~\cite{pe,pe87}. Schofield~\cite{ps66,ps68} elaborated a general derivation of the transport coefficients and 
thermal equilibrium correlation functions for a classical system having arbitrary number of microscopic conservation laws. 
This derivation gives both the structure of the
correlation functions in the hydrodynamic (long wavelength) region and a generalized definition of the transport coefficients for all wave-lengths
and frequencies. \\ A number of authors have given the formulation of nonlinear responses~\cite{ohta08,nlr1,nlr2,nlr3,nlr4,nlr5,nlr6}. It was shown that since in a nonlinear
system fluctuation sources and transport coefficients may considerably depend on a nonequilibrium state of the system, nonlinear nonequilibrium
thermodynamics should be a stochastic theory. From the other side the linearity of the theory itself was a source of many doubts.
The most serious criticism of the Kubo linear response theory was formulated by van Kampen~\cite{vk}. He argued strenuously that the standard derivation of the 
response functions are incorrect. In his own words, " the basic linearity assumption of linear response theory ... is completely unrealistic and
incompatible with basic ideas of statistical mechanics of irreversible processes". The main question raised by van Kampen concerned the logic of the
linear response theory, not the results. As van Kampen~\cite{vk} expressed it, "the task of statistical mechanics is not only to provide an expression
for this (transport) coefficient in term of molecular quantities, but also to understand how such a linear dependence arises from the microscopic
 equations of motion".\\
The van Kampen's objections~\cite{vk} to Kubo linear response theory can be reduced to the following points. In the linear 
response theory one solves the Liouville equation
to first order in the external field $\vec{E}$ (electric field). This is practically equivalent to following a perturbed trajectory in phase space in a vicinity
of the order of $|\vec{E}|$ of unperturbed trajectory. In the classical case trajectories are exponentially unstable and corresponding field $\vec{E}$
should be very small. Kubo's derivation  supposed non-explicitly that macroscopic linearity (Ohm law, \emph{etc}.) is the consequence of microscopic linearity, but
these two notions are not identical. Macroscopic linearity is the result of averaging over many trajectories and is not the same as linear deviations
from any one trajectory. In other words, van Kampen's argument was based on the observation that due to the Lyapunov instability of phase-space
trajectories, even a very small external field will rapidly drive any trajectory far away from the corresponding trajectory without field. Hence linear
response theory which is based on the proportionality of the trajectory separation with the external field could only be expected to hold for extremely
short times, of no physical interest.\\ 
Responses to van Kampen's objections were given by 
many authors~\cite{vliet08,kubo91,coh83,jac83,vv88,leb,cher93}. The main 
arguments of these responses, were based on the deep analysis of the statistical mechanical behavior of the many-body system under the external
perturbation. It was shown that in statistical mechanical calculations one deals with the probability distributions for the behavior of the many particles
rather than to the behavior of an individual particle.  An analysis of the structural stability of hyperbolic dynamics-averaging  and some other 
aspects of the dynamical behavior shows that the linear separation of trajectories goes on long enough for 
Green-Kubo integrals to decay. Moreover,
Naudts, Pule and Verbeure~\cite{npv79}  analyzed the long-time behavior of correlations between extensive variables  for spin-lattice systems and showed that the Kubo
formula, expressing the relaxation function in terms of of the linear response function, is exact in the thermodynamic limit.\\
It was mentioned above that in the 1950s and 1960s the fluctuation relations,  the so-called Green-Kubo relations~\cite{kuz07,vliet08,evse02,rondo08,sear08}, were derived 
for the causal transport coefficients that are defined by causal linear constitutive relations such as Fourier law of heat 
flow or Newton law of viscosity.
Later it was shown also that  it was possible to derive an exact
expression for linear transport coefficients which is valid for systems
of arbitrary temperature, $T$, and density. The Green-Kubo relations  give exact mathematical expression for transport
coefficients in terms of integrals of time correlation functions~\cite{kuz07,vliet08,evse02,rondo08,sear08}.  More precisely,
it was shown that linear transport coefficients are exactly related to the time dependence of
equilibrium fluctuations in the conjugate flux. For a more detailed discussion of these questions see Refs.~\cite{kuz07,vliet08,npv79,evse02,rondo08,sear08}\\
To summarize, close to equilibrium, linear response theory and linear irreversible thermodynamics provide a relatively complete treatment.
However, in systems where \emph{local thermodynamic equilibrium} has broken down, and thermodynamic properties are not the same local functions
of thermodynamic state variables such that they are at equilibrium, serious problems may appear~\cite{kuz07,vliet08}.
%
%
%
%
\section{The Nonequilibrium Statistical Operator Method  and Kinetic Equations}\label{nso}
%
The method of the  nonequilibrium statistical operator~\cite{zub,zub81,zub61,zub70,kal72} was reviewed already
in details~\cite{kuz07} by us.
In this section, we  remind very briefly    
the main ideas of the  nonequilibrium statistical operator (NSO) approach~\cite{zub,zub81,zub61,zub70,kal72,kuz07,kuz05} for the sake of a self-contained 
formulation. 
The precise definition of the nonequilibrium state is quite difficult
and complicated, and is not uniquely specified. Thus the
method of reducing the number of relevant variables was proposed. A large and important class
of transport processes can  reasonably be  modelled in terms of a reduced number of macroscopic
relevant variables~\cite{zub,macl89}. 
It was supposed that the  equations of motion for the "relevant" 
variables (the space- and time-dependent thermodynamic "coordinates" of a many-body nonequilibrium system), 
can be derived
directly from the Liouville equation. This can be done by defining a generalized canonical density
operator depending only upon present values of the thermodynamic "coordinates".
According to D. N. Zubarev~\cite{zub,zub81}, the NSO method
permits one to generalize the Gibbs ensemble method to the nonequilibrium case and to
construct a nonequilibrium statistical operator which  enables one to obtain the transport equations and
calculate the kinetic coefficients in terms of correlation functions, and which, in the case of
equilibrium, goes over to the Gibbs distribution. 
 The basic hypothesis is that after small
time-interval $\tau$ the nonequilibrium distribution is established. Moreover, it is supposed that
it is weakly time-dependent by means of its parameter only. Then the statistical operator $\rho$ for
$t \geq \tau$ can be considered as an "integral of motion" of the quantum Liouville equation
\begin{equation}\label{9.1}
 \frac{\partial \rho }{\partial t} + \frac{1}{i \hbar}[\rho, H ] = 0.
\end{equation}
Here $\partial \rho / \partial t$ denotes time differentiation with respect to the time variable
on which the relevant parameters $F_{m}$ depend. It is important to note once again that $\rho$ depends
on $t$ by means of $F_{m}(t)$ only. We may consider that the system is in thermal, material, and mechanical contact
with a combination of thermal baths and reservoirs  maintaining the given distribution of  parameters
$F_{m}$.  For example,  it can be the densities of energy, momentum, and particle  number for the
system which is macroscopically defined by given fields of temperature, chemical potential and
velocity. It is assumed that the chosen set of parameters is sufficient to characterize  macroscopically
the state of the system. The set of the relevant parameters are dictated by the external
conditions for the system under consideration and, therefore, the term $\partial \rho /\partial t$
appears as the result of the external influence upon the system;  this influence causes
that the system is non-stationary. \\ In order to describe the nonequilibrium process, it is also
necessary  to choose the reduced set of relevant operators $P_{m}$, where $m$ is the index (continuous 
or discrete). In the quantum case, all operators are considered to be in the Heisenberg representation
\begin{equation}\label{9.2}
 P_{m}(t) = \exp \left( \frac{iHt}{\hbar} \right)
 P_{m} \exp \left( \frac{-iHt}{\hbar} \right),
\end{equation}
where $H$ does not depend on the time. The relevant operators may be scalars or vectors. The equations
of motions for  $P_{m}$ will lead to the suitable "evolution equations"~\cite{zub,kuz07}. In the quantum
case
\begin{equation}\label{9.3}
\frac{\partial P_{m}(t)  }{\partial t} - \frac{1}{i \hbar}[P_{m}(t) , H ] = 0. 
\end{equation}
The time argument of the operator $P_{m}(t)$ denotes the Heisenberg
representation with the Hamiltonian $H$ independent of time.
Then we suppose that the state of the ensemble
is  described by a nonequilibrium statistical operator which is a functional of $P_{m}(t)$
\begin{equation}\label{9.4}
  \rho(t ) = \rho \{\ldots P_{m}(t)  \ldots \}.
\end{equation}
Then $\rho(t)$ satisfies the Liouville equation (\ref{9.1}). Hence the  quasi-equilibrium (local-equilibrium)
Gibbs-type distribution will have the form
\begin{equation}\label{9.5}
 \rho_{q} = Q^{-1}_{q} \exp \left(  - \sum_{m}F_{m}(t)P_{m}\right),
\end{equation}
where the parameters $F_{m}(t)$ have the sense of time-dependent thermodynamic parameters,
e.g., of temperature, chemical potential, and velocity (for the hydrodynamic stage), or the
occupation numbers of one-particle states (for the kinetic stage). The statistical
functional $Q_{q}$ is defined by demanding that the operator $\rho_{q}$ be normalized and  equal
to
\begin{equation}\label{9.6}
 Q_{q} = \textrm{Tr} \exp \left(  - \sum_{m}F_{m}(t)P_{m}\right).
\end{equation}
The kinetic equations are of great interest in the theory of transport
processes. In the NSO approach~\cite{zub,kuz07,kuz05}, the main quantities involved are
the following thermodynamically conjugate values:
\begin{equation}\label{9.16}
 \langle P_{m} \rangle = - \frac{\delta \Omega}{\delta F_{m}(t)};
\quad F_{m}(t)  = \frac{\delta S}{\delta \langle P_{m} \rangle }.   
\end{equation}
The generalized transport equations which describe the time evolution of variables $\langle P_{m} \rangle$ and $F_{m}$
follow from the equation of motion for the $ P_{m}$, averaged with the nonequilibrium statistical
operator (\ref{9.4}). It reads
\begin{equation}\label{9.17}
 \langle \dot {P}_{m}\rangle = - \sum_{n} \frac{\delta^{2} \Omega}{\delta F_{m}(t)\delta F_{n}(t)}\dot{F}_{n}(t);
\quad \dot{F}_{m}(t)  =  
\sum_{n} \frac{\delta^{2} S}{\delta \langle P_{m}\rangle \delta \langle P_{n}\rangle} \langle \dot {P}_{n} \rangle. 
\end{equation}
%
%
%
\section{Generalized Kinetic Equations and Electroconductivity}
%
%
%
%
\subsection{Basic formulas}
%
Let us consider a many-particle system in the quasi-equilibrium state. It is determined completely by the quasi-integrals
of motion which are the internal parameters of the system. In this and following sections we will use the notation $A_{j}$ for the relevant observables to distinguish it from the momentum
operator $\vec{P}$. Here, for the sake of simplicity, we shall mainly treat the simplest case of mechanical perturbations acting on the system.
The total Hamiltonian of the system under the influence of homogeneous external perturbation, depending on time as $ \sim \exp(i \omega t)$ is written in
the following form
\begin{equation}\label{10.1}
 {\mathcal H}(t) =  H + H_{F}(t), \quad  H_{F}(t) =  - \sum_{j} A_{j} F_{j}\exp(i \omega t).
\end{equation}
In the standard approach the statistical operator $\rho$  
 can be considered as an "integral of motion" of the quantum Liouville equation
\begin{equation}\label{10.2}
 \frac{\partial \rho (t) }{\partial t} + \frac{1}{i \hbar}[\rho (t), {\mathcal H}(t) ] = 0.
\end{equation}
Using the ideas of the method of the nonequilibrium statistical operator, as it was described above, we can write
\begin{equation}\label{10.3}
\rho(t)  = 
\varepsilon \int^{t}_{-\infty} dt_{1} 
e^{- \varepsilon (t - t_{1})} U(t,t_{1}) \rho(t_{1}) U^{\dag}(t,t_{1}).  
\end{equation}
The time-evolution operator $U(t,t_{1})$ satisfy the conditions
$$ \frac{\partial  }{\partial t}U(t,t_{1}) = \frac{1}{i \hbar}{\mathcal H}(t)U(t,t_{1}),$$
$$ \frac{\partial  }{\partial t_{1}}U(t,t_{1}) = - \frac{1}{i \hbar}{\mathcal H}(t)U(t,t_{1}), \quad U(t,t) = 1.$$
If we consider the special case in which $\rho(t_{1}) \rightarrow \rho_{0}$, where $\rho_{0}$ is an equilibrium solution of the
quantum Liouville equation (\ref{10.2}), then we can find the nonequilibrium statistical operator from the following equation:
\begin{equation}\label{10.4}
 \frac{\partial \rho_{\varepsilon}^{K} (t) }{\partial t} + \frac{1}{i \hbar}[\rho_{\varepsilon}^{K}, {\mathcal H}(t)] 
 = - \varepsilon ( \rho_{\varepsilon}^{K} - \rho_{0}).
\end{equation}
In the  $\lim_{\varepsilon \rightarrow 0^{+}}$ the nonequilibrium statistical operator will corresponds to the Kubo density matrix
\begin{equation}\label{10.5}
\rho_{\varepsilon}^{K}(t)  = \rho_{0}   
=  \frac{1}{i \hbar}\int^{t}_{-\infty} dt_{1} 
e^{- \varepsilon (t - t_{1})} U(t,t_{1}) [\rho_{0}, H_{F}(t_{1}) ]U^{\dag}(t,t_{1}).   
\end{equation}
An average of the observable $B_{j}$ are defined as
$$\lim_{\varepsilon \rightarrow 0^{+}} \textrm{Tr} (\rho_{\varepsilon}^{K}(t) B_{j} ) = \langle B_{j} \rangle_{t}. $$
The linear response approximation for the statistical operator is given by
\begin{eqnarray}\label{10.6}
\rho_{\varepsilon}^{K}(t)  = \rho_{0} \nonumber \qquad  \\
 - \frac{1}{i \hbar} \sum_{j} A_{j} F_{j}\exp(i \omega t)  \int^{0}_{-\infty} dt_{1} 
e^{( \varepsilon  + i \omega ) t_{1}} \exp \left( \frac{i H t_{1}}{\hbar} \right) [\rho_{0}, A_{j} ] \exp \left(\frac{ - i H t_{1}}{\hbar} \right). \qquad  
\end{eqnarray}
Using the obtained expression for the statistical operator the mean values of the relevant observables $B_{i}$ can be calculated. To
simplify notation only observables will be considered for which the mean value in the thermal equilibrium vanishes. In other words, in general the
$B_{i}$ will be replaced by $B_{i} - \langle B_{i} \rangle_{0},$ where $\langle B_{i}\rangle_{0} = \textrm{Tr} (\rho_{0} B_{i} ).$ 
We find
\begin{equation}\label{10.7}
\langle B_{i} \rangle_{t} \equiv \textrm{Tr} (\rho_{\varepsilon}^{K}(t) B_{i} ) = \sum_{j} L_{ij}(\omega ) F_{j}\exp(i \omega t), 
\end{equation}
where the linear response coefficients (linear admittances) are given by
\begin{equation}\label{10.8}
L_{ij}(\omega ) = \langle \dot{A}_{j}; B_{i} \rangle_{\omega - i \varepsilon}, \quad ( \varepsilon \rightarrow 0^{+}).
\end{equation}
This expression vanish for operators $A_{j}$ commuting with the Hamiltonian $H$ of the system, i.e. the Kubo expressions for $L_{ij}(\omega )$
vanish for all $\omega$ where for $\omega = 0$ the correct result is given by
\begin{equation}\label{10.9}
\langle B_{i} \rangle = \beta  \sum_{j} F_{j} \textrm{Tr} (\rho_{0} A_{j} B_{i} ) = \sum_{j} L_{ij}(\omega ) F_{j}\exp(i \omega t). 
\end{equation}
The scheme based on the NSO approach starts with the generalized Liouville equation
\begin{equation}\label{10.10}
 \frac{\partial \rho_{\varepsilon} (t) }{\partial t} + \frac{1}{i \hbar}[\rho_{\varepsilon}(t), {\mathcal H}(t) ] 
 = - \varepsilon ( \rho_{\varepsilon}(t) - \rho_{q}(t)).
\end{equation}
For the set of the relevant operators $P_{m}$ it is follow that
\begin{equation}\label{10.11}
\frac{d}{d t} \textrm{Tr} (\rho_{\varepsilon}(t) P_{m} ) = 0.
 \end{equation}
Here notation are
\begin{eqnarray}\label{10.12}
\rho_{q}(t) = Q^{-1} \exp[ - \beta (H - \mu N - \sum_{m}F_{m}(t)  P_{m})], \\ \nonumber \quad F_{m}(t) = F_{m}\exp(i \omega t),\quad
 P_{m} \rightarrow P_{m} - \langle P_{m} \rangle_{0}.
 \end{eqnarray}
To find the approximate evolution equations, the $\rho_{\varepsilon}(t)$ can be linearized with respect to the external fields $F_{j}$ and parameters
$F_{m}$
\begin{eqnarray}\label{10.13}
\rho(t) = \rho_{q}(t) - \exp(i \omega t)\rho_{0} \Bigl ( \int^{0}_{-\infty} dt_{1} 
e^{( \varepsilon  + i \omega ) t_{1}}  \int^{\beta}_{0} d\tau  \\ \nonumber [ - \sum_{j} F_{j} \dot{A}_{j}(t - i\tau) + \sum_{m} F_{m} \dot{P}_{m}(t - i\tau) +
i \omega \sum_{m} F_{m} P_{m}(t - i\tau)] \Bigr ). 
\end{eqnarray}
As a result we find
\begin{eqnarray}\label{10.14}
\sum_{n} F_{n} \left ( \langle \dot{P}_{n}; P_{m} \rangle_{\omega - i \varepsilon} + 
i \omega \langle P_{n}; P_{m} \rangle_{\omega - i \varepsilon} \right ) = \sum_{j} F_{j} \langle \dot{A}_{j}; P_{m} \rangle_{\omega - i \varepsilon}.
\end{eqnarray}
In other notation we obtain
\begin{eqnarray}\label{10.15}
\sum_{n} F_{n} \left ( (\dot{P}_{n}| P_{m})     +     \langle \dot{P}_{n}; \dot{P}_{m} \rangle_{\omega - i \varepsilon} + 
i \omega [  (P_{n}| P_{m}) +  \langle P_{n}; \dot{P}_{m} \rangle_{\omega - i \varepsilon}] \right ) \\ \nonumber
= \sum_{j} F_{j}\left (  (\dot{A}_{j}| P_{m})  +  \langle \dot{A}_{j}; \dot{P}_{m} \rangle_{\omega - i \varepsilon} \right ),  
\end{eqnarray}
where
\begin{eqnarray}\label{10.16}
\langle A; B \rangle_{\omega - i \varepsilon} = \int^{\infty}_{- \infty} dt e^{(\omega - i \varepsilon)t} (A(t) |B);
   \\ \nonumber
(A |B) = \int^{\beta}_{0} d\lambda
 \textrm{Tr} (\rho_{0}   A(-i\lambda) B  ), \quad \rho_{0} = \frac{1}{Q} e^{- \beta H}; \\ \nonumber
 \chi_{ij} = (A_{i}|A_{j}), \quad \sum_{k} \chi_{ik} (\chi^{-1})_{kj} = \delta_{ij}.    
\end{eqnarray}
Thus our generalized transport equation can be written in the following abbreviated form
\begin{equation}\label{10.17}
\sum_{n} F_{n} P_{nm}   = \sum_{j} F_{j}K_{jm}.
 \end{equation}
%
%
%
\subsection{Electrical conductivity}
%
%
The general formalism of the nonequilibrium statistical operator has been the starting point of many calculations
of transport coefficients in concrete physical systems. In the present section we consider some selected
aspects of the theory of electron conductivity in transition metals and disordered alloys~\cite{ck1,ck2,ck3}.
We  put $\hbar =1$  for simplicity of notation. 
Let us consider the \emph{dc}  electrical conductivity 
\begin{eqnarray}\label{10.25}
 \sigma =  \frac{e^{2}}{3m^{2}\Omega} \langle \vec{P}; \vec{P} \rangle = \frac{e^{2}}{3m^{2}\Omega}
  \int^{0}_{-\infty} dt e^{\varepsilon t} \int^{\beta}_{0}d\lambda
 \textrm{Tr} (\rho   P(t-i\lambda) \vec{P}  ), \quad \varepsilon \rightarrow 0
\end{eqnarray}
where $\vec{P}$ is the total momentum of the electrons. Representing $\vec{P}$ as a sum of operators $A_{i}$ 
(relevant observables) which can be chosen properly to describe the system considered (see below), the corresponding
correlation functions $\langle A_{i}; A_{j} \rangle $ can be calculated by the set of equations~\cite{ck1}
\begin{eqnarray}\label{10.26}
\sum_{k}\Bigl ( i\varepsilon \delta_{ik} - i\sum_{l} (\dot{A}_{l}|A_{l}) (\chi^{-1})_{lk} 
+ i\sum_{l} \Pi_{il}(\chi^{-1})_{lk} \Bigr ) \langle A_{k}; A_{j} \rangle = i \chi_{ij},
\end{eqnarray}
where
\begin{eqnarray}\label{10.27}
\chi_{ij} = (A_{i}|A_{j}), \quad \sum_{k} \chi_{ik} (\chi^{-1})_{kj} = \delta_{ij};   \\ \label{10.28}
(A_{i}|A_{j}) = \int^{\beta}_{0} d\lambda
 \textrm{Tr} (\rho   A_{i}(-i\lambda) A_{j}  ), \quad \rho = \frac{1}{Q} e^{- \beta H};   \\ 
 \Pi_{ij} = \langle \dot{A}_{i} \tilde{C} | C \dot{A}_{j} \rangle^{C}, \quad \dot{A}_{l} = i [H,A_{l}]. \label{10.29}
\end{eqnarray}
The operator $\tilde{C} = 1 - \tilde{P}$ is a projection operator~\cite{ck1} with $\tilde{P} = 
\sum_{ij} |A_{i}) (\chi^{-1})_{ij} (A_{i}|$ and $\langle \ldots \rangle^{C}$ denotes that in the time evolution
of this correlation function, operator $L = i [H, \ldots ]$ is to be replaced by $C L C$. By solving the set of equations
(\ref{10.26}) the correlation functions $\langle A_{i}; A_{j} \rangle$ which we started from are replaced by the correlation
functions $\Pi_{ij}$ (\ref{10.29}), and with a proper choice of the relevant observables these correlation functions can be calculated
in a fairly simple approximation~\cite{ck1}. It is, however, difficult to go beyond this first approximation and,
in particular, to take into account the projection operators. Furthermore this method is restricted to the calculation
of transport coefficients where the exact linear response expressions are known; generalization to thermal transport
coefficients is not trivial. In Ref.~\cite{kuz05}    we described   a kind of general formalism for the calculation
of transport coefficients which includes the approaches discussed above and which can be adapted to the problem investigated.\\
For simplicity of notation we restrict our consideration here on the influence of a stationary external electrical field.
In the linear response theory the density matrix of the system becomes
\begin{equation}\label{10.30}
\rho_{LR} = \frac{e \vec{E}}{m}  \int^{t}_{-\infty} dt' e^{\varepsilon (t' - t)} \int^{\beta}_{0} d\lambda
  \rho   \vec{P} (t' - t - i\lambda), \quad \varepsilon \rightarrow 0,
\end{equation}
where the time dependence in $P(t)$ is given by the total Hamiltonian of the system without the interaction term with
the electrical field. From another point of view we can say that there is a reaction of the system on the external
field which can be described by relevant observables such as shift of the Fermi body or a redistribution of the single
particle occupation numbers, \emph{etc}. Hence, for small external fields the system can be described in a fairly good
approximation by the quasi-equilibrium statistical operator of the form
\begin{equation}\label{10.31}
 \rho_{q} = \frac{1}{Q_{q}} e^{- \beta (H - \sum_{i} \alpha_{i} A_{i})},
\end{equation}
where the $A_{i}$ are the observables relevant for the reaction of the system and the $\alpha_{i}$ are parameters
proportional to the external field. Of course, the statistical operator (\ref{10.31}) is not a solution of the Liouville
equation, but an exact solution can be found  starting from (\ref{10.31}) as an initial condition:
\begin{equation}\label{10.32}
\rho_{s} = \rho_{q} - i \int^{o}_{-\infty} dt' e^{\varepsilon (t' - t)} 
\exp (iH_{s}(t' - t)) \rho_{q} \exp (-iH_{s}(t' - t)),
\end{equation}
where $H_{s} = H - e \vec{E}\sum_{i} \vec{r}_{i}.$  In order to determine the parameters $\alpha_{i}$ we demand that the mean values
of the relevant observables $A_{i}$ are equal in the quasi-equilibrium state $\rho_{q}$ and in the real state $\rho_{s}$,
i.e.
\begin{equation}\label{10.33}
\textrm{Tr} (\rho_{s} A_{i}) = \textrm{Tr} (\rho_{q} A_{i}).
\end{equation}
This condition is equivalent to the stationarity condition
$$\frac{d}{dt} \textrm{Tr} (\rho_{s} A_{i}) = 0. $$
For a sufficiently complete set of operators $A_{i}$ the condition (\ref{10.33}) ensures that $\rho_{q}$
describes the system with a sufficient accuracy. 
Linearizing Eq.(\ref{10.33}) with respect to the parameters $\alpha_{i}$ and the external field $E$, one obtains the set of
equations
\begin{equation}\label{10.34}
\sum_{j} \alpha_{j} \Bigl ( -i \textrm{Tr} (\rho[A_{j}, A_{i}]) + \langle \dot{A}_{j} ;   \dot{A}_{i} \rangle \Bigr ) =
\frac{e \vec{E}}{m} \Bigl ( ( \vec{P} |  A_{i}) +   \langle \vec{P}; \dot{A}_{i} \rangle         \Bigr ). 
\end{equation}
This set of equations can be shown to be equivalent to Eq.(\ref{10.26}); whereas  in the higher orders of interaction the equations
(\ref{10.34}) are more convenient to handle because the time dependence  is given here in an explicit form without 
any projection.\\ With the parameters $\alpha_{i}$ the current density is given by
\begin{equation}\label{10.35}
\vec{J} = \frac{e}{m \Omega} \textrm{Tr} (\rho_{q}\vec{P}) = \frac{e}{m \Omega} \sum_{j} \alpha_{j} (  A_{j} | \vec{P}).
\end{equation}
Supposing that the total momentum $\vec{P}$ of the electrons can be built up by the operators $A_{i}$,  it can be shown 
that the Eqs.(\ref{10.34}) and (\ref{10.35}) include the Kubo expression for the conductivity (see below). In order to solve the
system of equations (\ref{10.34}) a generalized variational principle can be formulated, but the reduction of the number of
parameters $\alpha_{i}$ by a variational ansatz corresponds to a new restricted choice of the relevant observable $A_{i}$.
%
\section{Resistivity of Transition Metal with  Non-spherical Fermi Surface}\label{nfermi}
%
The applicability of the transport equations  (\ref{10.34}) and (\ref{10.35})   derived above to a given problem depends strongly on the choice of the
relevant operators $A_{i}$. The first condition to be fulfilled is that the mean values of the occupation numbers of all
quasiparticles involved in the transport process should be time independent, i.e.
\begin{equation}\label{11.1}
\textrm{Tr} (\rho_{s}(\alpha_{i}) n_{k}) = \textrm{Tr} (\rho_{q} (\alpha_{i}) n_{k}), \quad \frac{d}{dt} \textrm{Tr} (\rho_{s}(\alpha_{i}) n_{k}) = 0.
\end{equation}
Of course, this condition is fulfilled trivially for $A_{i} \rightarrow n_{k}$, but in this case the 
equations (\ref{10.34}) 
cannot be solved in practice. In the most cases, however, there can be found a reduced set of operators sufficiently
complete to describe the reaction of the system on the external field. It can be shown that under certain conditions the
scattering process can be described by one relaxation time only and then the set of relevant operators reduces to the
total momentum of the electrons describing a homogeneous shift of the Fermi body in the Bloch (momentum) space. These conditions
are fulfilled for a spherical Fermi body at temperatures small in comparison to the degeneration temperature and for an
isotropic scattering mechanism where the scatterers remain in the thermal equilibrium. In this simple case the 
Eqs.(\ref{10.34})  are reduced to the so-called resistivity formula~\cite{ck1}. For non-spherical Fermi bodies
the set of relevant observables has to be extended in order to take into account not only its shift in the $k$-space
but also its deformation.
Our aim is to develop a theory of electron conductivity for the one-band model of a
transition metal. The model considered is the modified Hubbard model, which include the electron-electron
interaction as well as the electron-lattice interaction within MTBA. The non-spherical Fermi surface is taken into
account. \\ The studies of the electrical resistivity of many transition metals  revealed some peculiarities. It is
believed that these specific features are caused by the fairly complicated dispersion law of the carriers and the
existence of the two subsystems of the electrons, namely the broad $s-p$ band and relatively narrow $d$-band.
The Fermi surface  of transition metals is far from the spherical form. In addition,
the lattice dynamics and  dispersion relations of the phonons is much more complicated than in simple metals.
As a result, it is difficult to attribute the observed temperature dependence of the resistivity of transition metals
to definite scattering mechanisms.\\ Here we investigate the influence of the electron dispersion on the electrical
resistivity within a simplified but workable model. We consider an effective single-band model of  transition
metal with tight-binding dispersion relation of the electrons. We take into account the electron-electron and 
electron-lattice interactions within the extended Hubbard model. The electron-lattice interaction is described within
the MTBA. For the calculation of the electrical conductivity the generalized kinetic equations are used which were
derived by the nonequilibrium statistical operator method. In these kinetic equations the shift of the non-spherical
Fermi surface and its deformation by the external electrical field are taken into account explicitly. By using the weak scattering
limit the explicit expressions for the electrical resistivity are obtained and its temperature dependence is estimated.\\
We consider a transition metal model with one
non-spherical Fermi body shifted in the $k$-space and deformed by the external electrical field. Hence, the Fermi
surface equation $E(\vec{k}) = E_{F}$ is transformed into $\tilde{E}(\vec{k}) = E_{F}$, where
\begin{equation}\label{11.2}
\tilde{E}(\vec{k}) = E(\vec{k}) + m \vec{v}_{1}\frac{\partial E}{\partial \vec{k}}  +
m \sum_{i=1\sigma }^{n}    \vec{v}_{i} \Phi^{i}(\vec{k})\frac{\partial E}{\partial \vec{k}} + \ldots  
\end{equation}
The term proportional to $\vec{v}_{1}$ describes a homogeneous shift of the Fermi surface in the $k$-space and the
last terms allow for deformations of the Fermi body. The polynomials $\Phi^{i}(\vec{k})$ have to be chosen
corresponding to the symmetry of the crystal~\cite{gar,bro,al,pin} and in consequence of the 
equality $\tilde{E}(\vec{k} + \vec{G}) = \tilde{E}(\vec{k})$ (which is fulfilled in our tight-binding model) 
they should satisfy the relation
$$ \Phi^{i}(\vec{k}+ \vec{G}) = \Phi^{i}(\vec{k}),$$
where $\vec{G}$ is a reciprocal lattice vector.
Thus the relevant operators are given by
\begin{equation}\label{11.3}
A_{i} \rightarrow m \sum_{k\sigma }\Phi^{i}(\vec{k})\frac{\partial E}{\partial \vec{k}}n_{k\sigma}, \quad \Phi^{1} = 1.
\end{equation}
For our tight-binding model we are restricting by the assumption that $\Phi^{i} \rightarrow 0$ for $ i \geq 3$.
%
\subsection{Generalized kinetic equations}
%
The quantum-mechanical  many-body system is described by the statistical operator  $\rho_{s}$ obeying the 
modified Liouville equation of motion Eq.(\ref{10.10})
\begin{equation}\label{11.4}
 \frac{\partial \rho_{s}  }{\partial t} + \frac{1}{i \hbar}[\rho_{s}, H_{s} ] = 
 - \varepsilon (\rho_{s}  -  \rho_{q}),
\end{equation}
where $H_{s} = H + H_{E}$ is the total Hamiltonian of the system including the interaction with an external
electrical field $H_{E} = -e \vec{E}\sum_{i} \vec{r_{i}}$ and $\rho_{q}$ is the quasi-equilibrium statistical
operator. According to the NSO formalism the relevant operators $P_{m}$ should be selected. These operators include 
all the relevant observables which describe the reaction of the system on the external electrical field. 
These relevant operators satisfy the condition
\begin{equation}\label{11.5}
  \textrm{Tr} (\rho_{s} (t,0) P_{m} )= \langle P_{m}\rangle = \langle P_{m} \rangle_{q} ; \quad  \textrm{Tr}   \rho_{s}  = 1.  
\end{equation}
This condition is equivalent to the stationarity condition
\begin{equation}\label{11.6}
  \frac{d}{dt} \langle P_{m}\rangle =  0 \rightarrow \langle [ H_{s},P_{m}] \rangle  + 
  \langle [ H_{E},P_{m}] \rangle  = 0. 
\end{equation}
In the framework of the linear response theory the operators $\rho_{s}$ and $\rho_{q}$ in Eqs. (\ref{11.5}) and (\ref{11.6})
should be expanded to the first order in the external electrical field $\vec{E}$ and in the parameters $F_{m}$. Thus 
Eq. (\ref{11.6}) becomes~\cite{ck1}
\begin{equation}\label{11.7}
\sum_{n}F_{n} \Bigl ( -i \textrm{Tr} (\rho [ P_{n},P_{m}] ) + \langle \dot{P}_{n};\dot{P}_{m}\rangle  \Bigr )=  
\frac{e E}{m} \Bigl ( \textrm{Tr} (\rho   \vec{P}(-i\lambda) P_{m}  ) + \langle   \vec{P}  ;\dot{P}_{m}\rangle  \Bigr ),
\end{equation}
where
\begin{eqnarray}\label{11.8}
 \dot{P}_{m} = i[H_{s},P_{m}], \\
 \langle   A  ;B \rangle  =  \int^{0}_{-\infty}  dte^{\varepsilon t} \int^{\beta}_{0} d\lambda
 \textrm{Tr} (\rho   A(t-i\lambda) B ), \label{11.9}
\end{eqnarray}
$$ A(t) = \exp (iHt) A \exp (-iHt); \quad \rho = Q^{-1}\exp (-\beta H),$$ 
and $P$ is the total momentum of the electrons.\\ The equations (\ref{11.7}) are generalized kinetic equations in which
 the relaxation times and particle numbers are expressed via the correlation functions. It will be shown below that these
 equations can be reduced to the Kubo formula for the electrical conductivity provided a relation 
 $\vec{P}_{e} = \sum \vec{\alpha}_{i} P_{i}$ can be found. \\
The generalized kinetic equations  (\ref{11.7}) can be solved and the parameters $F_{m}$ can be determined by using
a variational principle. The current density is given by 
\begin{equation}\label{11.10}
\vec{j} = \frac{e}{m \Omega} \textrm{Tr} ( \rho_{s}\vec{P}_{e}) = \frac{e}{m \Omega} \textrm{Tr} ( \rho_{q}\vec{P}_{e}), 
\end{equation}
where the conditions (\ref{11.5}) have been used and   $\Omega$ is the volume of the system. Linearizing Eq.(\ref{11.10}) in
the parameter $F_{m}$ we find
\begin{equation}\label{11.11}
\vec{j} = \frac{e}{m \Omega} \sum_{m} F_{m} \int^{\beta}_{0} d\lambda
 \textrm{Tr} (\rho   P_{m}(t-i\lambda) \vec{P}_{e}  ) \equiv \frac{1}{R}\vec{E},
\end{equation}
where the proportionality of the $F_{m}$ to the external electrical field has been taken into account.
For the tight-binding model Hamiltonian  (\ref{4.97}),  (\ref{4.98}) the proper set of operators $P_{m}$ is given by
\begin{eqnarray}\label{11.12}
  P_{e} =  P_{1}  = m \sum_{k \sigma} \frac{\partial E}{\partial k} a^{\dag}_{k\sigma}a_{k\sigma}, \\
 P_{i}  = m \sum_{k \sigma} \Phi^{i}(\vec{k})\frac{\partial E}{\partial k} a^{\dag}_{k\sigma}a_{k\sigma}, 
 \quad (i=2 \ldots n).\label{11.13}
\end{eqnarray}
The parameters $F_{m}$ are replaced by the generalized drift velocities. Then the quasi-equilibrium statistical
operator $\rho_{q}$  take the form
\begin{eqnarray}\label{11.14}
\rho_{q} = \frac{1}{Q_{q}} \exp \Bigl ( -\beta [ H + m v_{1}\sum_{k \sigma} \frac{\partial E}{\partial k} a^{\dag}_{k\sigma}a_{k\sigma} + 
 \sum_{i=2 \ldots n}m v_{i}\Phi^{i}(\vec{k})\frac{\partial E}{\partial k} a^{\dag}_{k\sigma}a_{k\sigma}] \Bigr ). \qquad
\end{eqnarray}
It should be mentioned here that in general in $\rho_{q}$ the redistribution of the scatterers by collisions with electrons 
should be taken into consideration. For the electron-phonon problem, e.g., the phonon drag can be described by an additional
term $\vec{v}_{ph}\vec{P}_{ph}$ in Eq.(\ref{11.2}) where $\vec{v}_{ph}$ is the mean drift velocity  and 
$\vec{P}_{ph}$ is the total momentum of the phonons. Here it will be supposed for simplicity, that due to phonon-phonon
\emph{Umklapp} processes, \emph{etc}., the phonon subsystem remains near thermal equilibrium. In the same way the above consideration can be
generalized to many-band case. In the last case, the additional terms in (\ref{11.2}), describing shift and deformation of other
Fermi bodies should be taken into account.\\
For the tight-binding model Hamiltonian  (\ref{4.97}),  (\ref{4.98}) the time derivatives of the generalized momenta (generalized forces)
in Eq.(\ref{11.7}) are given by
\begin{equation}\label{11.15}
 \dot{P}_{n}  \rightarrow   \dot{P}_{j}  =  \dot{P}^{ee}_{j} +  \dot{P}^{ei}_{j}, 
\end{equation}
%
%
%
\begin{eqnarray}\label{11.16}
\dot{P}^{ee}_{j} = i[H_{ee},P_{j}] = \qquad    \\ 
\frac{i U m}{N} \sum_{k_{1}k_{2}}  \sum_{k_{3}k_{4}G}
\Bigl ( \Phi^{j}(\vec{k_{4}})\frac{\partial E}{\partial k_{4}} + \Phi^{j}(\vec{k_{2}})\frac{\partial E}{\partial k_{2}}
- \Phi^{j}(\vec{k_{3}})\frac{\partial E}{\partial k_{3}} - \Phi^{j}(\vec{k_{1}})\frac{\partial E}{\partial k_{1}}\Bigr )\cdot  \nonumber \\ 
a^{\dagger}_{k_{1}\uparrow}a_{k_{2}\uparrow}a^{\dagger}_{k_{3}\downarrow}
a_{k_{4}\downarrow} \delta (\vec {k}_{1} - \vec {k}_{2} + \vec {k}_{3} - \vec {k}_{4} + \vec {G}), \nonumber \\
\dot{P}^{ei}_{j} = m \sum_{k_{1}k_{2}}  \sum_{qG}  \sum_{\sigma \nu} g^{\nu}_{k_{1}k_{2}}
\Bigl ( \Phi^{j}(\vec{k_{2}})\frac{\partial E}{\partial k_{2}} -  \Phi^{j}(\vec{k_{1}})\frac{\partial E}{\partial k_{1}}
\Bigr ) \cdot \quad  \\ a^{\dagger}_{k_{2}\sigma}a_{k_{1}\sigma}(b^{\dagger}_{q\nu} + b_{-q\nu}) \delta (\vec {k}_{2} - \vec {k}_{1} + \vec {q} + \vec {G}). \nonumber
\label{11.17}
\end{eqnarray}
We confine ourselves by the weak scattering limit. For this case the total Hamiltonian $H$ in evolution
$ A(t) = \exp (iHt)A \exp (-iHt)$ we replace by the Hamiltonian of the free quasiparticles
$H^{0}_{e} + H^{0}_{i}$. With this approximation the correlation functions in Eq.(\ref{11.7}) can be calculated
straightforwardly. We find that
\begin{equation}\label{11.18}
\langle \dot{P}_{j};\dot{P}_{l}\rangle   \approx \langle \dot{P}^{ee}_{j};\dot{P}^{ee}_{l}\rangle    + 
\langle \dot{P}^{ei}_{j};\dot{P}^{ei}_{l}\rangle.   
\end{equation}
Restricting ourselves for simplicity to a cubic system, the correlation functions of the generalized forces are
given by
\begin{eqnarray}\label{11.19}
\langle \dot{P}^{ee}_{j};\dot{P}^{ee}_{l}\rangle = \\    \nonumber
 \frac{U^{2}m^{2}\beta \pi}{N^{2}}\sum_{k_{1}k_{2}}  \sum_{k_{3}k_{4}G}
A_{j}(k_{1},k_{2},k_{3},k_{4})A_{l}(k_{1},k_{2},k_{3},k_{4}) \cdot \\  \nonumber f_{k_{1}}(1 - f_{k_{2}})f_{k_{3}}(1 - f_{k_{4}})
\delta (E(k_{1}) - E(k_{2}) + E(k_{3}) - E(k_{4})) \\  \nonumber
\delta (\vec {k}_{1} - \vec {k}_{2} + \vec {k}_{3} - \vec {k}_{4} + \vec {G}),   \nonumber \\
\langle \dot{P}^{ei}_{j};\dot{P}^{ei}_{l}\rangle  = \\  \nonumber
  2 \pi m^{2}\beta \sum_{k_{1}k_{2}} \sum_{q \nu G}
( g^{\nu}_{k_{1}k_{2}})^{2} B_{j}(k_{1},k_{2})B_{l}(k_{1},k_{2})\cdot  \\  \nonumber  f_{k_{2}}(1 - f_{k_{1}}) N(q\nu)
\delta (E(k_{2}) - E(k_{1}) + \omega (q\nu)) \delta (\vec {k}_{2} - \vec {k}_{1} - \vec{q} + \vec {G}), \nonumber
\label{11.20}
\end{eqnarray}
where
\begin{eqnarray}\label{11.21}
A_{j}(k_{1},k_{2},k_{3},k_{4}) = \Bigl ( \Phi^{j}(\vec{k_{4}})\frac{\partial E}{\partial k_{4}} + \Phi^{j}(\vec{k_{2}})\frac{\partial E}{\partial k_{2}}
- \Phi^{j}(\vec{k_{3}})\frac{\partial E}{\partial k_{3}} - \Phi^{j}(\vec{k_{1}})\frac{\partial E}{\partial k_{1}}\Bigr ), \qquad\\
B_{j}(k_{1},k_{2}) = \Bigl ( \Phi^{j}(\vec{k_{2}})\frac{\partial E}{\partial k_{2}} -  \Phi^{j}(\vec{k_{1}})\frac{\partial E}{\partial k_{1}}
\Bigr ) \qquad  \label{11.22}
\end{eqnarray}
and
$$ f(E(k)) = f_{k} = [ \exp \beta (E(k) - E_{F}) + 1 ]^{-1},$$ 
$$ N(\omega (q\nu)) = N(q\nu)  = [ \exp \beta \omega (q\nu) - 1 ]^{-1}.$$
The correlation functions $\langle \vec{P}_{1};\dot{\vec{P}}_{l}\rangle $ vanish in the weak scattering limit. 
The generalized electron numbers in Eq.(\ref{11.13}) become
\begin{equation}\label{11.23}
N_{l} = \frac{1}{m} \textrm{Tr} \Bigl ( \rho \vec{P}_{1}(i\lambda);\vec{P}_{l} \Bigr ) = 
m \beta \sum_{k} \Phi^{l}(\vec{k})\frac{\partial E}{\partial \vec{k}} f_{k}(1 - f_{k}).
\end{equation}
%
%
\subsection{Temperature dependence of R}
%
Let us consider the low-temperature dependence of the electrical resistivity obtained above. For this region
we have
$$ \lim_{T \rightarrow 0} \beta f_{k}(1 - f_{k}) \rightarrow \delta (E(k) - E_{F}). $$
Thus the generalized electron numbers $N_{l}$ in Eq.(\ref{11.23})  do not depend on temperature, and the temperature
dependence of $R$ in Eq.(\ref{11.11}) is given by the correlation functions (\ref{11.19}) and (\ref{11.20}). For the term arising from
the electron-electron scattering we find
\begin{eqnarray}\label{11.24}
\langle \dot{P}^{ee}_{j};\dot{P}^{ee}_{l}\rangle = \beta \int_{0} \int \int^{E_{max}} dE(k_{1})dE(k_{2})dE(k_{3})
F^{1}_{jl}(E(k_{1}), E(k_{2}), E(k_{3}))\cdot \nonumber \\ 
f_{k_{1}}(1 - f_{k_{2}})f_{k_{3}}[1 - f(E(k_{1}) - E(k_{2}) + E(k_{3}))], \qquad
\end{eqnarray} 
where
\begin{eqnarray}\label{11.25}
F^{1}_{jl}(E(k_{1}), E(k_{2}), E(k_{3})) = \frac{U^{2}m^{2}\pi}{N^{2}}  \frac{\Omega^{3}}{(2\pi)^{9}}\sum_{G}
\int d^{2}S_{1}\int d^{2}S_{2} \int d^{2}S_{3} \cdot \nonumber \\ \nonumber 
\frac{A_{j}(k_{1},k_{2},k_{3},k_{1} - k_{2} + k_{3} + G) A_{l}(k_{1},k_{2},k_{3},k_{1} - k_{2} + k_{3} + G)}
{| \frac{\partial E}{\partial k_{1}}| | \frac{\partial E}{\partial k_{2}}| | \frac{\partial E}{\partial k_{3}}|} \cdot\nonumber \\
\delta (E(k_{1}) - E(k_{2}) + E(k_{3}) - E(k_{1} - k_{2} + k_{3} + G)).
\end{eqnarray} 
With the substitution
$$ x = \beta (E(k_{1}) - E_{F} ), \quad  y = \beta (E(k_{2}) - E_{F} ), \quad  z = \beta (E(k_{3}) - E_{F} ),$$
the expression (\ref{11.24}) reads
\begin{eqnarray}\label{11.26}
\langle \dot{P}^{ee}_{j};\dot{P}^{ee}_{l}\rangle = \beta^{-2} \int_{-\beta E_{F}} \int \int^{\beta E_{max} - E_{F}} 
 \frac{1}{1 + \exp(x)}  \frac{1}{1 + \exp(-y)} \frac{1}{1 + \exp(z)} \cdot \nonumber \\
  \frac{dx dy dz}{1 + \exp (- x + y - z)} 
F^{1}( \frac{x}{\beta} + E_{F},  \frac{y}{\beta} + E_{F},  \frac{z}{\beta} + E_{F} ) \nonumber \\
 = \beta^{-2} A^{ee}_{jl}. \qquad
\end{eqnarray} 
It is reasonably to conclude from this expression, that in the limits
$$ \lim \beta E_{F} \rightarrow \infty, \quad     \lim \beta (E_{max} - E_{F}) \rightarrow \infty~, $$
the electron-electron correlation function for low temperatures becomes proportional to $T^{2}$ for any polynomial
$\Phi^{j}(\vec{k})$. \\
For the electron-phonon contributions to the resistivity the temperature dependence is given by the Bose distribution
function of phonons $N(q \nu)$. Because of the quasi-momentum conservation law $\vec{q} = \vec{k}_{2} - \vec{k}_{1} - \vec{G}$
the contribution of the electron-phonon \emph{Umklapp} processes freezes out at low temperatures as 
$[ \exp \beta \omega(q_{min}) - 1 ]^{-1}$, where $q_{min}$ is the minimal distance between the closed Fermi surfaces
in the extended zone scheme. For electron-phonon normal  processes as well as for electron-phonon \emph{Umklapp} processes
in metals with an open Fermi surface the quasi-momentum conservation law can be fulfilled for phonons with 
$q \rightarrow 0$ being excited at low temperatures solely. To proceed further we use the relation
$\omega(q\nu) = v^{\nu}_{0} ( \vec{q}/q  )$ 
and the periodicity of the quasiparticle dispersion relation with the reciprocal lattice vector $\vec{G}$. Taking into 
account the relation $ \Phi^{i}(\vec{k}+ \vec{G}) = \Phi^{i}(\vec{k}),$  we find to the first non-vanishing order in $q$
\begin{eqnarray}\label{11.27}
g^{\nu}_{k,k+q+G} \approx  q ( \frac{\vec{q} }{q} \frac{\partial}{\partial k'}  ) g^{\nu}_{k,k'} \vert _{k'=k},\\
B_{j} (k,k+q+G) \approx q ( \frac{\vec{q} }{q} \frac{\partial}{\partial k'}  ) B_{j}(k,k')\vert _{k'=k}, \label{11.28} \\
\delta (E(k+q+G) - E(k) + \omega(q\nu)) \approx \frac{1}{q} \delta (\frac{\vec{q} }{q} \frac{\partial E}{\partial k}  + v^{\nu}_{0}).
\label{11.29}
\end{eqnarray}
Hence, we have
\begin{eqnarray}\label{11.30}
\langle \dot{P}^{ei}_{j}; \dot{P}^{ei}_{l} \rangle \cong  \qquad \\ \nonumber
m^{2} \beta  \frac{\Omega^{2}}{(2\pi)^{9}} \sum_{\nu}
\int^{q_{max}}_{0} q^{5} dq \int \sin (\theta_{q}) d\theta_{q} \int d\varphi_{q}  
[ \exp (\beta v^{\nu}_{0} q) - 1 ]^{-1} F^{2}_{jl} (\theta_{q},\varphi_{q}),
\end{eqnarray}
where
\begin{eqnarray}\label{11.31}
F^{2}_{jl} = \int dk \Bigl ( (\frac{\vec{q} }{q} \frac{\partial}{\partial k'}  ) g^{\nu}_{k,k'} \vert _{k'=k} \Bigr )^{2}
\Bigl ( (\frac{\vec{q} }{q} \frac{\partial}{\partial k}  ) B_{l}(k,k') \vert _{k'=k} \Bigr )\cdot \nonumber \\
\Bigl ( (\frac{\vec{q} }{q} \frac{\partial}{\partial k}  ) B_{j}(k,k') \vert _{k'=k} \Bigr )
f_{k}(1 - f_{k})\delta (\frac{\vec{q} }{q} \frac{\partial E}{\partial k}  + v^{\nu}_{0}).
\end{eqnarray}
In the $k$-integral in Eq.(\ref{11.30}) the integration limits have to be chosen differently for normal and \emph{Umklapp} processes.
With the substitution $x = \beta v^{\nu}_{0} q$ we find
\begin{eqnarray}\label{11.32}
\langle \dot{P}^{ei}_{j}; \dot{P}^{ei}_{l} \rangle =  \qquad \\ \nonumber
 \beta^{-5} \frac{m \Omega^{2}}{(2\pi)^{9}} \sum_{\nu} \frac{1}{(v^{\nu}_{0})^{6}}
\int^{\beta v^{\nu}_{0}q_{max}}_{0} dx \frac{x^{5}}{e^{x} - 1}  \int \sin (\theta_{q}) d\theta_{q} \int d\varphi_{q}  
 F^{2}_{jl} (\theta_{q},\varphi_{q}) = A^{ei}_{jl} T^{5}. 
\end{eqnarray}
Thus we can conclude that the electron-phonon correlation function is proportional to $T^{5}$ for any polynomial
$\Phi^{i}_{k}$. It is worthy to note that for an open Fermi surface this proportionality follows for normal and 
\emph{Umklapp} processes either. For a closed Fermi surface the \emph{Umklapp} processes freeze out at sufficiently low temperatures,
and the electron-phonon normal  processes contribute to the electrical resistivity only. With the aid of the 
Eqs.(\ref{11.26}) and (\ref{11.32}) the generalized kinetic equations (\ref{11.7}) becomes
\begin{equation}\label{11.33}
\sum_{i=1 }^{n} v_{i}
 \Bigl ( A^{ee}_{ij} T^{2} +  A^{ei}_{ij} T^{5}  \Bigr ) = e E N_{j}.
\end{equation}
For simplicity we restrict our consideration to two parameters $v_{1}$ and $v_{2}$ describing the homogeneous shift and
one type of deformation of the Fermi body. Taking into consideration more parameters is straightforward but does not
modify very much qualitative results. Finally, the expression for the electrical resistivity becomes~\cite{ck1,ck2}
\begin{equation}\label{11.34}
R = \frac{\Omega}{3e^{2}} \frac{(A^{ee}_{11} T^{2} +  A^{ei}_{11} T^{5})(A^{ee}_{22} T^{2} +  A^{ei}_{22} T^{5}) - 
(A^{ee}_{12} T^{2} +  A^{ei}_{12} T^{5})^{2}}{N^{2}_{1}(A^{ee}_{22} T^{2} +  A^{ei}_{22} T^{5}) + 
N^{2}_{2}(A^{ee}_{11} T^{2} +  A^{ei}_{11} T^{5}) - 2N_{1} N_{2}(A^{ee}_{12} T^{2} +  A^{ei}_{12} T^{5})}. \qquad
\end{equation}
In general, a simple dependence $ R \sim T^{2}$ or $ R \sim T^{5}$ can be expected only if one of the scattering
mechanisms  dominate. For example, when $  A^{ee}_{ij} \approx 0$, we find
\begin{equation}\label{11.35}
R = \frac{\Omega}{3e^{2}} \frac{(A^{ei}_{11}A^{ei}_{22} - A^{ei}_{12}) T^{5} }
{N^{2}_{1} A^{ei}_{22} + N^{2}_{2} A^{ei}_{11} - 2N_{1} N_{2} A^{ei}_{12}}. 
\end{equation}
If, on the other hand, the deformation of the Fermi body is negligible $(v_{2} = 0 )$, then from Eq.(\ref{11.34}) it
follows that
\begin{equation}\label{11.36}
R = \frac{\Omega}{3e^{2}N^{2}_{1}} (A^{ee}_{11} T^{2} +  A^{ei}_{11} T^{5}).
\end{equation}
It is interesting to note that a somewhat similar in structure to expression (\ref{11.34}) have been used to describe the
resistivity of the so-called strong-scattering metals. In order to improve our formula for the resistivity derived above
the few bands and 
the interband scattering (e.g. $s-d$ scattering) as well as the phonon drag effects should be taken into account.
%
%
\subsection{Equivalence of NSO approach and Kubo formalism}
%
Equivalence of the generalized kinetic equations to the Kubo formula for the electrical resistivity can be
outlined as following. Let us consider the generalized kinetic equations
\begin{equation}\label{11.37}
\sum_{n}F_{n} \Bigl ( -i \textrm{Tr} (\rho [ P_{n},P_{m}] ) + \langle \dot{P}_{n};\dot{P}_{m}\rangle  \Bigr )=  
\frac{e E}{m} \Bigl ( \textrm{Tr} (\rho   \vec{P_{e}}(-i\lambda) P_{m}  ) + \langle   \vec{P_{e}}  ;\dot{P}_{m}\rangle  \Bigr ).
\end{equation}
To establish the correspondence of these equations with the Kubo expression for the electrical resistivity,  it
is necessary to express the operators of the total electron momentum $\vec{P_{e}}$ and the current density $\vec{j}$ 
in terms of the operators $P_{m}$. In the other words, we suppose that there exists a suitable set of coefficients
$a_{i}$ with the properties
\begin{equation}\label{11.38}
 \vec{P_{e}} =  \sum_{i}\vec{a}_{i} P_{i}, \quad \vec{j} = \frac{e}{m\Omega} \vec{P_{e}}.
\end{equation}
We get, by integrating Eq.(\ref{11.37}) by parts, the following relation
\begin{equation}\label{11.39}
\sum_{n}F_{n} \Bigl (   \int^{\beta}_{0} d\lambda   Tr (\rho   P_{n}(-i\lambda)P_{m}  )
- \varepsilon \langle P_{n};P_{m}\rangle  \Bigr ) =  
\frac{e E}{m}   \langle  \vec{P_{e}}; P_{m}\rangle.  
\end{equation}
Supposing the correlation function $\langle P_{n};P_{m}\rangle$ to be finite and using Eq.(\ref{11.38}) we find in 
the limit $\varepsilon \rightarrow 0$
\begin{equation}\label{11.40}
\sum_{i} \sum_{n} a_{i} F_{n}   \int^{\beta}_{0} d\lambda   Tr (\rho   P_{n}(-i\lambda)P_{i}  ) 
 =  \frac{eE}{m}   \langle  \vec{P_{e}}; \vec{P}_{e}\rangle.  
\end{equation}
From the equality (\ref{11.40}) it follows that
\begin{equation}\label{11.41}
\sum_{n} F_{n}   \int^{\beta}_{0} d\lambda   Tr (\rho   P_{n}(-i\lambda)P_{e}  )  
 =  \frac{eE}{m}   \langle  \vec{P_{e}}; \vec{P}_{e}\rangle.  
\end{equation}
Let us emphasize again that the condition
\begin{equation}\label{11.42}
  \lim_{\varepsilon \rightarrow 0} \varepsilon \langle P_{n};P_{m}\rangle = 0
\end{equation}
is an additional one for a suitable choice of the operators $P_{m}$. It is, in the essence, a certain  
 boundary condition for the kinetic equations (\ref{11.37}). Since the Kubo expression for the electrical conductivity
\begin{equation}\label{11.43}
  \sigma \sim \langle  \vec{P_{e}}; \vec{P}_{e}\rangle \sim \langle P_{n};P_{m}\rangle  
\end{equation}
should be a finite quantity, the condition (\ref{11.42}) seems quite reasonable. To make a following step, we must
take into account Eq.(\ref{11.10}) and  (\ref{11.11}). It is easy to see that the r.h.s. of the Eq.(\ref{11.41}) is proportional
to the current density. Finally, we reproduce the Kubo expression (\ref{10.25}) (for cubic systems)
\begin{equation}\label{11.44}
  \sigma = \frac{j}{E} =  \frac{e^{2}}{3m^{2}\Omega} \langle  \vec{P_{e}}; \vec{P}_{e}\rangle,  
\end{equation}
where the proportionality of the $F_{m}$ to the external electrical field has been taken into account.\\
It will be instructive to consider  a concrete problem to clarify some points discussed above.
There are some cases when the calculation of  electrical resistivity is more convenient to be performed within
the approach of the generalized transport equations (\ref{10.34}) and  (\ref{10.35}) than within the Kubo formalism for the
conductivity. We can use these two approaches as two complementary calculating scheme, depending on its
convenience to treat the problem considered~\cite{ck1}. To clarify this, let us start from the condition
\begin{equation}\label{11.45}
\textrm{Tr} (\rho_{LR} B_{i}) = \textrm{Tr} (\rho_{s} B_{i}),
\end{equation}
where the density matrices $\rho_{LR}$ and $\rho_{s}$ were given by equations (\ref{10.30}) and  (\ref{10.32}). For operators $B_{i}$ which can be
represented by linear combinations of the relevant observables $A_{i}$ the equation (\ref{11.45}) is fulfilled exactly
where for other operators the equations (\ref{11.45}) seems to be plausible if the relevant observables have been chosen
properly. The conditions (\ref{11.45}) make it possible to determine a set of parameters which can be used in approximate
expressions for the correlation functions. In simple cases the conditions (\ref{11.45}) even allow to calculate the
correlation functions in (\ref{10.34}) without resorting to another technique. As an example we consider the one-band Hubbard
model (\ref{4.45})
\begin{equation}\label{11.46}
H = H_{e} + H_{ee}
\end{equation}
in the strongly-correlated limit~\cite{kuz09,rnc} $|t|/U \ll 1.$ It well known that in this limit the band splits into two
sub-bands separated by the correlation energy $U.$ In order to take into account the band-split we have to project the
one-electron operators onto the sub-bands. The relevant operators will have the form
\begin{equation}\label{11.47}
\vec{P}^{\alpha \beta} = m \sum_{k\sigma} \frac{\partial E}{\partial \vec{k}} n_{k\sigma}^{\alpha \beta},  
\end{equation}
where
\begin{equation}\label{11.48}
n_{k\sigma}^{\alpha \beta} = N^{1} \sum_{ij}e^{i\vec{k}(\vec{R}_{i} - \vec{R}_{j}) } 
a^{\dag}_{i\sigma} n_{i-\sigma}^{\alpha }a_{j\sigma}  n_{j-\sigma}^{\beta }
\end{equation}
with the projection operators
\begin{eqnarray}\label{11.49}
n_{i-\sigma}^{\alpha } = 
\begin{cases}
n_{i-\sigma} & {\rm if}  \quad  \alpha = + ,\cr
 (1 - n_{i-\sigma}) &  {\rm if}  \quad  \alpha = +~.
\end{cases}
\end{eqnarray}
Here $\vec{P}^{\alpha \alpha}$ is the operator of the total momentum of the electrons in the sub-band $\alpha$, and
$\vec{P}^{\alpha \beta}, \quad  (\alpha \neq \beta)$ describes kinematical transitions between the sub-bands. It can
be shown that correlation functions $\langle \vec{P}^{\alpha \beta} ; \vec{P}^{\gamma \delta}    \rangle$ and
$\langle \vec{P}^{\alpha \alpha} ; \vec{P}^{\beta \beta}    \rangle$  vanish for $\alpha \neq \beta$ in the limit
$|t|/U \ll 1.$ To make an estimation it is necessary to decouple the higher correlation functions in $|t|/U $ and
take into account nearest neighbor hopping terms only.  Then the correlation functions 
$\langle \vec{P}^{\alpha \alpha} ; \vec{P}^{\alpha \alpha}    \rangle$ $(\alpha = \pm)$ can be calculated directly by
means of Eq.(\ref{11.45}), where $B_{i} \rightarrow \vec{P}^{\alpha \alpha}$.  The conductivity becomes
\begin{equation}\label{11.50}
\sigma = W T^{-1} \sum_{\sigma} \sum_{\alpha} \langle n_{\sigma}^{\alpha } \rangle^{-1/2} 
\langle n_{\sigma}^{\alpha } n_{-\sigma}^{\alpha } \rangle (\langle n_{\sigma}^{\alpha } \rangle 
- \langle n_{\sigma}^{\alpha } n_{-\sigma}^{\alpha } \rangle ),
\end{equation}
where
\begin{equation}\label{11.51}
W =    \frac{e^{2}}{3m^{2}\Omega k}\frac{ 1}{\sqrt{2z} } \frac{1}{|t|} 
\sum_{\vec{k}} \left ( \frac{\partial E}{\partial \vec{k}} \right )^{2}
\end{equation}
and 
$$ \langle W \rangle = \textrm{Tr} (\rho W), \quad \rho = \frac{1}{Q} \exp ( - \beta H).$$
Here $z$ is the number of nearest neighbours and $t$ the nearest neighbour hopping matrix element. With the well-known
expressions for the mean values $\langle n_{\sigma}^{\alpha } n_{-\sigma}^{\alpha } \rangle $ we find the conductivity
in dependence on the electron number in the form~\cite{ck1}
\begin{equation}\label{11.52}
\sigma = W T^{-1} \frac{1}{2} \frac{n(1 - n)}{\sqrt{1 - n/2}}, \quad ( 0 \leq n < 1),
\end{equation}
\begin{equation}\label{11.53}
\sigma = W T^{-1}\frac{1}{\sqrt{2}} \exp (- U/2kT), \quad (  n = 1),
\end{equation}
\begin{equation}\label{11.54}
\sigma = W T^{-1} \sqrt{2} \frac{(1 - n/2)(1 - n)}{\sqrt{n}}, \quad ( 2 \geq n > 1).
\end{equation}
Thus, it was shown in this and in the preceding sections that the formalism of the generalized kinetic
equations has certain convenient features and its own specific in comparison with Kubo formalism. The derived expressions are compact
and easy to handle with.
%
\subsection{High-temperature resistivity and  MTBA}
%
At high temperatures the temperature dependence of the electrical resistivity $R$ of some transition metals and
 highly resistive metallic systems such as A15 compounds may deviate substantially from the linear dependence, which
follows from the Bloch-Gruneisen law. These strong deviations from the expected behavior with a tendency to flatten to
a constant resistivity value was termed by {\em resistivity  saturation}~\cite{fw76} and have been studied both experimentally and
theoretically by many authors~\cite{wood,rwc,fl73,pba78,pba79,pba80,pba81,rapp,sl82,cs84,pba00,han,cg02}. The phenomenon 
of  {\em resistivity saturation}  describes a less-than-linear
rise in \emph{dc} electrical resistivity $R$ when temperature $T$ increases. It was found that this effect is common in
transition metal compounds (with pronounced $d$-band structure) when $R$ exceeds $ \sim 80 \, \mu\Omega cm$, and that 
$R$ seems bounded above by a value $R_{max} \sim 150 \, \mu \Omega cm$ which varies somewhat with material. 
In Ref.~\cite{han}, in particular,
it was formulated that the electrical resistivity, $R,$  of a metal is usually interpreted in 
terms of the mean free path (the average distance, $l$, an electron travels before 
it is scattered). As the temperature is raised, the resistivity increases and 
the apparent mean free path is correspondingly reduced. In this semi-classical 
picture, the mean free path cannot be much shorter than the distance, $a$, between 
two atoms. This has been confirmed for many systems and was considered to be a 
universal behavior. Recently, some apparent exceptions were found, 
including alkali-doped fullerenes and high-temperature 
superconductors~\cite{ohun}. However, there remains the possibility that these systems are 
in exotic states, with only a small fraction of the conduction electrons 
contributing to the conductivity; the mean free path would then have to be 
correspondingly larger to explain the observed resistivity. The authors of Ref.~\cite{pba00}  performed  a 
model calculation of electron conduction in alkali-doped fullerenes, in which 
the electrons are scattered by intramolecular vibrations. The resistivity at 
large temperatures implies $l \sim a$, demonstrating that there is no fundamental 
principle requiring $l > a$. At high temperatures, the semi-classical picture 
breaks down, and the electrons cannot be described as quasiparticles. Recent review
of theoretical and experimental investigations in this field was given in Refs.~\cite{ohun,pba,gunn}  
(for discussion of the electronic thermal conductivity at high temperatures see Ref.~\cite{vaf06}).\\
The nature of saturation phenomenon of electrical resistivity is not fully understood. Resistivity of a metallic
system as a function of temperature reflects an overall electron-phonon interaction effects as well as certain
contribution effects of disorder~\cite{gu1,gu2,gu3}. There have been some attempts to explain the saturation phenomenon
in the framework of of the Boltzmann transport theory using special assumptions concerning the band structure, \emph{etc}.
The influence of electron-phonon scattering on electrical resistivity at high temperatures was investigated 
in Refs.~\cite{sl82,cs84}   in the framework of the Fr\"{o}hlich Hamiltonian for the electron-phonon interaction. 
In Ref.~\cite{sl82}  authors calculated a temperature-dependent self-energy to the lowest non-vanishing order of the 
electron-phonon interaction.\\
However, as it was shown above, for transition metals and their disordered alloys the modified tight-binding approximation
is more adequate. Moreover the anisotropic effects are described better within MTBA. Here we consider a single-band
model of transition metal with the Hamiltonian
\begin{equation}\label{11.55}
H = H_{e} + H_{i} + H_{ei}.
\end{equation} 
The electron subsystem is described by the Hubbard model (\ref{4.45}) in the Hartree-Fock approximation 
\begin{equation}\label{11.56}
H_{e} =  \sum_{k\sigma}E(k\sigma)a^{\dagger}_{k\sigma}a_{k\sigma}, 
\quad E(k\sigma) = E(k) + \frac{U}{N}\sum_{p} \langle n_{k-\sigma} \rangle.
\end{equation} 
For the tight-binding electrons in crystals we 
use $E(k) = 2 \sum_{\alpha}t^{0}(\vec R_{\kappa}) \cos (\vec {k}\vec R_{\kappa})$, where $t^{0}(\vec R_{\kappa})$ is the
hopping integral between nearest neighbours, and $\vec R_{\kappa} ( \kappa = x, y. z)$ denotes the lattice vectors
in a simple lattice in an inversion centre. 
For the electron-phonon interaction we use the  Hamiltonian  (\ref{4.98})
\begin{equation}\label{11.57}
H_{ei} = \sum_{\sigma}\sum_{kq}
V(\vec k, \vec k + \vec q)Q_{\vec q}a^{+}_{k+q\sigma}a_{k\sigma}, \quad Q_{\vec q} = 
\frac{1}{ \sqrt{2 \omega(q)}} (b_{q} + b^{\dagger}_{-q}), 
\end{equation}
where
\begin{equation}\label{11.58}
V(\vec k, \vec k + \vec q) =
\frac{iq_{0}}{( N M )^{1/2}}\sum_{\kappa \nu}
t^{0}(\vec R_{\kappa})   \frac{\vec R_{\kappa} \vec{e}_{\nu}(\vec q)}{|\vec R_{\kappa}|}[\sin \vec R_{\kappa} \vec k
- \sin \vec R_{\kappa} (\vec k + \vec q)].
\end{equation}
The  one-electron hopping $t^{0}(\vec R_{\kappa})$ is the overlap integral between a given site $\vec R_{m}$ and one
of the two nearby sites lying on the lattice axis $\vec R_{\kappa}$.
Operators $b^{\dagger}_{q}$ and $  b_{q}$ are creation and annihilation phonon operators and
$\omega(q)$ is the acoustical phonon frequency. 
$N$ is the
number of unit cells in the crystal and $M$ is the ion mass. The $\vec
e_{\nu}(\vec q)$ are the polarization vectors of the phonon modes.\\
For the ion subsystem we have
\begin{equation}\label{11.59}
H_{i}  = \sum_{q} \omega(\vec q) (b^{\dagger}_{q}b_{q}  + 1/2).
\end{equation}
For the resistivity calculation we use the following formula~\cite{sl82}
\begin{equation}\label{11.60}
R =  \frac{\Omega}{3e^{2} \mathcal{N}^{2}}\frac{\langle \vec{F}; \vec{F} \rangle }{1 + (1/3m \mathcal{N}) \langle \vec{P}; \vec{F} \rangle }. 
\end{equation}
Here $\mathcal{N}$ is the effective number of electrons in the band considered
\begin{equation}\label{11.61}
\mathcal{N} = \frac{1}{3m} \int^{\beta}_{0} d \lambda Tr (\rho \vec{P}(-i \hbar \lambda)  \vec{P})
\end{equation}
and $\vec{P}$ is the total momentum operator
\begin{equation}\label{11.62}
\vec{P} = \frac{m}{\hbar} \sum_{\vec{k}} \left ( \frac{\partial E(\vec{k} \sigma)}{\partial \vec{k}} \right ) n_{k\sigma}, \quad
n_{k\sigma} = a^{\dag}_{k\sigma}a_{k\sigma}.
\end{equation}
The total force $\vec{F}$ acting on the electrons is giving by
\begin{equation}\label{11.63}
\vec{F} =  \frac{i}{\hbar} [ H, \vec{P}] = - \frac{i m}{\hbar} \sum_{\vec{k}\vec{q}\sigma} V(\vec{k}, \vec{k} + \vec{q})
( \vec{v}_{k+q,\sigma} - \vec{v}_{k,\sigma})Q_{q}a^{\dag}_{k+q\sigma}a_{k\sigma},
\end{equation}
with the velocity defined as $\vec{v}_{k,\sigma} = \partial E(\vec{k} \sigma)/\hbar \partial \vec{k}$. It is convenient to 
introduce a notation 
$$\frac{V(\vec{k}, \vec{k} + \vec{q})}{\sqrt{2 \omega(q)}} = \frac{i \Lambda F_{q}}{\sqrt{\Omega}}.$$
Correlation functions in Eq.(\ref{11.60}) can be expressed in terms of the double-time thermodynamic Green functions
\begin{equation}\label{11.64}
\langle \vec{F}; \vec{F} \rangle  = 
\frac{2 \pi i}{\hbar} \langle \langle \vec{F}|\vec{P} \rangle \rangle_{-i \hbar \varepsilon},  
\end{equation}
\begin{equation}\label{11.65}
\langle \vec{P}; \vec{F} \rangle  = 
\frac{2 \pi m i}{\hbar} \langle \langle \vec{P}| \vec{F} \rangle \rangle_{-i \hbar \varepsilon},
\end{equation}
\begin{eqnarray}\label{11.66}  
\langle \langle \vec{F}| A \rangle \rangle_{-i \hbar \varepsilon} = \nonumber \\
\frac{1}{2 \pi} \int^{\infty}_{-\infty} dt e^{\varepsilon t} \theta (-t) \textrm{Tr} \left (\rho [A, F(t)]    \right),
\end{eqnarray}
where $A$ represents either the momentum operator $\vec{P}$ or the position operator $\vec{R}$ 
with $\vec{P} = im [H,\vec{R}]/\hbar.$ \\
We find the following relation
\begin{eqnarray}\label{11.67}
\langle \vec{F}; \vec{F} \rangle  = \nonumber \\
\frac{2 \pi m}{\hbar^{2}} \sum_{\vec{k}\vec{q}\sigma}   
\frac{V(\vec{k}, \vec{k} + \vec{q})}{\sqrt{2 \omega(q)}} ( \vec{v}_{k+q,\sigma} - \vec{v}_{k,\sigma})
\langle \langle a^{+}_{k+q\sigma}a_{k\sigma}(b_{q} + b^{\dagger}_{-q}) | \vec{P} \rangle \rangle_{-i \hbar \varepsilon}~.  
\end{eqnarray}
Thus we obtain
\begin{eqnarray}\label{11.68}
\langle \langle \vec{F}; \vec{A} \rangle \rangle_{-i \hbar \varepsilon}  =  \nonumber \\
- i\sum_{\vec{k}\vec{q}\sigma}
\frac{V(\vec{k}, \vec{k} + \vec{q})}{\sqrt{2 \omega(q)}} ( \vec{v}_{k+q,\sigma} - \vec{v}_{k,\sigma})  
\Bigl ( \langle \langle a^{+}_{k+q\sigma}a_{k\sigma}b_{q}| \vec{A} \rangle \rangle_{-i \hbar \varepsilon} -  \\ \nonumber
\langle \langle a^{+}_{k+q\sigma}a_{k\sigma}b^{\dagger}_{-q}) | \vec{A} \rangle \rangle_{-i \hbar \varepsilon}  
\Bigr )~.  \nonumber
\end{eqnarray}
Calculation of the higher-order Green functions gives
\begin{eqnarray}\label{11.69}
( E(k+q \sigma) - E(k\sigma) - \Omega_{q } - i \hbar \varepsilon) 
\langle \langle a^{\dag}_{k+q\sigma}a_{k\sigma}B_{q}| \vec{A} \rangle \rangle_{-i \hbar \varepsilon} = \nonumber \\
T_{kq}(A) + \sum_{\vec{q'}} \Bigl ( \frac{V(\vec{k} - \vec{q'}, \vec{k} )}{\sqrt{2 \omega(q')}}
\langle \langle a^{\dag}_{k+q\sigma}a_{k-q'\sigma}(b_{q'} - b^{\dagger}_{-q'}) B_{q}| \vec{A} \rangle \rangle_{-i \hbar \varepsilon} - \nonumber \\
\frac{V(\vec{k} + \vec{q}, \vec{k} + \vec{q} + \vec{q'})}{\sqrt{2 \omega(q')}}
\langle \langle a^{\dag}_{k+q+q'\sigma}a_{k\sigma}(b_{q'} - b^{\dagger}_{-q'}) B_{q}| \vec{A} \rangle \rangle_{-i \hbar \varepsilon}
\Bigr ) \nonumber \\
- \sum_{\vec{k'}}\frac{V(\vec{k'}, \vec{k'} - \vec{q} )}{\sqrt{2 \omega(q)}}
\langle \langle a^{\dag}_{k+q\sigma}a_{k\sigma}a^{\dag}_{k'-q\sigma}a_{k'\sigma}| \vec{A} \rangle \rangle_{-i \hbar \varepsilon}
\end{eqnarray}
with notation
$$ \Omega_{q } = \hbar \omega(q) \rightarrow B_{q} = b_{q}; \quad E_{q} = - \hbar \omega(q) \rightarrow B_{q} = b^{\dag}_{-q},$$
$$T_{kq}(\vec{P}) = \frac{im}{2\pi} ( \vec{v}_{k+q,\sigma} - \vec{v}_{k,\sigma}) \langle a^{\dag}_{k+q\sigma}a_{k\sigma}
B_{q} \rangle, $$
$$T_{kq}(\vec{R}) = - \frac{1}{2\pi} \sum_{\vec{q'}}   \frac{\partial}{\partial q'}  
\langle a^{\dag}_{k+q+q' \sigma}a_{k+q'\sigma}B_{q} \rangle \delta_{q',0},$$
$$\mathcal{N} = \frac{m}{3 \hbar} \sum_{kq'\sigma}\delta_{q',0} \vec{v}_{k,\sigma}  \frac{\partial}{\partial q'} 
\langle a^{\dag}_{k+q'\sigma}a_{k+q'\sigma} \rangle.$$
We find also
\begin{eqnarray}\label{11.70}
\langle \langle a^{\dag}_{k+q\sigma}a_{k-q'\sigma}b_{q}b_{q'}| A \rangle \rangle_{-i \hbar \varepsilon} =  \qquad\\ 
- \langle b^{\dag}_{q'}b_{q'} \rangle_{0} \frac{V(\vec{k}, \vec{k} - \vec{q'} )}{\sqrt{2 \omega(q')}}
\frac{\langle \langle a^{\dag}_{k+q\sigma}a_{k\sigma}b_{q}| A \rangle \rangle_{-i \hbar \varepsilon}}
{( E(k+q \sigma) - E(k-q'\sigma) - E_{q' } - \hbar\omega(q') -i \hbar \varepsilon)},\nonumber
\end{eqnarray}
\begin{eqnarray}\label{11.71}
\langle \langle a^{\dag}_{k+q+q'\sigma}a_{k\sigma}b_{q}b_{q'}| A \rangle \rangle_{-i \hbar \varepsilon} =  \qquad\\
- \langle b^{\dag}_{q'}b_{q'} \rangle_{0} \frac{V(\vec{k}+\vec{q}+\vec{q'}, \vec{k} + \vec{q} )}{\sqrt{2 \omega(q')}}
\frac{\langle \langle a^{\dag}_{k+q\sigma}a_{k\sigma}b_{q}| A \rangle \rangle_{-i \hbar \varepsilon}}
{( E(k+q+q' \sigma) - E(k\sigma) - E_{q' } - \hbar\omega(q') -i \hbar \varepsilon)}.\nonumber
\end{eqnarray}
Here the symmetry relations
$$V(\vec{k} - \vec{q'}, \vec{k} ) = V^{*}(\vec{k}, \vec{k} - \vec{q'}); 
\quad V(\vec{k}+ \vec{q}+\vec{q'},  \vec{k}+ \vec{q} ) = V^{*}(\vec{k}+ \vec{q},   \vec{k}+ \vec{q}+\vec{q'})$$
were taken into account.\\
Now the Green function of interest can be determined by introducing the self-energy~\cite{sl82}
\begin{eqnarray}\label{11.72}
\langle \langle a^{\dag}_{k+q\sigma}a_{k-q'\sigma}b_{q}B_{q}| A \rangle \rangle_{-i \hbar \varepsilon} = \nonumber \\
- \langle b^{\dag}_{q'}b_{q'} \rangle_{0} \frac{T_{kq}(A )}
{( E(k+q \sigma) - E(k-q'\sigma) - E_{q} - M_{kq\sigma}(E_{q},-i \hbar \varepsilon ) -i \hbar \varepsilon)}~.
\end{eqnarray}
The self-energy is given by
\begin{eqnarray}\label{11.73}
M_{kq\sigma}(E_{q},-i \hbar \varepsilon ) = \nonumber \\
\sum_{q'} \frac{1}{2\omega(q')} \Bigl ( \langle b^{\dag}_{q'}b_{q'} \rangle [
\frac{|V(\vec{k}, \vec{k} - \vec{q'})|^{2}}{ E(k+q \sigma) - E(k-q'\sigma) - E_{q'} - \hbar\omega(q') -i \hbar \varepsilon} + \nonumber \\
\frac{|V(\vec{k}+ \vec{q},   \vec{k}+ \vec{q}+\vec{q'})|^{2}}{  E(k+q+q' \sigma) - E(k\sigma) - E_{q'} - \hbar\omega(q') -i \hbar \varepsilon} ] \nonumber \\
+ \langle b_{q'} b^{\dag}_{q'}\rangle [
\frac{|V(\vec{k}, \vec{k} - \vec{q'})|^{2}}{ E(k+q \sigma) - E(k-q'\sigma) - E_{q' } + \hbar\omega(q') -i \hbar \varepsilon} + \nonumber \\
\frac{|V(\vec{k}+ \vec{q},   \vec{k}+ \vec{q}+\vec{q'})|^{2}}{  E(k+q+q' \sigma) - E(k\sigma) - E_{q'} + \hbar\omega(q') -i \hbar \varepsilon} ] \Bigr )~.
\end{eqnarray}
In equation (\ref{11.73}) the energy difference $(E(k+q \sigma) - E(k-q\sigma) - E_{q }))$ is that for the scattering process of electrons on
phonons, while emission or absorption of one phonon is possible, corresponding to $E_{q}$. These scattering processes are contained in the usual
Boltzmann transport theory leading to the Bloch-Gruneisen law. The self-energy $M_{kq\sigma}$ describes multiple scattering corrections to the
Bloch-Gruneisen behavior to second order in $V$, which depends on the temperature via the phonon occupation numbers.\\ Furthermore, it is assumed that the averages
of occupation numbers for phonons in the self-energy and for electrons in the effective particle number, are replaced by the Bose and Fermi distribution
functions, respectively
\begin{eqnarray}\label{11.74}
\langle b^{\dag}_{q}b_{q} \rangle = N_{q}, \quad N_{q} = [\exp(\beta\hbar\omega(q)) - 1]^{-1}, \\
\langle a^{\dag}_{k\sigma}a_{k\sigma} \rangle = f_{k}, \quad f_{k} = [\exp(\beta E(k\sigma) - E_{F}) + 1 ]^{-1}.
\label{11.75}
\end{eqnarray}
This corresponds to neglecting the influence of multiple scattering corrections on the phonon and electron distribution functions.\\ In order to calculate the
expectation values in the inhomogeneities, Eqs. (\ref{11.72})   and (\ref{11.73}), the spectral theorem~\cite{dnz60,tyab} should be used. In the lowest non-vanishing order
of the electron-phonon interaction parameter $V$ we obtain
\begin{eqnarray}\label{11.76}
\langle a^{\dag}_{k+q\sigma}a_{k\sigma}B_{q} \rangle = \nonumber \\ 
\frac{V(\vec{k}+ \vec{q},   \vec{k})}{\sqrt{2 \omega(q)}}f_{k+q}(1 - f_{k})\nu_{q}(E_{q})  
\frac{[\exp(\beta( E(k+q\sigma) - E(k\sigma) - E_{q})] - 1 }{E(k+q\sigma) - E(k\sigma) - E_{q}}, 
\end{eqnarray}
with
\begin{eqnarray}\label{11.77}
\nu_{q}(E_{q}) = \frac{1}{1 - \exp(- \beta E_{q})} = 
\begin{cases} 
{1 + N_{q}} & {\rm if} \quad  B_{q} = b_{q}; \quad E_{q}  = \hbar \omega(q)   \cr
N_{q} & {\rm if} \quad  B_{q} = b^{\dag}_{-q}; \quad  E_{q} = - \hbar \omega(q). 
\end{cases}
\end{eqnarray}
Applying the approximation scheme discussed above we have found the following expressions for the Green function with $ A = P$
\begin{eqnarray}\label{11.78}
\langle \langle a^{\dag}_{k+q\sigma}a_{k\sigma}B_{q}| \vec{P} \rangle \rangle_{-i \hbar \varepsilon} = \nonumber \\
\frac{i m}{2\pi}\frac{V(\vec{k}+ \vec{q},   \vec{k})}{\sqrt{2 \omega(q)}}( \vec{v}_{k+q,\sigma} - \vec{v}_{k,\sigma}) 
\frac{P^{1}_{kq\sigma}(E_{q})}{\Omega_{kq\sigma}(E_{q}) - M_{kq\sigma} - i \hbar \varepsilon},
\end{eqnarray}
and for the Green function with $ A = R$
\begin{eqnarray}\label{11.79}
\langle \langle a^{\dag}_{k+q\sigma}a_{k\sigma}B_{q}| \vec{R} \rangle \rangle_{-i \hbar \varepsilon} = \nonumber \\
\frac{\hbar}{2\pi}\frac{V(\vec{k}+ \vec{q},   \vec{k})}{\sqrt{2 \omega(q)}}( \vec{v}_{k+q,\sigma} - \vec{v}_{k,\sigma}) 
\frac{P^{2}_{kq\sigma}(E_{q}) - P^{3}_{kq\sigma}(E_{q})}{\Omega_{kq\sigma}(E_{q}) - M_{kq\sigma} - i \hbar \varepsilon}~.
\end{eqnarray}
We have introduced in the above equations the following notation:
\begin{equation}\label{11.80}
P^{(1)}_{kq\sigma}(E_{q}) = f_{k+q}(1 - f_{k}) \nu_{q}(E_{q})  \gamma_{1}(\Omega_{kq\sigma} (E_{q})),
\end{equation}
\begin{equation}\label{11.81}
P^{(2)}_{kq\sigma}(E_{q}) = (v_{k+q\sigma} - v_{k\sigma} )f_{k+q}(1 - f_{k}) \nu_{q}(E_{q}) \Bigl ( \gamma_{2}(\Omega_{kq\sigma} (E_{q})) - 
\frac{\beta \exp (\beta \Omega_{kq\sigma} (E_{q}))}{\Omega_{kq\sigma} (E_{q})} \Bigr ),
\end{equation} 
\begin{equation}\label{11.82}
P^{(3)}_{kq\sigma}(E_{q}) = f_{k+q}(1 - f_{k}) \nu_{q}(E_{q}) \beta \gamma_{1}(\Omega_{kq\sigma} (E_{q}))[f_{k}v_{k\sigma} - (1 - f_{k+q})v_{k+q\sigma}],
\end{equation}
with
\begin{equation}\label{11.83}
\gamma_{n}(\Omega_{kq\sigma} (E_{q})) = \frac{\beta \exp (\beta \Omega_{kq\sigma} (E_{q}))- 1}{(\Omega_{kq\sigma} (E_{q}))^{n}},
\end{equation}
and
\begin{equation}\label{11.84}
\Omega_{kq \sigma} (E_{q})  = E(k+q\sigma) - E(k\sigma) - E_{q}.
\end{equation}
For the effective particle number we find
\begin{equation}\label{11.85}
\mathcal{N} = \frac{2}{3} m \beta \sum_{k} (v_{k})^{2}f_{k}(1 - f_{k}).
\end{equation}
Before starting of calculation of the resistivity it is instructive to split the self-energy into real and imaginary part ($\varepsilon \rightarrow 0$)
\begin{equation}\label{11.86}
\lim_{\varepsilon \rightarrow 0} M_{kq\sigma}(\hbar\omega(q)  \pm  i \hbar \varepsilon ) = {\rm Re} M_{kq\sigma}(\hbar\omega(q)) \mp 
i {\rm Im} M_{kq\sigma}(\hbar\omega(q))
\end{equation}
and perform an interchange of variables $ k+q \rightarrow k; q \rightarrow -q.$ Now for the relevant correlation functions we obtain the expressions
\begin{equation}\label{11.87}
\langle \vec{F}; \vec{F} \rangle  = \frac{2 m^{2}}{\hbar}\sum_{kq\sigma}
 \frac{|V( \vec{k},\vec{k}+ \vec{q})|^{2}}{\sqrt{2 \omega(q)}} (v_{k+q\sigma} - v_{k\sigma} )^{2}P^{(1)}_{kq\sigma}(\hbar\omega(q)) S_{kq\sigma}(\hbar\omega(q)),
\end{equation}
\begin{eqnarray}\label{11.88}
\langle \vec{P}; \vec{F} \rangle  =  \qquad \\  \nonumber - m^{2}\sum_{kq\sigma}
 \frac{|V( \vec{k},\vec{k}+ \vec{q})|^{2}}{\sqrt{2 \omega(q)}} (v_{k+q\sigma} - v_{k\sigma} ) 
 \Bigl ( P^{(2)}_{kq\sigma}(\hbar\omega(q))  + P^{(3)}_{kq\sigma}(\hbar\omega(q)) \Bigr ) S_{kq\sigma}(\hbar\omega(q)),
\end{eqnarray}
where
\begin{equation}\label{11.89}
S_{kq\sigma}(\hbar\omega(q)) = 
\frac{{\rm Im} M_{kq\sigma}(\hbar\omega(q))}{(\Omega_{kq\sigma} (\hbar\omega(q)) - 
{\rm Re} M_{kq\sigma}(\hbar\omega(q)))^{2} + ({\rm Im} M_{kq\sigma}(\hbar\omega(q)))^{2}}~.
\end{equation}
In order to obtain Eq.(\ref{11.89}) the following symmetry relation for the self-energy was used
\begin{equation}\label{11.90}
M_{kq\sigma}(\hbar\omega(q) - i\hbar \varepsilon) = - M_{kq\sigma}(\hbar\omega(q) + i\hbar \varepsilon).
\end{equation}
The inspection of both the correlation functions $\langle \vec{F}; \vec{F} \rangle$ and $\langle \vec{P}; \vec{F} \rangle $ shows that
its include two dominant parts. The first one is the scattering part $S_{kq\sigma}(\hbar\omega(q)),$ which contains all the 
information about the scattering processes. The second part describes the occupation possibilities before and after the scattering
processes $(P^{(1)}_{kq\sigma}(\hbar\omega(q)), P^{(2)}_{kq\sigma}(\hbar\omega(q)), P^{(3)}_{kq\sigma}(\hbar\omega(q))  ),$ and
includes both the Fermi and Bose distribution functions. The approximation procedure described above neglects the multiple scattering 
corrections in these factors for the occupation possibilities.\\
For further estimation of the correlation functions the quasi-elastic approximation can be used. In this case, in the energy difference
$\Omega_{kq\sigma} (\hbar\omega(q)),$ the phonon energy $\hbar\omega(q)$ can be 
neglected  $\Omega_{kq\sigma} (\hbar\omega(q)) \simeq \Omega_{kq\sigma} (0)$. The phonon wave number $\vec{q} $ only is taken
into account via the electron dispersion relation. Furthermore, for the Bose distribution function it was assumed that
\begin{equation}\label{11.91}
\langle b^{\dag}_{q} b_{q}\rangle  = \langle b_{q} b^{\dag}_{q} \rangle  \simeq  (\beta \hbar\omega(q))^{-1}.
\end{equation}
This approximation is reasonable at temperatures which are high in comparison to the Debye temperature $\Theta_{D}$.\\
It is well known~\cite{zi,cjac10}  from Bloch-Gruneisen theory that the quasi-elastic approximation does not disturb the temperature dependence of the electrical
resistivity at high or low temperatures. The absolute value of the resistivity is changed, but the qualitative  picture of the power law of the
temperature dependence of the resistivity is not influenced.\\ In the framework of the quasi-elastic approximation the scattering contribution 
can be represented in the form
\begin{equation}\label{11.92}
S_{kq\sigma} = 
\frac{{\rm Im} M^{e}_{kq\sigma}}{(\Omega_{kq\sigma} (0) - 
{\rm Re} M^{e}_{kq\sigma})^{2} + ({\rm Im} M^{e}_{kq\sigma}))^{2}}~.
\end{equation}
Here the real and imaginary parts of the self-energy have the following form
\begin{eqnarray}\label{11.93}
{\rm Re} M^{e}_{kq\sigma} = \frac{k T}{\hbar} \sum_{p} \Bigl ( \frac{|V( \vec{k},\vec{k}+ \vec{p})|^{2}}{ \omega(p)^{2}} P
(\frac{1}{E(k+q \sigma) - E(k-p \sigma)} ) +  \nonumber \\ \frac{|V( \vec{k} + \vec{q},\vec{k}+ \vec{q} + \vec{p})|^{2}}{ \omega(p)^{2}} P
(\frac{1}{E(k+q+p\sigma) - E(k\sigma)} ) \Bigr ),
\end{eqnarray}
\begin{eqnarray}\label{11.94}
{\rm Im} M^{e}_{kq\sigma} = \frac{\pi k T}{\hbar} \sum_{p} \Bigl ( \frac{|V( \vec{k},\vec{k}+ \vec{p})|^{2}}{ \omega(p)^{2}}  
\delta (E(k+q \sigma) - E(k-p \sigma) ) +  \nonumber \\ \frac{|V( \vec{k} + \vec{q},\vec{k}+ \vec{q} + \vec{p})|^{2}}{ \omega(p)^{2}}  
\delta (E(k+q+p\sigma) - E(k\sigma)  ) \Bigr ).
\end{eqnarray}
The occupation possibilities given by $P^{(n)}_{kq\sigma}$ can be represented in the quasi-elastic approximation as~\cite{sl82}
\begin{eqnarray}\label{11.95}
P^{(1)e}_{kq\sigma}  \simeq  \frac{k T}{\hbar \omega(q)} \delta (E_{F} - E(k\sigma) ); \\
P^{(2)e}_{kq\sigma} = P^{(3)e}_{kq\sigma} = 0.
\label{11.96}
\end{eqnarray}
In this approximation the momentum-force correlation function  disappears $\langle \vec{P}; \vec{F} \rangle  \simeq 0$.
Thus we have
\begin{eqnarray}\label{11.97}
R \simeq \frac{\Omega}{3e^{2} N^{2}} \langle \vec{F}; \vec{F} \rangle ,  \\
\langle \vec{F}; \vec{F} \rangle = \frac{2 m^{2} k T}{\hbar} \sum_{k q \sigma}  \frac{|V( \vec{k},\vec{k}+ \vec{q})|^{2}}{ \omega(q)^{2}}
 (v_{k+q\sigma} - v_{k\sigma} )^{2} \delta (E_{F} - E(k\sigma) )\cdot \nonumber \\
\frac{{\rm Im} M^{e}_{kq\sigma}}{(E(k+q \sigma) - E(k\sigma) - 
{\rm Re} M^{e}_{kq\sigma})^{2} + ({\rm Im} M^{e}_{kq\sigma}))^{2}}~. 
\label{11.98}
\end{eqnarray}
The explicit expression for the electrical resistivity was calculated in Ref.~\cite{sl82}. The additional simplifying assumptions have been made
to achieve it. For the electrons and phonons the following simple dispersion relations were taken
$$ E(k) = \frac{\hbar^{2}k^{2}}{2m^{*}}; \quad \omega(q) = v_{0}|\vec{q}|, \quad V( \vec{k},\vec{k}+ \vec{q}) \sim \sqrt{|\vec{q}|}.$$
It was shown that the estimation of    $P^{(1)e}_{kq\sigma}$  is given by
\begin{eqnarray}\label{11.99}
P^{(1)e}_{kq\sigma}  \simeq  \frac{T k  m^{*}}{v_{0} \hbar^{3} k_{F}q} \delta (k_{F} - k )~.
\end{eqnarray}
Then, for the electrical resistivity the following result was found
\begin{eqnarray}\label{11.100}
R  \simeq  \frac{3 V^{2} m^{*}q_{D}}{2e^{2}k^{5}_{F} \hbar}  \frac{T}{\Theta_{D}} \int^{q_{D}}_{0} dq \int^{1}_{-1} dz q^{4}\cdot \nonumber \\
\frac{{\rm Im} M^{e}_{k_{F} q z}}{[(\hbar^{2}q/m^{*} )(z k_{F} + q/2) - 
{\rm Re} M^{e}_{k_{F} q z}]^{2} + [{\rm Im} M^{e}_{k_{F} q z})]^{2}}.
\end{eqnarray}
This result shows that the usual Bloch-Gruneisen theory of the electrical resistivity can be corrected by including the self-energy in the
final expression for the resistivity. The Bloch-Gruneisen theory can be reproduced in the weak scattering limit using the relation
\begin{eqnarray}\label{11.101}
\lim_{{\rm Re}({\rm Im}) M \rightarrow 0} \frac{{\rm Im} M^{e}_{k_{F} q z}}{[(\hbar^{2}q/m^{*} )(z k_{F} + q/2) - 
{\rm Re} M^{e}_{k_{F} q z}]^{2} + [{\rm Im} M^{e}_{k_{F} q z})]^{2}} \nonumber \\ = \frac{\pi m^{*}}{\hbar^{2}q k_{F}} \delta (z +  \frac{q}{2k_{F} } )~.
\end{eqnarray}
Inserting Eq.(\ref{11.101}) in the resistivity expression Eq.(\ref{11.100}) gives the electrical resistivity $R_{w}$ in the weak scattering limit, showing a
linear temperature dependence
\begin{equation}\label{11.102}
R_{w} \simeq \frac{3 \pi}{8} \frac{V^{2}( m^{*})^{2}q^{5}_{D}}{e^{2}k^{6}_{F} \hbar^{3}} \frac{T}{\Theta_{D}}.
\end{equation}
In this form, the resistivity formula contains two main parameters that influence substantially. The first is the Debye temperature $\Theta_{D}$
characterizing the phonon system, and, the second, the parameter $\alpha = (V m^{*} )^{2}$ describing the influence of both the electron system
and the strength of the electron-phonon interaction. The numerical estimations~\cite{sl82} were carried out for $Nb$ and gave the magnitude of the
saturation resistivity 207 $\mu\Omega cm.$ \\ In the theory described above the deviation from linearity in the high-temperature region of
the resistivity may be caused by multiple scattering corrections. The multiple scattering processes which describe the scattering processes of
electrons on the phonon system by emission or absorption of more than one phonon in terms of self-energy corrections become more and more
important with increasing temperature. As was shown above, even for simple dispersion relation of electrons and phonons within one-band model the
thermally induced saturation phenomenon occurs. For the anisotropic model within MTBA the extensive numerical calculations are necessary. \\ In
subsequent paper~\cite{cs84}, Christoph and Schiller have considered the problem of the microscopic foundation of the empirical 
formula~\cite{fw76} (parallel resistor model)
\begin{equation}\label{11.103}
\frac{1}{R(T)} = \frac{1}{R_{SBT}(T)} + \frac{1}{R_{max}(T)}
\end{equation}
within the framework of the transport theory of Christoph and Kuzemsky~\cite{ck1}. The parallel resistor formula describing the saturation 
phenomenon of electrical resistivity in systems with strong electron-phonon interaction was derived. In Eq.(\ref{11.103}) $R_{SBT}(T)$ is the resistivity
given by the semiclassical Boltzmann transport theory $R_{SBT}(T) \sim T$ and the saturation resistivity $R_{max}$ corresponds to the maximum
metallic resistivity~\cite{wies}. The higher-order terms in the electron-phonon interaction were described by a self-energy which 
was determined self-consistently. They found for the saturation resistivity the formula~\cite{cs84}
\begin{equation}\label{11.104}
R_{max} =
\frac{3 \pi^{3}}{32} \frac{\hbar}{e^{2}} \frac{q^{4}_{D}}{k^{5}_{F}} \frac{1}{|P (q_{D}/2k_{F})|}.
\end{equation}
Within the frame of this approach, the saturation behavior of the electrical resistivity was explained by the
influence of multiple scattering processes described by a temperature-dependent damping term of one-electron energies. In the standard picture,
the conventional linear temperature dependence of the resistivity $R \sim T ( T \gg \Theta_{D})$ is explained by taking into account that the number
of phonons is proportional to the temperature and, moreover, assuming that the electron momentum is dissipated in single-phonon scattering processes
only. For an increasing number of phonons, however, the multiple scattering processes become more important and the single scattering event becomes
less effective. This argument coincides in some sense with the Yoffe-Regel criterion~\cite{gu1,gu2,gu3} stating that an increase in the number of scatterers
does not result in a corresponding increase of the resistivity if the the mean free path of the electrons becomes comparable with the lattice
distance. Indeed, the saturation resistivity (\ref{11.104}) coincides roughly with the inverse minimal metallic conductivity which can be derived 
using this criterion.
%
%
\section{Resistivity of Disordered Alloys}
%
In the present section a theory of electroconductivity in disordered transition metal alloys with the proper microscopic treatment of
the non-local electron-phonon interaction is considered. It was established long ago that any deviation from perfect periodicity will lead to 
a resistivity contribution which will depend upon the spatial extent and lifetime of the disturbance measured in relation to the conduction
electron mean free path and relaxation time. It is especially important to develop a theory for the resistivity of concentrated alloys, because
of its practical significance.
The electrical resistivity of disordered  metal alloys and its temperature
coefficient is of a considerable practical and theoretical interest~\cite{dug,wat,viss,moi,rep02,rep03,vysrep05,rep05,ram98,pba00,cote,jans,zuc,foh}. The work in this field has been considerably 
stimulated by Mooij paper~\cite{moi}
where it has been shown that the temperature coefficient of the resistivity of disordered  alloys becomes negative if their residual resistivity
exceeds a given critical value. To explain this phenomenon, one has to go beyond the weak-scattering limit and to take into account the interference
effects between the static disorder scattering and the electron-phonon scattering~\cite{zuc,foh,chen,mark,harr,gjon,gir,bel}.\\ In the 
weak-scattering limit~\cite{nick} the contributions
of impurity and phonon scattering add to the total resistivity without any interference terms (Matthiessen rule). For disordered systems many physical
properties can be related to the configuration-averaged Green functions~\cite{gon}. There were formulated a few methods for calculation of these
averaged Green functions. It vas found that the single-site CPA (coherent potential approximation)~\cite{sov,dwt,vke,veli,kru} provides a convenient and
accurate approximation for it~\cite{levi,abc,fuk,koho,ashe,sher,achen,bre}. The CPA is a self-consistent 
method~\cite{levi,abc,fuk,koho,ashe,sher,achen,bre} that predicts alloy electronic properties, interpolating between those of the pure constituents
over the entire range of concentrations and scattering strengths. The self-consistency condition is introduced by requiring that the coherent
potential, when placed at each lattice site of the ordered lattice, reproduces all the average properties of the actual crystal. The coherent-potential
approximation has been developed within the framework of the multiple-scattering description of disordered systems~\cite{gon}. A given scatterer in
the alloy can be viewed as being embedded in an effective medium with a complex energy-dependent potential whose choice is open and can be 
made self-consistently such that the average forward scattering from the real scatterer is the same as free propagation in the effective medium.
The strong scattering has been first considered 
by Velicky~\cite{veli} in the framework of the single-site CPA  using the Kubo-Greenwood formula. These results have been extended later
to a more general models~\cite{levi,abc,fuk,koho,ashe,sher,achen,bre,kozo,vojt,arpa,papa,gon1,gon2,gon3}. \\ 
The first attempt to include the electron-phonon scattering  in the CPA calculations of the resistivity 
was given by Chen \emph{et al.}~\cite{chen}  A model was introduced in which phonons were treated phenomenologically while electrons were described in CPA. The
electron-phonon interaction was described by a local operator. Chen, Weisz and Sher~\cite{chen} (CWS) have performed a model calculation on the temperature
dependence of the electronic density of states and the electrical conductivity of disordered binary alloys, based on CPA solutions by introducing
thermal disorder in the single-band model. They found that the effect of thermal disorder is to broaden and smear the static-alloy density of states.
The electrical conductivity in weak scattering alloys always decreases with temperature. However, in the strong-scattering case, the temperature
coefficient of conductivity can be negative, zero, or positive, depending on the location of the Fermi energy. Brouers and Brauwers~\cite{bb75}
have extended the calculation to an $s-d$ two-band model that accounts for the general behavior of the temperature dependence of the electrical
resistivity in concentrated transition metal alloys. In Ref.~\cite{sher}  a generalization of CWS theory~\cite{chen} was made by including the effect of uniaxial strain on the
temperature variation of the electronic density of states and the electrical conductivity of disordered concentrated binary alloys. The validity of the
adiabatic approximation in strong-scattering alloys was analyzed by Chen, Sher and Weisz~\cite{achen}. It was shown that  the electron screening
process in the moving lattice may be modified by lattice motion in disordered alloys. Were this modification significant, not only the effective
Hamiltonian but also the whole adiabatic approximation would need to be reconsidered.\\
A consistent theory of the electroconductivity in  disordered transition metal alloys with the proper
microscopic treatment of the electron-phonon interaction was carried out by Christoph and Kuzemsky~\cite{ck3}. They used 
the approach of paper~\cite{wk82}, where  a self-consistent microscopic
theory for the calculation of one-particle Green functions for the electron-phonon problem in disordered transition metal alloys was
developed. However, this approach cannot be simply generalized to the calculation of two-particle Green functions needed for the calculation of the
conductivity by the Kubo formula. Therefore, for the sake of simplicity, in their study Christoph and Kuzemsky~\cite{ck3} neglected the influence of disorder on the phonons.
Thus, in the model investigated here, in contrast to the GWS approach~\cite{chen}, the dynamics of the phonons is taken into account microscopically,
but they are treated as in a virtual reference crystal.\\ For a given configuration of atoms the total Hamiltonian of the electron-ion system in the
substitutionally disordered alloy can be written in the form~\cite{wk82,ck3}
\begin{equation}\label{12.1}
H = H_{e} + H_{i} + H_{ei},
\end{equation} 
where
\begin{equation}\label{12.2}
H_{e} = \sum_{i\sigma}\epsilon_{i} a^{\dag}_{i\sigma} a_{i\sigma}  +  \sum_{ij\sigma} t_{ij}a^{\dag}_{i\sigma} a_{j\sigma}
\end{equation} 
is the one-particle Hamiltonian of the electrons. For our main interest is the description of the electron-phonon interaction, we can
suppose that the electron-electron correlation in the Hubbard form has been taken here in the Hartree-Fock approximation in analogy with
Eq.(\ref{11.56}).\\ For simplicity, in this paper the vibrating ion system will be described by the usual phonon Hamiltonian
\begin{equation}\label{12.3}
H_{i}  = \sum_{q\nu} \omega(\vec q \nu) ( b^{\dagger}_{q\nu} b_{q\nu}  + 1/2 ).
\end{equation}
The electron-phonon interaction term is taken in the following form~\cite{wk82}
\begin{equation}\label{12.4}
H_{ei} = \sum_{ij} \sum_{\alpha \sigma}T^{\alpha}_{ij}( u^{\alpha}_{i} - u^{\alpha}_{j})a^{\dag}_{i\sigma} a_{j\sigma},
\end{equation} 
where $u^{\alpha}_{i}  (\alpha = x,y,z )$ is the ion displacement from the equilibrium position $ \vec{R}_{i}. $\\ In terms of phonon operators
this expression can be rewritten in the form
\begin{equation}\label{12.5}
H_{ei} = \sum_{i \neq j} \sum_{q \nu \sigma}A_{q \nu}( ij )( b_{q\nu} +  b^{\dagger}_{-q\nu} ) a^{\dag}_{i\sigma} a_{j\sigma},
\end{equation} 
where
\begin{equation}\label{12.6}
A_{q \nu}(ij) = \frac{q_{0}}{\sqrt{2 \langle M \rangle N \omega(\vec q \nu)} }t^{0}_{ij}  \frac{\vec R_{j} - \vec R_{i}}{|\vec R_{j} - \vec R_{i}|}  \vec{e}_{\nu}(\vec q)
\Bigl ( e^{i \vec q \vec R_{i}} - e^{i \vec q \vec R_{j}} \Bigr ).
\end{equation} 
Here $\omega(\vec q \nu)$ are the acoustic phonon frequencies, $\langle M \rangle$ is the average ion mass, $\vec{e}_{\nu}(\vec q)$ are the polarization 
vectors of the phonons, and $q_{0}$ is the Slater coefficient originated in the exponential radial decrease of the tight-binding 
electron wave function. It is convenient to rewrite this expression in the form
\begin{equation}\label{12.7}
H_{ei} = \sum_{i \neq j} \sum_{q }A_{q}(ij)( b_{q } +  b^{\dagger}_{-q } ) a^{\dag}_{i} a_{j},
\end{equation} 
where the spin and phonon polarization indices are omitted for brevity.\\ The electrical conductivity will be calculated starting with
the Kubo expression for the \emph{dc} conductivity:
\begin{equation}\label{12.8}
\sigma(i\varepsilon) = - \langle \langle \vec{J} | \vec{P} \rangle \rangle_{i\varepsilon}, \quad ( \varepsilon \rightarrow 0^{+}),
\end{equation} 
where $\vec{P} = e \sum_{i}\vec{R}_{i}a^{\dag}_{i} a_{i} $ and $\vec{R}_{i}$ is the position vector; $ m/e \vec{J} = m/e \dot{\vec{P}}$ is
the current operator of the electrons. It has the form
\begin{equation}\label{12.9}
\vec{J} = - i e  \sum_{ij} (\vec{R}_{i} - \vec{R}_{j}) t_{ij} a^{\dag}_{i} a_{j}.
\end{equation} 
Then the normalized conductivity becomes
\begin{equation}\label{12.10}
\sigma^{\alpha \beta} = \frac{ie^{2}}{\Omega}  \sum_{ij}  \sum_{l} (\vec{R}_{i} - \vec{R}_{j})^{\alpha}\vec{R}_{l}^{\beta} 
t_{ij} \langle \langle  a^{\dag}_{i} a_{j} |  a^{\dag}_{l} a_{l} \rangle \rangle_{i\varepsilon}~,
\end{equation} 
where $\Omega$ is the volume of the system.
It should be emphasized here that $\sigma^{\alpha \beta}$depends on the configuration of the alloy.
A realistic treatment of disordered alloys must involve a formalism to deal with one-electron Hamiltonian  that include both diagonal and
off-diagonal randomness~\cite{levi,abc,fuk,koho,ashe,sher,achen,bre}. In the present study, for the sake of simplicity, we restrict ourselves to
a diagonal disorder. Hence we can rewrite hopping integral $t_{ij}$ as
\begin{equation}\label{12.11}
t_{ij} = \frac{1}{N}  \sum_{k} E(\vec{k}) \exp [i \vec{k} (\vec{R}_{i} - \vec{R}_{j})]~.  
\end{equation} 
Thus to proceed it is necessary to find the Green function $G_{ij,lm} =  \langle \langle  a^{\dag}_{i} a_{j} |  a^{\dag}_{l} a_{m} \rangle \rangle$.
It can be calculated by the equation of motion method. Using the Hamiltonian (\ref{12.1}) we find by a differentiation with respect to the left hand side
\begin{eqnarray}\label{12.12}
\sum_{nr} H_{ij,rn} G_{nr,lm}(\omega) = \langle a^{\dag}_{i} a_{m} \rangle \delta_{lj} - \langle a^{\dag}_{l} a_{j} \rangle \delta_{mi} + \nonumber \\
\sum_{q n} \Bigl ( A_{q }( j - n ) e^{i \vec{q}\vec{R}_{j} } 
\langle \langle  a^{\dag}_{i} a_{n}( b_{q } +  b^{\dagger}_{-q } ) |  a^{\dag}_{l} a_{m} \rangle \rangle - \nonumber \\
A_{q }( n - i ) e^{i \vec{q}\vec{R}_{n} } 
\langle \langle  a^{\dag}_{n} a_{j}( b_{q } +  b^{\dagger}_{-q } ) |  a^{\dag}_{l} a_{m} \rangle \rangle \Bigr ),
\end{eqnarray}
where
\begin{equation}\label{12.13}
H_{ij,rn} =  (\omega - \epsilon_{n} + \epsilon_{r}) \delta_{ni}\delta_{rj} - t_{jr} \delta_{ni}   +  t_{ni} \delta_{rj}.
\end{equation} 
We define now the zeroth-order Green functions $G^{0}_{ij,lm}$ that obey the following equations of motion
\begin{eqnarray}\label{12.14}
\sum_{nr} H_{ij,rn} G^{0}_{nr,lm} = \langle a^{\dag}_{i} a_{m} \rangle \delta_{lj} - \langle a^{\dag}_{l} a_{j} \rangle \delta_{mi}, \\
\sum_{nr} H_{rn,lm} G^{0}_{ij,nr} = \langle a^{\dag}_{i} a_{m} \rangle \delta_{lj} - \langle a^{\dag}_{l} a_{j} \rangle \delta_{mi},
\label{12.15}
\end{eqnarray}
where Eq.(\ref{12.15}) has been obtained by a differentiation with respect to the right hand side of $G^{0}_{ij,lm}.$ Using these definitions it can
be shown that
\begin{eqnarray}\label{12.16}
\sum_{nr} (\langle a^{\dag}_{s} a_{n} \rangle \delta_{rt} - \langle a^{\dag}_{r} a_{t} \rangle \delta_{sn}  ) G_{nr,lm} (\omega) = \nonumber \\
\sum_{ij} (\langle a^{\dag}_{i} a_{m} \rangle \delta_{lj} - \langle a^{\dag}_{l} a_{j} \rangle \delta_{mi}  ) G^{0}_{st,ji} (\omega) + \nonumber \\
\sum_{ijn} \sum_{q} \Bigl ( A_{q }( j - n ) e^{i \vec{q}\vec{R}_{j} } 
\langle \langle  a^{\dag}_{i} a_{n}( b_{q } +  b^{\dagger}_{-q } ) |  a^{\dag}_{l} a_{m} \rangle \rangle - \nonumber \\
A_{q }( n - i ) e^{i \vec{q}\vec{R}_{n} } 
\langle \langle  a^{\dag}_{n} a_{j}( b_{q } +  b^{\dagger}_{-q } ) |  a^{\dag}_{l} a_{m} \rangle \rangle \Bigr )G^{0}_{st,ji} (\omega)~.
\end{eqnarray}
The r.h.s. higher order Green functions can be calculated in a similar way. To proceed we approximate the electron-phonon Green function as
\begin{equation}\label{12.17}
\langle \langle  a^{\dag}_{n} a_{r} b^{\dagger}_{q } b_{q }|  B \rangle \rangle \simeq N(q) \langle \langle  a^{\dag}_{n} a_{r}|  B \rangle \rangle. 
\end{equation} 
Here $N(q)$ denotes the Bose distribution function of the phonons.\\
As a result we find
\begin{eqnarray}\label{12.18}
\sum_{nr} (\langle a^{\dag}_{s} a_{n} \rangle \delta_{rt} - \langle a^{\dag}_{r} a_{t} \rangle \delta_{sn}  ) 
\langle \langle  a^{\dag}_{n} a_{r} b_{q }| a^{\dag}_{l} a_{m} \rangle \rangle  = \nonumber \\
\omega (q) \sum_{ij} \langle \langle  a^{\dag}_{i} a_{j} b_{q }| a^{\dag}_{l} a_{m} \rangle \rangle G^{0}_{st,ji} (\omega) - \nonumber \\
\sum_{ijn} (1 + N(q)) \Bigl ( A_{- q }( j - n ) e^{- i \vec{q}\vec{R}_{j} } G_{in,lm}(\omega) - \nonumber \\
 A_{- q }( n - i ) e^{- i \vec{q}\vec{R}_{n} } G_{nj,lm}(\omega) \Bigr )G^{0}_{st,ji} (\omega) - \nonumber \\
\sum_{ij} \sum_{np} A_{- q }( n - p ) e^{- i \vec{q}\vec{R}_{n} } \Bigl ( \langle a_{p} a^{\dag}_{i} \rangle G_{nj,lm}  - 
 \langle a^{\dag}_{p} a_{j} \rangle G_{ip,lm} \Bigr ) G^{0}_{st,ji} (\omega)
\end{eqnarray}
and a similar equation for $\langle \langle  a^{\dag}_{n} a_{r} b_{-q }| a^{\dag}_{l} a_{m} \rangle \rangle. $\\ In the above equations the
Green functions $G$ and $G^{0}$ as well as the mean values $\langle a^{\dag}_{i} a_{j} \rangle $ which can be expressed by one-particle
Green functions depend on the atomic configuration. For the configuration averaging (which we will denote by $\overline{G}$)  we use 
the simplest approximation
\begin{equation}\label{12.19}
\overline{G \cdot G} \sim \overline{G} \cdot \overline{G},
\end{equation} 
i.e. in all products the configurational-dependent quantities will be averaged separately. Taking into account Eqs. (\ref{12.14}) and (\ref{12.19}), the
averaged zeroth-order Green function  $\overline{G^{0}_{ij,lm}}$ is given by the well-known CPA solution  for  two-particle Green function  in
disordered metallic alloy~\cite{veli}
\begin{equation}\label{12.20}
\overline{G^{0}_{ij,lm}} (\omega) = \frac{1}{N^{2}}\sum_{k_{1}k_{2}} e^{i\vec{k}_{1} (\vec{R}_{m} - \vec{R}_{i})}
e^{i\vec{k}_{2} (\vec{R}_{j} - \vec{R}_{l})} F_{2}(\vec{k}_{1},\vec{k}_{2}),
\end{equation} 
where $F_{2}(\vec{k}_{1},\vec{k}_{2})$ is given by
\begin{eqnarray}\label{12.21}
 F_{2}(\vec{k}_{1},\vec{k}_{2}) \approx i (E(\vec{k}_{2}) - E(\vec{k}_{1})) \int d \omega \frac{\partial f}{\partial \omega}
 [{\rm Im}( \frac{1}{\omega - \Sigma (\omega) - E(\vec{k}_{1})} )]^{2}, \\ \nonumber
 {\rm for} \quad |E(\vec{k}_{1}) - E(\vec{k}_{2}) | \ll | \Sigma (E(\vec{k}_{1})) |, \\
F_{2}(\vec{k}_{1},\vec{k}_{2}) \approx \frac{f(E(\vec{k}_{1})) -  f(E(\vec{k}_{2}))}{E(\vec{k}_{1}) - E(\vec{k}_{2})}, \\ \nonumber
{\rm for} \quad |E(\vec{k}_{1}) - E(\vec{k}_{2}) | \gg | \Sigma (E(\vec{k}_{1})) |. \label{12.22}
\end{eqnarray} 
Here $\Sigma (\omega)$ denotes the coherent potential and $f(\omega)$ is the Fermi distribution function. The configurational averaged terms 
$\overline{\langle a^{\dag}_{s} a_{n} \rangle} $ are given by
\begin{eqnarray}\label{12.23}
\overline{\langle a^{\dag}_{s} a_{n} \rangle}   = \sum_{k} e^{i\vec{k} (\vec{R}_{n} - \vec{R}_{s})}F_{1}(\vec{k}), \\ \nonumber
F_{1}(\vec{k}) = - \frac{1}{\pi} \int d \omega f(\omega){\rm Im}\Bigl ( \frac{1}{\omega - \Sigma (\omega) - E(\vec{k})} \Bigr ).
\end{eqnarray} 
After the configurational averaging equations (\ref{12.16}) and (\ref{12.18}) can be solved by Fourier transformation and we find
\begin{equation}\label{12.24}
\overline{G_{ij,lm}(\omega)}  = \frac{1}{N^{2}}\sum_{k_{1}k_{2}} \sum_{k_{3}k_{4}}e^{-i\vec{k}_{1}\vec{R}_{i}} e^{i\vec{k}_{2}\vec{R}_{j}}
e^{-i\vec{k}_{3}\vec{R}_{l}} e^{i\vec{k}_{4}\vec{R}_{m}} G(\vec{k}_{1},\vec{k}_{2};\vec{k}_{3},\vec{k}_{4}),
\end{equation} 
where
\begin{eqnarray}\label{12.25}
G(\vec{k}_{1},\vec{k}_{2};\vec{k}_{3},\vec{k}_{4}) \equiv G(\vec{k}_{1},\vec{k}_{2}) = 
F_{2}(\vec{k}_{1},\vec{k}_{2}) \delta(\vec{k}_{4},\vec{k}_{1}) \delta(\vec{k}_{3},\vec{k}_{2}) - 
\frac{F_{2}(\vec{k}_{1},\vec{k}_{2})}{F_{1}(\vec{k}_{1}) - F_{1}(\vec{k}_{2})} \cdot \nonumber \\
\sum_{q} \Bigl ( \frac{X(\vec{q},\vec{k}_{2} )G(\vec{k}_{1},\vec{k}_{2}) + 
Y(\vec{q},\vec{k}_{1},\vec{k}_{2} )G(\vec{k}_{1} - \vec{q},\vec{k}_{2} - \vec{q})}
{[F_{1}(\vec{k}_{1}) - F_{1}(\vec{k}_{2} - \vec{q}) - \omega(\vec{q})F_{2}(\vec{k}_{1},\vec{k}_{2} - \vec{q})]
(F_{2}(\vec{k}_{1},\vec{k}_{2} - \vec{q}))^{-1}}  \nonumber \\ 
+ \frac{X_{1}(\vec{q},\vec{k}_{1}, \vec{k}_{2}) G(\vec{k}_{1} - \vec{q},\vec{k}_{2} - \vec{q})   - 
Y_{1}(\vec{q},\vec{k}_{1} )G(\vec{k}_{1},\vec{k}_{2})}
{[F_{1}(\vec{k}_{1} - \vec{q}) - F_{1}(\vec{k}_{2}) - \omega(\vec{q})F_{2}(\vec{k}_{1} - \vec{q},\vec{k}_{2})]
(F_{2}(\vec{k}_{1} - \vec{q},\vec{k}_{2}))^{-1}}  \nonumber \\ 
- 2 {\rm  \quad terms \quad with} \quad \omega(\vec{q}) \rightarrow - \omega(\vec{q}),\quad  N(q) \rightarrow (-1 - N(q))\Bigr ), \qquad
\end{eqnarray} 
and
\begin{equation}\label{12.26}
A(q,k)  = \frac{1}{N}\sum_{k}e^{-i\vec{k}(\vec{R}_{i} - \vec{R}_{j})} A_{q }( i - j ).
\end{equation} 
Here the following notation were introduced
\begin{eqnarray}\label{12.27}
X(\vec{q},\vec{k}_{2} ) = A(\vec{q},\vec{k}_{2} - \vec{q})A(-\vec{q},\vec{k}_{2})(F_{1}(\vec{k}_{2} - \vec{q}) - 1 - N(q)),\\
\label{12.28}
Y(\vec{q},\vec{k}_{1},\vec{k}_{2} ) = A(\vec{q},\vec{k}_{2} - \vec{q})A(-\vec{q},\vec{k}_{1})(F_{1}(\vec{k}_{1}) + N(q)),\\
\label{12.29}
X_{1}(\vec{q},\vec{k}_{1}, \vec{k}_{2}) = A( \vec{q},\vec{k}_{1})A(- \vec{q},\vec{k}_{2} - \vec{q})(1 + N(q) - F_{1}(\vec{k}_{2})),\\
\label{12.30}
Y_{1}(\vec{q},\vec{k}_{1} ) = A( \vec{q},\vec{k}_{1})A(- \vec{q},\vec{k}_{1} - \vec{q})(F_{1}(\vec{k}_{1} - \vec{q}) +  N(q)).
\end{eqnarray} 
Equation (\ref{12.25}) is an integral equation for the Green function $G(\vec{k}_{1},\vec{k}_{2})$ to be determined.\\
The structural averaged conductivity can be obtained, in principle, by using Eq.(\ref{12.10}), where the 
Green function $\langle \langle  a^{\dag}_{i} a_{j} |  a^{\dag}_{l} a_{l} \rangle \rangle$ is to be replaced by $\overline{G_{ij,ll}(\omega)}$
as given by Eq.(\ref{12.24}). It is, however, more convenient to start with the Kubo formula in the following form~\cite{ck3}
\begin{equation}\label{12.31}
\sigma = \frac{ie^{2}}{\Omega}   \lim_{p \rightarrow 0} \sum_{k} \frac{1}{p^{2}} \left( \frac{\partial E(k)}{\partial k}p \right)
\langle \langle  a^{\dag}_{k} a_{k + p} | \eta_{- p}  \rangle \rangle_{i\varepsilon},
\end{equation} 
where $\eta_{- p} =  \sum_{k} a^{\dag}_{k} a_{k - p}$ is the electron density operator. To find the 
Green function $\langle \langle  a^{\dag}_{k} a_{k + p} | \eta_{- p}  \rangle \rangle$, the integral equation (\ref{12.25}) has to be 
solved. In general, this can be done only numerically, but we can discuss here two limiting cases explicitly. At first we consider the 
weak-scattering limit being realized for a weak disorder in the alloy, and second, we investigate the temperature coefficient of the conductivity for a 
strong potential scattering.\\
In the weak-scattering limit the CPA Green function is given by the expression
\begin{eqnarray}\label{12.32}
 F_{2}(\vec{k}_{1},\vec{k}_{2}) \approx i (E(\vec{k}_{2}) - E(\vec{k}_{1}))  \frac{d f}{d E(k_{1})} \cdot
 \frac{1}{\Sigma (E(k_{1}))},
  \\ \nonumber
 {\rm for} \quad |E(\vec{k}_{2}) - E(\vec{k}_{1}) | \ll | \Sigma (E(\vec{k}_{1})) |.
\end{eqnarray} 
Corresponding to this limit the following solution ansatz for the Green function $G(k, k+p)$ can be used
\begin{equation}\label{12.33}
G(k, k+p) = \langle \langle  a^{\dag}_{k} a_{k + p} | \eta_{- p}  \rangle \rangle_{i\varepsilon} \simeq i \Bigl ( \frac{\partial E(k)}{\partial k}p \Bigr )
\frac{d f}{d E(k)} \cdot \frac{1}{\Sigma (E(k)) + \gamma(E(k))},
\end{equation} 
where $\gamma$ describes the contribution of the electron-phonon scattering to the coherent potential. Taking into account that 
in the weak-scattering limit $|\Sigma | \ll \omega(\vec{q})$, the terms $F_{2}(k, k-q)$ in the right hand side denominators of Eq.(\ref{12.25}) can be
replaced by the expression (\ref{12.32}), and then the integral equation (\ref{12.25}) becomes for $ \lim p \rightarrow 0$
\begin{eqnarray}\label{12.34}
i   \frac{\partial f(E(k))}{\partial E(k)} \left( \frac{\partial E(k)}{\partial k}p \right)\frac{1}{\Sigma (E(k)) + \gamma(E(k))} \simeq 
i   \frac{\partial f(E(k))}{\partial E(k)} \left( \frac{\partial E(k)}{\partial k}p \right)\frac{1}{\Sigma (E(k))} - \nonumber \\ \frac{1}{\Sigma (E(k))} \frac{1}{N} \sum_{q}
A( \vec{q},\vec{k} - \vec{q})A(- \vec{q},\vec{k})[ \frac{Z_{1}(k,q) + Z_{2}(k,q)}{E(k) - E(k - q) - \omega(q) + i \varepsilon} + \nonumber \\
\frac{Z_{3}(k,q) - Z_{4}(k,q)}{E(k) - E(k - q) + \omega(q) + i \varepsilon} - \nonumber \\
2 {\rm  \quad terms \quad with} \quad \omega(\vec{q}) \rightarrow - \omega(\vec{q}),\quad  N(q) \rightarrow (-1 - N(q))]. \qquad
\end{eqnarray} 
Here the following notation were introduced
\begin{eqnarray}\label{12.35}
Z_{1}(k,q) = (f(E(k-q)) - 1 - N(q))\frac{d f}{d E(k)} \cdot \frac{( \frac{\partial E(k)}{\partial k}p )}{\Sigma (E(k)) + \gamma(E(k))}, \qquad\\
\label{12.36}
    Z_{2}(k,q) = (f(E(k)) + N(q))\frac{d f}{d E(k-q)}\cdot  \frac{( \frac{\partial E(k-q)}{\partial (k-q)}p )}{\Sigma (E(k-q)) + \gamma(E(k-q))},\qquad \\
\label{12.37}
Z_{3}(k,q) = (1 - f(E(k))+ N(q))\frac{d f}{d E(k-q)} \cdot \frac{( \frac{\partial E(k-q)}{\partial (k-q)}p )}{\Sigma (E(k-q)) + \gamma(E(k-q))}, \qquad \\
\label{12.38}
Z_{4}(k,q) = (f(E(k-q)) + N(q))\frac{d f}{d E(k)} \cdot \frac{( \frac{\partial E(k)}{\partial k}p )}{\Sigma (E(k)) + \gamma(E(k))}.\qquad
\end{eqnarray} 
Approximating the self-energy terms $\Sigma (E(k))$ and $ \gamma(E(k))$ by $\Sigma (E_{F}) \equiv \Sigma$ and $\gamma (E_{F}) \equiv  \gamma$,
respectively, the terms proportional to $\Sigma$  cancel and $\gamma$ can be calculated by
\begin{eqnarray}\label{12.39}
\gamma \frac{d f}{d E(k)} \cdot \Bigl ( \frac{\partial E(k)}{\partial k}p \Bigr )  = - \frac{\pi}{N} 
\sum_{q}A(\vec{q},\vec{k} - \vec{q})A(- \vec{q}, \vec{k})\cdot \qquad  \\
\Bigl ( [(f(E(k)) + N(q)) \frac{d f}{d E(k-q)} \cdot  \Bigl ( \frac{\partial E(k-q)}{\partial (k-q)}p \Bigr ) - \nonumber \\
(1 - f(E(k-q))+ N(q)) \frac{d f}{d E(k-q)} \cdot \Bigl ( \frac{\partial E(k)}{\partial k}p \Bigr )] \delta (E(k) - E(k - q) - \omega(q))\Bigr ) - \nonumber \\
\frac{\pi}{N} 
\sum_{q}A(\vec{q},\vec{k} - \vec{q})A(- \vec{q}, \vec{k}) \cdot \nonumber \\
\Bigl (
[(f(E(k-q)) + N(q)) \frac{d f}{d E(k)}  \cdot  \Bigl ( \frac{\partial E(k)}{\partial k}p \Bigr ) -  \nonumber \\
(1 - f(E(k))+ N(q)) \frac{d f}{d E(k-q)}  \cdot \Bigl ( \frac{\partial E(k-q)}{\partial (k-q)}p \Bigr )] \delta (E(k) - E(k - q) + \omega(q)) \Bigr ).  \nonumber
\end{eqnarray} 
Using the approximations
$$\frac{\partial E(k)}{\partial k} \simeq \frac{1}{m^{*}} k, \quad  A(\vec{q},\vec{k} - \vec{q})A(- \vec{q}, \vec{k}) \simeq A^{2}q, 
\quad q \rightarrow 0,$$
where effective mass $m^{*} = m^{*} (E_{F})$, we find
\begin{equation}\label{12.40}
\gamma = \beta \frac{\Omega}{2 \pi N}   \frac{A^{2}m^{*}}{2 (2m^{*}E_{F})^{3/2}} \int dq q^{4}\omega(q) N(q)\Bigl (1 + N(q)\Bigr )
\end{equation} 
and
\begin{equation}\label{12.41}
\gamma \sim  
\begin{cases}
 T^{5} & {\rm if}  \quad T \ll \theta_{D},  \cr
T  & {\rm if} \quad T \gg \theta_{D}.
\end{cases}     
\end{equation} 
For a binary alloy $A_{x}B_{1-x}$ with concentrations of the constituents $c_{A}$ and $c_{B}$ and the corresponding atomic energies
$\epsilon_{A}$ and $\epsilon_{B}$, in the weak-scattering limit the coherent potential is given by~\cite{chen}
\begin{equation}\label{12.42}
\Sigma = c_{A}c_{B} (\epsilon_{A} - \epsilon_{B} )^{2} D(E_{F}).
\end{equation} 
Then the conductivity becomes
\begin{equation}\label{12.43}
\sigma = \frac{e^{2}}{3(2 \pi)^{3}} \int  d k \Bigl (\frac{\partial E(k)}{\partial k}\Bigr )^{2} \frac{d f}{d E(k)}\cdot \tau,
\end{equation} 
where
\begin{equation}\label{12.44}
\tau^{-1} = \Sigma + \gamma~,
\end{equation} 
in correspondence with the Matthiessen, Nordheim and Bloch-Gruneisen rules~\cite{zi}.\\
Now we estimate temperature coefficient of the conductivity for a strong potential scattering.  For a strongly disordered alloy the
electron-phonon interaction can be considered as a small perturbation and the Green functions $G(k, k')$ on the right hand side of equation (\ref{12.25})
can be replaced by CPA Green functions $F(k, k').$ For simplicity, on the right hand side of equation (\ref{12.25}) we take into consideration only terms
proportional to the Bose distribution function giving the main contribution to the temperature dependence of the conductivity. Then
$\langle \langle  a^{\dag}_{k} a_{k_{1}} | \eta_{k - k_{1}}  \rangle \rangle$ becomes $(k_{1} = k + p \simeq k)$
\begin{eqnarray}\label{12.45}
\langle \langle  a^{\dag}_{k} a_{k_{1}} | \eta_{k - k_{1}}  \rangle \rangle = F_{2}(k, k_{1}) \Bigl ( 1 - \frac{2}{F_{1}(k ) - F_{1}( k_{1})}
\sum_{q}A(\vec{q},\vec{k} - \vec{q})A(- \vec{q}, \vec{k})N(q) \cdot  \nonumber \\    \{
 \frac{F_{2}(k, k-q)[F_{2}(k-q, k_{1}-q) - F_{2}(k, k_{1})] (F_{1}(k ) - F_{1}(k-q) )}{[F_{1}(k ) - F_{1}(k-q)]^{2} - \omega^{2}(q)F^{2}_{2}(k, k-q)} \nonumber \\
+  \frac{F_{2}(k-q, k)[F_{2}(k-q, k_{1}-q) - F_{2}(k, k_{1})] (F_{1}(k-q ) - F_{1}(k) )}{[F_{1}(k-q ) - F_{1}(k )]^{2} - \omega^{2}(q)F^{2}_{2}(k-q, k)} \} 
 \Bigr ). \qquad  
\end{eqnarray} 
Neglecting at low temperatures the terms $\omega^{2}(q)F^{2}_{2}(k_{1}-q, k_{1}) \sim q^{4}$ as compared to $[F_{1}(k_{1} ) - F_{1}(k_{1}-q )]^{2} \sim q^{2}$
and using equation (\ref{12.21}) for $\omega(q) \ll |\Sigma |$, we find for small $q$ and $p  \rightarrow 0$
\begin{equation}\label{12.46}
\langle \langle  a^{\dag}_{k} a_{k + p} | \eta_{-p}  \rangle \rangle \simeq F_{2}(k, k+p)[1 + \Bigl (\frac{d F_{1}}{d E(k)}\Bigr )^{-2}\sum_{q}
(\Delta(q,k-q) - \Delta(q,k))].
\end{equation} 
Here $\Delta(q,k)$ is the temperature-dependent correction terms to the CPA Green function are given by
\begin{equation}\label{12.47}
\Delta(q,k) = 2 A^{2} q N(q) \Bigl ( \int d \omega \frac{d f(\omega)}{d \omega} [ {\rm Im} (\frac{1}{\omega - \Sigma(\omega) - E(k)} )]^{2} \Bigr )^{2}.
\end{equation} 
For temperatures $k_B T \ll E_{F}$ we can write
\begin{equation}\label{12.48}
 \int d \omega \frac{d f(\omega)}{d \omega} S (\omega, E(k) ) \cong - S (E_{F}, E(k) )
\end{equation} 
and the conductivity becomes
\begin{equation}\label{12.49}
\sigma = \sigma_{CPA} + \Delta \sigma(T),
\end{equation} 
where
\begin{equation}\label{12.50}
 \sigma_{CPA} = \frac{e^{2}}{\Omega} \sum_{k}  \Bigl (\frac{\partial E(k)}{\partial k}\Bigr )^{2} [ {\rm Im} \Bigl (\frac{1}{E_{F} - \Sigma(E_{F}) - E(k)} \Bigr )]^{2}
\end{equation} 
is the standard CPA expression for the conductivity and
\begin{eqnarray}\label{12.51}
\Delta \sigma(T) = \frac{2e^{2} A^{2}}{\Omega} \sum_{k}  \Bigl (\frac{\partial E(k)}{\partial k}\Bigr )^{2}
\sum_{q}q N(q) \cdot \nonumber \\   \Bigl (    
[ {\rm Im}\left(\frac{1}{E_{F} - \Sigma(E_{F}) - E(k-q)} \right)]^{4} - [ {\rm Im} \left( \frac{1}{E_{F} - \Sigma(E_{F}) - E(k)} \right)]^{4} \Bigr ).  
\end{eqnarray} 
Introducing the effective mass of the electrons with $E(k) \simeq E_{F},$ the temperature-dependent correction to the conductivity becomes
\begin{equation}\label{12.52}
\Delta \sigma(T) \cong \frac{2e^{2} A^{2}}{\Omega} \frac{1}{(m^{*})^{2}} \sum_{k}\sum_{q} q^{3}  N(q) [ {\rm Im} \left(\frac{1}{E_{F} - \Sigma(E_{F}) - E(k-q)} \right)]^{4}.
\end{equation} 
Here the quantity $\Delta \sigma(T)$ is positive definite and increasing with increasing temperature. Hence, in strongly disordered alloys where the electron-phonon 
scattering is weak as compared with the disorder scattering the temperature coefficient of the resistivity is negative. It should be mentioned,
however, that the concrete temperature dependence of the correction term (\ref{12.52}) is a crude estimation only because in the derivation of  (\ref{12.52})
the influence of the disorder on the lattice vibrations has been neglected.\\
One more remark is appropriate for the above consideration. For the calculation of transport coefficients in
disordered $3d$ systems the classical approaches as the Boltzmann equation become useless if the random fluctuations
of the potential are too large~\cite{ram98,jans}. The strong potential fluctuations force the  electrons into localized
states. In order to investigate the resistivity of metallic alloys near the metal-insulator transition~\cite{ram98,jans} the corresponding
formula for the resistivity can be deduced along the line described above. For a binary transition metal alloy the
corresponding Hamiltonian is given by
\begin{equation}\label{12.53}
H = \sum_{i}\epsilon_{i} a^{\dag}_{i} a_{i}  +  \sum_{ij} t_{ij}a^{\dag}_{i} a_{j}
\end{equation} 
(with $\epsilon_{i} = \epsilon_{A}, \epsilon_{B}$ depending on the occupation of the lattice site $i$). A corresponding
integral equation for the Green function $\langle \langle  a_{j} |  a^{\dag}_{i}  \rangle \rangle_{\omega}$
can be written down. Using a simple ensemble averaging procedure and approximating the averaged Green function by
the expression
\begin{equation}\label{12.54}
\overline{\langle \langle  a_{j} |  a^{\dag}_{i}  \rangle \rangle_{\omega}} \approx \frac{1}{N} \sum_{k}
\exp[i k (R_{i} - R_{j})] \frac{1}{\omega - \epsilon_{k}} \exp \left(- \alpha (\epsilon_{k})|R_{i} - R_{j}| \right),
\end{equation} 
the integral equation transforms into an equation for the parameter $\left(\alpha (\epsilon_{k})\right)^{-1}$
which is proportional to the averaged mean free path of the electrons. It can be shown then, by solving this equation
for electrons at the Fermi surface $E_{F}$, that $\left(\alpha (E_{F})\right)^{-1}$ and the conductivity
$\sigma $ drop in a discontinued way from $\left(\alpha\right)^{-1}_{min}$ and $\sigma_{min}, $  respectively, to zero as the
potential fluctuations exceed a critical value. Note that $\left(\alpha\right)^{-1}_{min}$ is of the order $1/d,$ where
$d$ is the lattice parameter.
%
\section{Discussion}
%
In the foregoing sections we have discussed some selected statistical mechanics approaches to the calculation of the
 electrical conductivity in metallic systems
like transition metals and their disordered alloys within a model approach.\\
Electrons in metals are scattered by impurities and phonons.  The theory of transport processes for ordinary metals
was based on the consideration of various types of scattering mechanisms and, as a rule, has used the Boltzmann
equation approach. The aim of the present review was to describe an alternative approach to the calculation of
electroconductivity, which can be suitable for transition metals and their disordered alloys. 
There is an important aspect of this consideration. The approximations
used here are the tight-binding and modified tight-binding, which are admittedly not ideally precise but does
give (at least as the first approximation) reasonable qualitative results for paramagnetic transition metals and their disordered alloys.
We studied the electronic conduction in a model of transition metals and their disordered alloys utilizing the method of
generalized kinetic equations. The reasonable and workable expressions for the electrical conductivity were established 
and analyzed. 
We discussed briefly  various approaches for computing electrical conductivity as well.\\
We hope that these considerations have been done with sufficient
details to bring out their scope and workability. In this paper, we have  
considered   the idealized  Hubbard model   which
is  the simplest (in the sense of formulation, but not solution)
and most popular model of correlated lattice fermions~\cite{kuz09,rnc}.
We believe that
this technique can be applied to other model systems (e.g. multi-band Hubbard model, 
periodic Anderson model, etc.).
As it is seen, this treatment has some advantages in comparison with
the standard methods of computing electrical conductivity within the
Boltzmann equation approach, namely, the very compact form. The physical picture 
of   electron-electron and electron-phonon
scattering processes in the interacting many-particle systems is
clearly seen at every stage of calculations, which is not the
case with the standard methods. This  picture of interacting many-particle system 
on a lattice is far richer and gives more possibilities for the
analysis of phenomena which can actually  take place in real metallic systems.
We believe that our approach offers a convenient way for  approximate considerations
of the resistivity   of the   correlated electron systems on a lattice.
We believe that in view of the great difficulty of developing a first-principles microscopic theory
of transport processes in solids, the present approach is a useful alternative for description the influence of
electron-electron, electron-phonon and disorder scattering effects on the transport properties of 
transition metals and their disordered alloys.\\ 
In the confines of a review of this nature many topics of great practical and theoretical interest have
necessarily to be omitted  (see e.g. Refs.~\cite{aus07,gra1,gra2}). In recent years the field of mesoscopic physics is developed 
rapidly~\cite{jans,ihn04,dat05,du08,bu09}.
It deals with systems under experimental conditions where several quantum length scales for electrons are comparable.
The physics of transport processes in such systems is rich of quantum effects, which is typically characterized
by interplay of quantum interference and many-body interactions. It would be of interest to generalize the present approach
to quantum transport phenomena.\\ 
In conclusion, the foregoing analysis suggests that the method of the  generalized kinetic equations is an efficient 
and useful formalism for the studying of some selected transport processes in metallic systems.
\end{document}